\documentclass[reprint,superscriptaddress,amsmath,amssymb,aps,longbibliography]{revtex4-2}

\usepackage[utf8]{inputenc}
\usepackage{graphicx}

\usepackage{dsfont}
\usepackage{textcomp}
\usepackage{multirow}

\begin{document}

\title{Phase Transitions in Random Circuit Sampling}
\author{Google Quantum AI and Collaborators}

\begin{abstract}
Undesired coupling to the surrounding environment destroys long-range correlations on quantum processors and hinders the coherent evolution in the nominally available computational space. This incoherent noise is an outstanding challenge to fully leverage the computation power of near-term quantum processors~\cite{Preskill2018quantumcomputingin}. It has been shown that benchmarking Random Circuit Sampling (RCS) with Cross-Entropy Benchmarking (XEB) can provide a reliable estimate of the effective size of the Hilbert space coherently available~\cite{boixo2018characterizing,neill2018blueprint,arute_supremacy_2019,wu2021,zhu2022,liu2022validating,PhysRevLett.131.110601}. The extent to which the presence of noise can trivialize the outputs of a given quantum algorithm, i.e. making it spoofable by a classical computation, is an unanswered question. Here, by implementing an RCS algorithm we demonstrate experimentally that there are two phase transitions observable with XEB, which we explain theoretically with a statistical model. The first is a dynamical transition as a function of the number of cycles and is the continuation of the anti-concentration point in the noiseless case.
The second is a quantum phase transition controlled by the error per cycle; to identify it analytically and experimentally, we create a weak link model which allows varying the strength of noise versus coherent evolution.
Furthermore, by presenting an RCS experiment with 67 qubits at 32 cycles, we demonstrate that the computational cost of our experiment is beyond the capabilities of existing classical supercomputers, even when accounting for the inevitable presence of noise. Our experimental and theoretical work establishes the existence of transitions to a stable computationally complex phase that is reachable with current quantum processors.
\end{abstract}

\maketitle

The computational complexity of quantum systems arises from the exponential growth of the  Hilbert space dimension with system size. On near-term quantum processors whose practical complexity is limited by noise, Random Circuit Sampling (RCS) has emerged as the most suitable candidate for a beyond-classical demonstration, as it allows for quantum correlations to spread at maximum speed~\cite{boixo2018characterizing,mi2021information}. 
The interplay between computational complexity and noise is highlighted by recent RCS experiments, starting with a 53-qubit Sycamore quantum processor in 2019~\cite{arute_supremacy_2019}. Ever since, similar experiments with expanded system sizes and reduced noise have been reported~\cite{wu2021, zhu2022,shaw2023benchmarking,bluvstein2023logical}, while classical algorithms have also advanced substantially~\cite{markov2008simulating,boixo2017simulation,gray2021hyper,huang2020classical,pan2022solving,kalachev2021classical}. This intensifying quantum-classical competition motivates two questions: are there well-defined boundaries for the region where the exponentially large Hilbert space is, in fact, leveraged by a noisy quantum processor? More importantly, can we establish an experimental observable that directly probes these boundaries?

In this work, we provide direct insight to these two questions using RCS on a 2D grid of superconducting qubits.
We demonstrate that the interplay between quantum dynamics and noise can drive the system into distinct phases, whose boundaries are resolved using finite-size studies with cross-entropy benchmarking (XEB)~\cite{boixo2018characterizing,neill2018blueprint,arute_supremacy_2019,liu2021benchmarking}. Reaching the desired phase of maximized complexity requires a noise rate per cycle below a critical threshold whose value is determined by the growth rate of quantum correlations.

\begin{figure}
    \centering
    \includegraphics[]{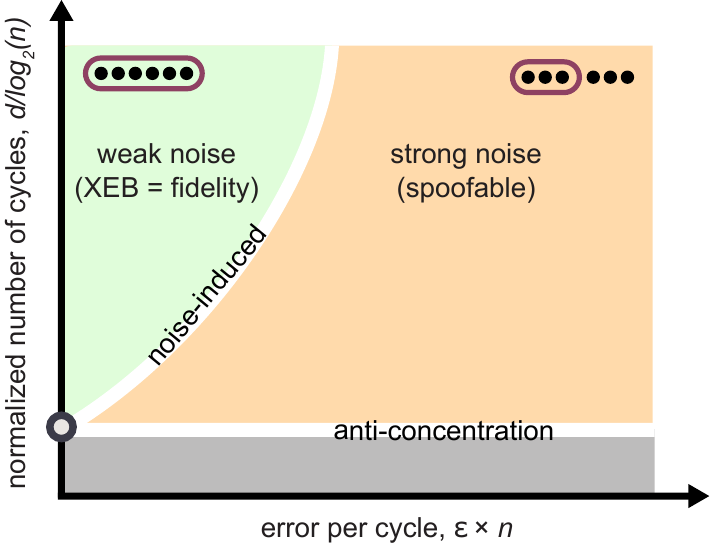} 
    \caption{\textbf{Phase transitions in Random Circuit Sampling:} One phase
transition goes between a concentrated output distribution of bitstrings from RCS at low number of cycles to a broad or anti-concentrated distribution. There is a second phase transition in a noisy system. Strong enough error per cycle induces a phase transition from a regime where correlations extend to the full system to a regime where the system may be approximately represented by the  product  of  multiple  uncorrelated  subsystems. We will show that in the strong noise regime, full-system XEB fails to give a faithful estimate of the underlying fidelity, and can be spoofable.}
    \label{fig:fig1}
\end{figure}

The structure of these phases is schematically illustrated in Fig.~\ref{fig:fig1}. Driven by the circuit number of cycles or depth, the system first goes through a dynamical phase transition, where the output distribution is no longer concentrated in a fraction of bitstrings. This suggests that the system becomes sufficiently delocalized in the computational basis for linear XEB to become a good estimator of system fidelity. Anti-concentration is a key ingredient of mathematical arguments on the complexity of simulating noiseless RCS~\cite{boixo2018characterizing,dalzell2022,bouland2019complexity,movassagh2019quantum}. Nevertheless, we will show that this is a necessary but not sufficient condition for global entanglement\,(see SM Sec. H), which maximizes computational cost.  

\begin{figure*}
    \centering
    \includegraphics{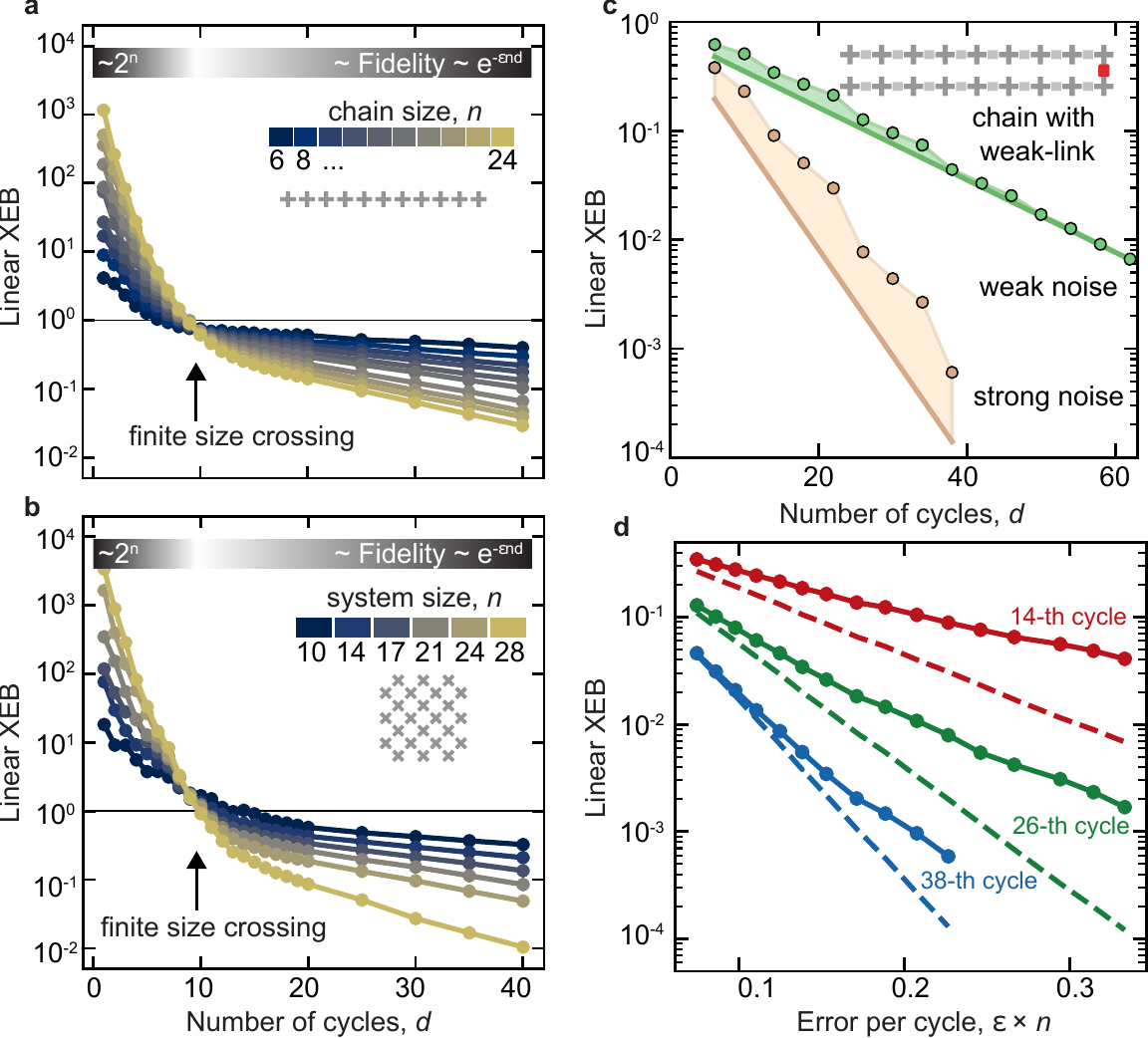} 
    \caption{\textbf{Phase transitions in the Linear Cross-entropy:} At low number of cycles, XEB grows with the size of the system. In a noiseless device, XEB will converge to 1 with the number of cycles. In the presence of noise XEB becomes an estimator of the system fidelity. In \textbf{a} and \textbf{b}, we observe experimentally a dynamical phase transition at a fixed number of cycles between these two regimes in 1D and 2D respectively. The random circuits here use Haar random single-qubit gates and an iSWAP-like gate as an entangler. In \textbf{c} and \textbf{d} we probe experimentally a noise induced phase transition using a weak-link model (see main text), where the weak link is applied every 12 cycles (discrete gate set, see main text). In \textbf{c} we show the two different regimes: in the weak noise regime, XEB converges to the fidelity, whereas in the strong noise regime, XEB remains higher than predicted by the digital error model. In \textbf{d}, we induce errors to scan the transition from one regime to the other.} 
    \label{fig:fig2}
\end{figure*}

The second is a transition driven by noise, specifically error rate per cycle $\epsilon \times n$, where $\epsilon$ is the error per gate and $n$ is the number of qubits. As illustrated in Fig.~\ref{fig:fig1}, the behavior of quantum correlations falls into two regimes: when the error rate per cycle is large, the wavefunction of the system could be approximately represented by multiple uncorrelated subsystems.
This leaves the quantum system open to spoofing by classical algorithms that represent only part of the system at a time~\cite{markov2018quantum,arute_supremacy_2019,zlokapa2020boundaries,gao2021limitations,aharonov2022polynomial}. In the regime where the error rate per cycle is sufficiently low, correlations eventually span the entire system restoring its computational complexity, and the experiment can not be spoofed (see SM Sec. F). The boundary between these two phases is determined by the competition between $\epsilon \times n$ and the convergence of the system to the ergodic state~\footnote{See Ref.\cite{ware2023sharp} for a related numerical study in the case of all-to-all connectivity.}.

We find that XEB is a proper observable to resolve the aforementioned regimes experimentally. Specifically linear XEB is measured as
\begin{equation}
    {\rm XEB} = \left\langle 2^n p_{\text{sim}}(s) - 1\right\rangle_{\rm s}\;,\label{eq:xeb-def}
\end{equation}
where $n$ is the number of qubits, $p_{\text{sim}}(s)$ is the ideal (simulated) probability of bitstring $s$ and the average is over experimentally observed bitstrings $s$. We measure XEB as a function of the number of cycles or depth $d$ for different system sizes to resolve the dynamical phase transition (Fig.~\ref{fig:fig1}). The experimental results are shown in Fig.~2a and Fig.~2b for 1D and 2D systems, respectively. We find that XEB increases with system size for small $d$, where the system wavefunction is concentrated in a fraction of basis states. However, for large $d$, XEB decays exponentially and approximates the circuit fidelity. At intermediate number of cycles we observe a crossing point where all the measured XEB curves intersect at a single point where the value of XEB is approximately independent of the system size. This behavior can be  understood in the following way: at $d=0$ the system is in a single basis state and $p_{\text{sim}}(s)$ is non-zero for a single bitstring. Here XEB equals $ 2^n - 1$, i.e. increases with $n$ exponentially. This exponential growth with $n$ is preserved at short times with a dependence $\exp(n e^{-d})$ (see SM Sec E and Ref.~\cite{dalzell2022}). At a later time the trend switches and it decays exponentially as $F^d$ where $F\approx\exp(-\epsilon n)$ is the fidelity per cycle (i.e. the digital error model~\cite{arute_supremacy_2019}).
There is a crossing point between these two trends corresponding to the number of cycles where the exponents $ e^{-d}$ and $\epsilon d$ are approximately equal.
This estimate does not distinguish between a crossover and a phase transition. The detailed theory presented in SM Sec. E shows that this is indeed a phase transition associated with the role of the boundary.

Having identified the minimum number of cycles at which XEB approximates the system fidelity, we now formulate an experimental protocol for locating the transition between the strong and weak noise regimes (Fig.~\ref{fig:fig1}). 
A conceptually simple setup that highlights the underlying physics for this transition is the so-called weak-link model, where two subsystems of size $n/2$ are coupled via an entangling gate applied every $T$ cycles. In the limit where $T = \infty$, i.e. no weak link is applied, the subsystems are uncoupled and the overall system converges to a product state $\rho_{A} \otimes \rho_{B}$, where $\rho_A$ ($\rho_B$) is the pure ergodic state of each subsystem. Adding noise, we assume the so-called depolarizing channel noise model for the density matrix of each subsystem $A/B$: $F^{d/2} \rho_{A/B} + (1 - F^{d/2}) I_{A/B}/2^{n/2}$, where $I_A$ ($I_B$) is the identity matrix.
Direct substitution of this density matrix into Eq.~(\ref{eq:xeb-def}) gives ${\rm XEB} = F^d +2F^{d/2}$, using XEB $=1$ for $\rho_{A/B}$ and XEB $= 0$ for $I_{A/B}/2^n$.

\vspace{1mm}
At finite yet large $T$ each subsystem approaches the ergodic state in less than $T$ cycles. In the average evolution of linear XEB over random circuits~\cite{mi2021information,gao2021limitations} (see SM~D), the application of each gate between the two systems entangles them with probability $1 - \lambda$. Entanglement evolves the subsystem ergodic state $\rho_A$ (or $\rho_B$) towards the overall ergodic state $\rho_{AB}$. The probability $\lambda$ depends on the entangling gate and is $1/4$ for the iSWAP-like gates employed in our experiment. Therefore, a simplified model for linear XEB is
\begin{equation}
{\rm XEB} \approx 2\lambda^{d/T} F^{d/2} + F^d.\label{eq:xeb-wl-model}
\end{equation}

We probe this behavior by measuring XEB experimentally as a function of $d$, as shown in Fig.~\ref{fig:fig2}c. Here we have employed a noise-injection protocol that effectively changes gate fidelities in our quantum circuits (see SM~C2) and show results corresponding to different noise levels. We use the discrete set of single-qubit gates chosen randomly from $Z^p X^{1/2} Z^{-p}$ with $p \in \{-1, -3/4, -1/2, \ldots, 3/4\}$. We observe that in the weak noise regime, XEB converges to the expected fidelity of the entire system, $F^d$. This is because $F$ is sufficiently high such that $F^d$ dominates the contribution to XEB. On the other hand, XEB is significantly above $F^d$ in the strong noise regime owing to the dominant contribution of $2\lambda^{d/T} F^{d/2}$ to XEB. These results are preliminary indications of the two different noise induced phases and exemplify the competition between the exponential decay of global correlations $\propto \epsilon n$, and the entangling rate between subsystems ($\propto 1/T$ in this example).
 
 \begin{figure*}
    \centering
    \includegraphics[scale=0.95]{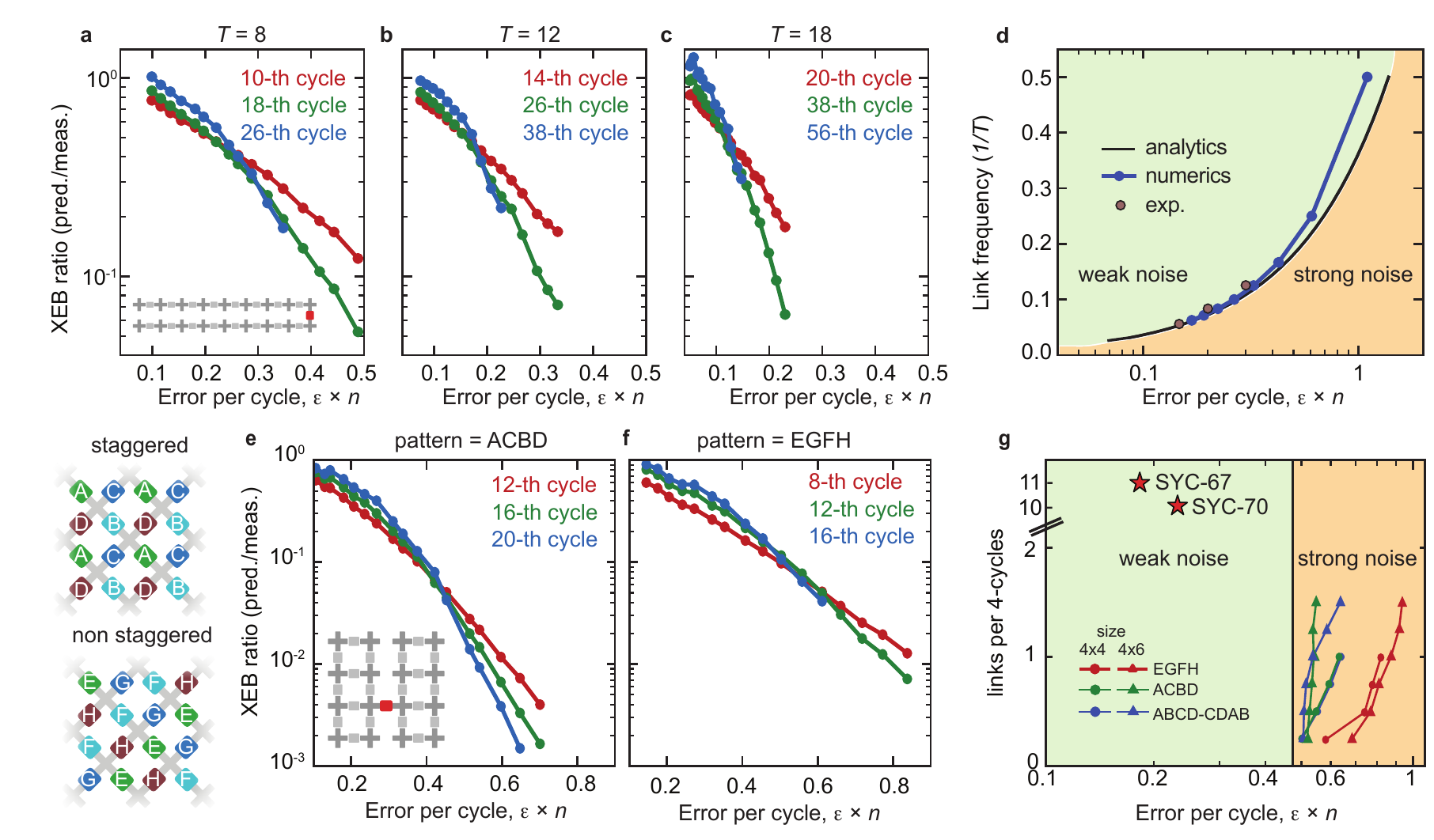} 
    \caption{\textbf{Noise induced phase transition:} \textbf{a}, \textbf{b}, \textbf{c}: Experimental noise induced phase transition as a function of the error per cycle for several periods $T$ of the weak-link model\,(discrete gate set). As $T$ increases, the critical value of noise or error per cycle become lower. \textbf{d}: Phase diagram of the transition with analytical, numerical and experimental data. The experimental data is extracted from the crossing of the largest number of cycles scan. \textbf{e}, \textbf{f}: Experimental transition in 2D with different patterns: staggered ACBD and non-staggered EGFH respectively\,(discrete gate set). \textbf{g}: Numerical phase diagram of the 2D phase transition. We show the critical point for different sizes and patterns. 
    The qubits are arranged as shown in the inset of panel\,\textbf{e}. For fixed size we vary the number of bridges (such as the red coupler in panel\,\textbf{e}) up to the point when all bridges are applied (4 and 6 for the $4\times4$ and $4\times6$ systems respectively), denoted as links per 4-cycles in the panel\,\textbf{g}. For all the patterns, we delimit a lower bound on the critical error rate of 0.47 errors per cycle to separate the region of strong noise where XEB fails to characterize the underlying fidelity and global correlations are subdominant. The experiments of Fig.~\ref{fig:fig4}, SYC-67, and of SM Sec.~C1, SYC-70, are represented by red stars and are well within the weak noise regime.} 
    \label{fig:fig3}
\end{figure*}
 
The transition between the two different noise induced phases is more clearly seen by fixing $d$ to a few values past the dynamical phase transition. We then vary the effective noise level (i.e. $F$) and measure XEB at these fixed numbers of cycles (Fig.~\ref{fig:fig2}d). We observe that XEB exhibits a nontrivial scaling distinct from $F^d$. In particular, we see that the rate of decay with respect to errors decreases at higher error rates. This is also consistent with the fact that $2\lambda^{d/T} F^{d/2}$ dominates at high errors.
 
To experimentally locate the critical value of error per cycle (or equivalently, $F$) where the dynamical exponent of XEB changes, we define a modified order parameter $F^d / \rm{XEB}$, which asymptotes to a distinct value of 1 (0) in the weak (strong) noise regime. The transition between the two limits becomes a discontinuity when $d\rightarrow \infty$, indicating a phase transition for finite $\epsilon n \approx \kappa_c$, where the critical value $\kappa_c$ may be a function of $\lambda$ and $T$. In the transition region we can observe the finite size critical behavior where $F^d / \rm{XEB}$ is approximately a function of $(\epsilon n - \kappa_c)d$. This is revealed in the order parameter for different number of cycles $d$ crossing at a single point, as can be verified from Eq.~(\ref{eq:xeb-wl-model}) and numerically for the circuits used in the experiment. The experimentally obtained $F^d / \rm{XEB}$, shown in Figs.~\ref{fig:fig3}a-c and \ref{fig:fig3}e-f, indeed manifests the expected critical behavior: for $\epsilon n \lesssim \kappa_c$ the order parameter increases as $d$ increases, whereas for $\epsilon n \gtrsim \kappa_c$ the order parameter decreases as $d$ increases. At the critical point $\epsilon n \approx \kappa_c$, the data sets cross and the order parameter is approximately independent of $d$.  We attribute the slight drift in the crossing point between different depths to potential systematic errors in the experimental estimation of $F$.

Next, we explore the mechanism underlying the noise induced phase transition. The period $T$ in the weak-link model effectively controls the rate of coherent entanglement between the two subsystems and consequently the critical error rate per cycle associated with the transition. The results are illustrated in Figs.~\ref{fig:fig3}a-c, where the period $T$ takes values of 8, 12 and 18 cycles~\footnote{The data shown is mostly for large values of the period since the phase transition point moves to larger noise for small periods. Hence, in order to measure the phase transition experimentally we needed to go to a larger error rate and therefore smaller XEB values. This would require very high precision and hence too many samples.}. We observe that the critical noise rate $\epsilon n$ decreases as we increase the value of $T$. This is intuitively expected: by increasing $T$, we reduce the entanglement generation rate between the two subsystems, making the entire system more susceptible to noise induced correlation localization. To construct the corresponding phase diagram, we extract the value of critical noise rate and plot them against $1/T$ — the link frequency in Fig.~\ref{fig:fig3}d. They are in good agreement with numerical simulations with noise, marking the phase boundary between the weak and strong noise regimes. When comparing with the functional form $\epsilon n \simeq 4/T \log2$ predicted by the analytical weak-link model, we observe appreciable deviations when the link frequency is approaching $1/2$, which corresponds to the regular 1D chain. The deviation is due to the fact that the weak-link model is no longer  applicable in this regime. A more accurate description is provided by a generalization of this model in SM Sec. E1. 

With the goal of demonstrating and verifying the beyond-classical performance, it is critical to construct a similar phase diagram for the full 2D structure to ensure that our system is in the required weak noise regime, which in turn guarantees that XEB is a proper estimate of the fidelity. However, performing the same experiment and analysis at the level of 67 qubits is classically intractable. Instead, we combine experiments with numerical analysis to give a proper bound. 
The experimental results are shown in Fig.~\ref{fig:fig3}e-f for a $4 \times 4$ square grid of qubits and two different circuit structures, whereby the two-qubit gates are applied either in a staggered (Fig.~\ref{fig:fig3}e) or a non-staggered (Fig.~\ref{fig:fig3}f) fashion. Similar to 1D, the 16-qubit system is divided into two halves that are connected by a single iSWAP-like gate applied every 4 cycles. For both circuit structures, we observe a similar crossing between $F^d / \rm{XEB}$ measured at three different cycles, with a higher value of $\epsilon n$ observed for the non-staggered patterns. 

\begin{figure*}
    \centering
    \includegraphics[scale=0.95]{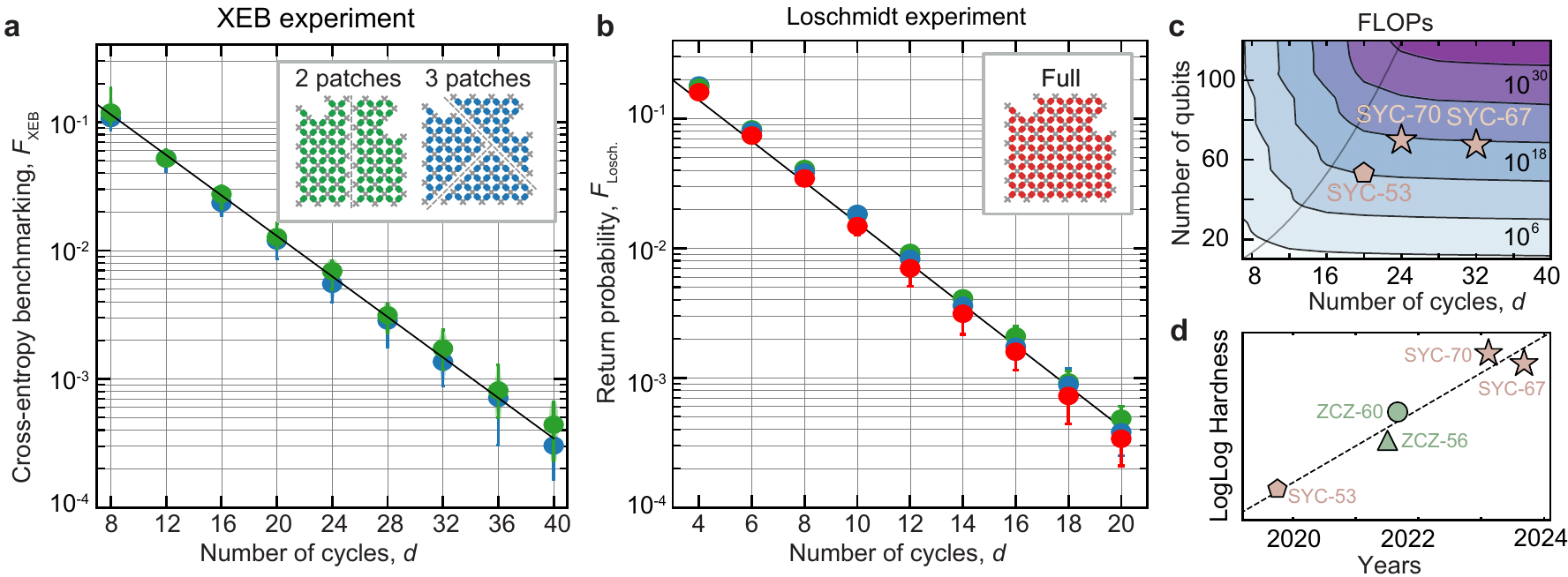}  
    \caption{\textbf{Demonstration of  a classically intractable computation:} \textbf{a}: Verification of RCS fidelity with logarithmic XEB. The full device is divided into two (green) or three (blue) patches to allow for XEB fidelity estimation with modest computational cost.  We use the discrete gate set of single-qubit gates chosen randomly from $Z^p X^{1/2} Z^{-p}$ with $p \in \{-1, -3/4, -1/2, \ldots, 3/4\}$.  For each number of cycles, 20 circuit instances are sampled with 100 thousand shots each. The solid lines indicate the estimated XEB from the digital error model.
    \textbf{b}: Verification of RCS fidelity with Loschmidt echo. The inversion is done by reversing the circuit and inserting single qubit gates. In this case, the Loschmidt echo number of cycles is doubled.
    \textbf{c}: Estimated hardness as a function of the number of qubits and the number of cycles for a set of circuits. As a working definition of hardness we consider the estimated FLOP count --number of multiplications and additions-- needed to compute the probability of a single bitstring assuming no memory constraints. 
    \textbf{d}: Evolution of the computational hardness of the RCS experiments with the dashed line representing a doubly-exponential growth to guide the eye. }
    \label{fig:fig4}
\end{figure*}

We numerically evaluate critical noise rates for systems of different sizes and circuit structures, including both the staggered and non-staggered patterns and the ABCD-CDAB pattern used in Ref.~\cite{arute_supremacy_2019}. 
Here we use Haar random single-qubit gates, see SM Sec.~D.
As illustrated in Fig.~\ref{fig:fig3}g, 2D gate patterns introduce qualitatively different behavior for quantum correlations under noise. Compared with 1D, the critical noise rates in 2D systems demonstrate a much weaker dependence on the applied link frequency, assuming a narrow range of  $\epsilon n$ between 0.5 to 0.6. This reduced sensitivity comes from the fact that in 2D, the dynamics are dominated by the bulk  within each subsystem instead of the weak link itself. This weak dependence is also observed on the system size. When increased from $4 \times 4$ (dots) to $4 \times 6$ (triangles), we found that the critical noise rate remains within the same range. 
The vertical line shown in Fig.~\ref{fig:fig3}g gives a lower bound for the noise induced phase transition. Furthermore, the noise induced phase transition for the discrete gate set used in the experiment occurs at higher noise rates, see SM Sec.~F2.
We compare this lower bound with the 67-qubit and 70-qubit RCS experiments that we will present next. We obtain the value of the error per cycle by fitting the exponential decay of the fidelity, see Fig.~\ref{fig:fig4}. It is evident that these experiments fall well within the weak noise regime, satisfying the requirement to fully utilize the computational capacity of the noisy quantum processors.

We can describe both phase transitions mentioned above with a map to a model in statistical mechanics\,(see SM Sec.~E). On the one hand these transitions correspond to qualitatively different configurations. On the other hand, as shown in SM Sec.~E, all regimes correspond to an ordered phase and differ by a small number of excitations. The noise induced phase transition is driven by a control parameter (analogous to a magnetic field in an Ising model) that scales with the system size (number of qubits). This is loosely alike to Freederiks transitions in liquid crystals \cite{degenne1993physics}
~\footnote{Note that the transition discussed here is qualitatively different from the quantum to classical transition discussed in Ref.~\cite{Aharonov2000}. The transition discussed here is a competition between the finite rate of convergence to the overall ergodic state and the fidelity per cycle. The transition in Ref.~\cite{Aharonov2000} is a competition between local interactions and the error rate per qubit.}.

In Fig.~\ref{fig:fig4}, we demonstrate beyond-classical RCS by performing an experiment on a 67-qubit Sycamore chip. These random circuits follow the same 2-dimensional ABCD-CDAB pattern as Ref.~\cite{arute_supremacy_2019}, and the single-qubit gates are chosen randomly from the discrete set $Z^p X^{1/2} Z^{-p}$ with $p \in \{-1, -3/4, -1/2, \ldots, 3/4\}$.
We show in SM~B the fidelity of the elementary operations of the random circuit. On average, we achieve a readout error of $1.3(0.5)\times 10^{-2}$, a dressed two-qubit Pauli error rate of $5.5(1.3)\times 10^{-3}$ that can be further decomposed into a single-qubit Pauli error rate of $1.0(0.5)\times 10^{-3}$, and an intrinsic two-qubit simultaneous error rate of $3.5(1.4)\times 10^{-3}$ \footnote{An intrinsic Pauli error rate of $3.5\times 10^{-3}$ corresponds to an average fidelity of $99.72 \% $}.
We validate the digital error model by looking at patched variations of the random circuit\,(inset in Fig.~\ref{fig:fig4}a), where slices of two-qubit gates have been removed, creating patched circuits for which each patch XEB can be verified at modest computational cost. The total fidelity is then the product of the patch fidelities. Computing XEB over full circuits is currently an intractable classical task.
We thus estimate the fidelity after 32 cycles using the discrete error model, obtaining 0.1\%. We have collected over 70 million sample bitstrings for a single circuit at this depth. In SM~C1 we report fidelities for another XEB experiment, SYC-70, done on 70 qubits and with 24 cycles. In Fig.~\ref{fig:fig4}b, we verify these extracted fidelities with Loschmidt echo, where we use the same circuit and its inversion to return to the initial state. We observe good agreement with the XEB experiment and digital error model.

Finally, we estimate the equivalent classical computational cost of RCS with the tensor network contraction method~\cite{markov2008simulating,boixo2017simulation,chen2018classical,markov2018quantum,villalonga2019flexible,gray2021hyper,huang2020classical,pan2022solving,kalachev2021classical,kalachev2021classical,liu2022validating}. 
With this method, Ref.~\cite{pan2022solving} classically sampled  the RCS experiment performed in 2019~\cite{arute_supremacy_2019} in 15 hours using 512 GPUs.
Ref.~\cite{kalachev2021classical} also performed this task with a similar cost.
Furthermore, another team computed in Ref.~\cite{liu2022validating} the corresponding XEB confirming the predictions of Ref.~\cite{arute_supremacy_2019}, which is a harder computational task than noisy sampling.
Hence, these remarkable improvements in classical algorithms significantly raised the beyond-classical threshold. 
For completeness, in SM Sec.~H we also study Matrix Product States (MPS), a popular tensor network variational representation of 1D quantum states with limited entanglement~\cite{white1993density,vidal2003efficient,ayral2022density}. We find that given current supercomputer memory constraints MPS fails to reach the experimental fidelity, and offers worse performance than tensor network contraction.

\begin{table}[]
    \begin{tabular}{|c|c|c|c|c|}
        \hline
        \multirow{2}{*}{\textbf{Exp.}} & \textbf{1 amp.}  & \multicolumn{3}{|c|}{\textbf{1 million noisy samples}} \\
        \cline{2-5}
                                            & FLOPs             & FLOPs             & XEB fid.   & Time        \\ \hline \hline
        SYC-53 \cite{arute_supremacy_2019}  & $6 \times 10^{17}$ & $2 \times 10^{17}$ & $2  \times 10^{-3}$   & 6 s                \\ \hline
        ZCZ-56 \cite{wu2021}                & $6 \times 10^{19}$ & $6 \times 10^{19}$ & $6  \times 10^{-4}$   & 20 min             \\ \hline
        ZCZ-60 \cite{zhu2022}               & $1 \times 10^{21}$ & $1 \times 10^{23}$ & $3  \times 10^{-4}$   & 40 days            \\ \hline
        SYC-70                              & $5 \times 10^{23}$ & $6 \times 10^{25}$ & $2  \times 10^{-3}$   & 50 yr              \\ \hline
        \multirow{3}{*}{SYC-67}             & \multirow{3}{*}{$2 \times 10^{23}$} & $2 \times 10^{37}$ & \multirow{3}{*}{$1 \times 10^{-3}$}   & $1 \times 10^{13}$ yr             \\
        & & $2 \times 10^{28}$ & & $1 \times 10^{4}$ yr$^{*\,}$
                  \\
        & & $2 \times 10^{25}$ & & $12$ yr$^{**}$ \\\hline
    \end{tabular}
    \caption{
        \textbf{Estimated computational cost of simulation:}
        The second column shows the FLOP count needed for the computation of a single output amplitude  assuming no memory constraints.
        The last three columns refer to the cost of the simulation of noisy sampling of 1 million bitstrings.
        We use the specifications of Frontier for our estimates, with a theoretical peak performance of $1.685 \times 10^{18}$ single precision FLOP per second.
        We assume a 20\% FLOP efficiency~\cite{gray2021hyper,huang2020classical,pan2022solving} and account for the low target fidelity of the simulation in the computational cost~\cite{markov2018quantum,villalonga2019flexible,gray2021hyper,huang2020classical}.
        Each single precision complex FLOP requires 8 machine FLOPs. 
        SYC-67 refers to the result from Fig. \ref{fig:fig4}, and we include the cost estimated assuming distributing memory over all RAM$^{*}$ or even all secondary storage$^{**}$, ignoring realistic bandwidth constraints. All other rows use tensor contraction algorithms embarrassingly  parallelizable  over  each  GPU~\cite{chen2018classical,huang2020classical,pan2022solving}.  
        SYC-70 refers to the results from SM Sec. G.
    }
    \label{table:cost}
\end{table}

We report improvements in tensor network contraction techniques, which result in lower estimated computational costs for simulated RCS (see SM~G).
In Fig.~\ref{fig:fig4}c we show the FLOP count (the number of multiplications and additions) as a function of number of qubits and cycles required to compute a single amplitude at the output of a random circuit without memory constraints. The FLOP count in this context is rigorously defined as the minimum contraction cost per amplitude~\cite{boixo2017simulation}. 
This serves as a proxy lower bound for the hardness of both sampling and verification. For a fixed number of qubits and increasing number of cycles, there is a notional crossover in the scaling of the FLOP count from exponential to linear.
Given a noisy experimental setup, this implies an optimal number of cycles for the trade off between computational hardness and fidelity: beyond the crossover, fidelity decreases faster than the hardness increases.
The crossover number of cycles is consistent with a scaling $\sqrt{n}$, as indicated with a dashed line.
Note that this is related to the number of cycles at which ``typical'' entanglement is achieved (see SM~H) and is a stronger requirement than the anti-concentration of the output distribution.
For both 67 and 70 qubits, 24 cycles is deep enough to saturate the exponential growth of this FLOP count. Fig.~\ref{fig:fig4}d shows the growth of the FLOP count without memory constraints, or computational ``hardness'', over the last few years.

A practical estimate of the computational resources to simulate RCS needs to take into account the finite FLOPS (FLOPs per second) computational efficiency of a supercomputer as well as its memory constraints and other limitations such as finite bandwidth. Table~\ref{table:cost} shows estimates of the runtime for the approximate simulation of the largest instances of RCS from Refs.~\cite{arute_supremacy_2019,wu2021,zhu2022} and the current work when using the state-of-the-art methods discussed in SM~G.
In these estimates, we consider sampling 1 million uncorrelated bitstrings at a fidelity similar to that of the experiment using the current top-performing supercomputer, Frontier, with $1.7 \times 10^{18}$ single precision FLOPS of theoretical peak performance spread across GPUs with 128 GB of RAM each.
This requires the computation of 10 million approximate probability amplitudes of uncorrelated bitstrings to be used in rejection sampling~\cite{markov2018quantum}.
Despite remarkable progress achieving tensor contraction algorithms embarrassingly parallelizable over each GPU~\cite{chen2018classical,huang2020classical}, we find that this technique breaks down for the SYC-67 much deeper circuits with 32 cycles given the tight memory constraints.
As a result, an estimated lower bound for the sampling cost becomes substantially more prohibitive.
Assuming a distributed use of all RAM$^{*}$ and under the unrealistic assumption of negligible bandwidth constraints the computational cost is around $6 \times 10^4$ years, see Table~\ref{table:cost}.
In the untested case that we expand working memory to all secondary storage$^{**}$ and still ignoring bandwidth we obtain an estimate of $12$ years.

In conclusion, our experiment provides direct insights on how quantum dynamics interacts with noise. The observed phase boundaries lay out quantitative guidance to the regimes where noisy quantum devices can properly leverage their computational power. In addition, we present new RCS results with an estimated fidelity of $1.5 \times 10^{-3}$ at 67 qubits and 32 cycles or 880 entanglement gates, corresponding to more than doubling the circuit volume over \cite{arute_supremacy_2019} for the same fidelity. The fact that global correlations dominate XEB in the weak noise phase protects RCS against ``spoofing'' attacks, in contrast to Boson-Sampling~\cite{aaronson2011computational}, where all known metrics for recent experiments~\cite{zhong2020,zhong2021,madsen2022} are dominated by local correlations~\cite{villalonga2021efficient}. Looking forward, despite the success of RCS in quantifying the available coherent resources, finding practical applications for near-term noisy quantum processors still remains an outstanding challenge. Certified randomness generation~\cite{aaronson2018,bassirian2021certified,aaronson2023certified} could be a promising candidate for such an application\,(see SM I).

\subsection*{Author contribution:} 
Y.~Chen and S.~Boixo led the overall project. A.~Morvan and X.~Mi performed the experiment. A.~Bengtsson contributed to readout developments. X.~Mi and P.~V.~Klimov contributed to gate developments. B.~Villalonga developed and ran the tensor network contraction optimizations. S.~Mandr\`a developed and ran the MPS simulations. A.~Morvan, Z.~Chen, S.~Hong, C.~Erickson, P.~V.~Klimov and I.~K.~Drozdov contributed to large-system calibration and stability improvements. I.~Aleiner and K.~Kechedzhi developed theories and performed numerical simulations of phase transitions. R. Movassagh proved the XEB for approximate tensor representations. J.~Chau, G.~Laun, R.~Movassagh, L. T.A.N. Brand\~ao and R. Peralta worked on certified randomness. A.~Asfaw provided technical program management to the project. All authors contributed to building the hardware and software infrastructures and writing the manuscript. 

\subsection*{Acknowledgment:}
We acknowledge insightful discussions with Carl Miller on certified randomness and Kevin Jeffery Sung for his work on randomness extractors.  S.~Mandr\`a is partially supported by the Prime Contract No. 80ARC020D0010 with the NASA Ames Research Center and acknowledges funding from DARPA under IAA 8839. 

\newpage
\onecolumngrid

\vspace{1em}
\begin{flushleft}
{\small Google Quantum AI and Collaborators}

\bigskip
{\small
\renewcommand{\author}[2]{#1$^\textrm{\scriptsize #2}$}
\renewcommand{\affiliation}[2]{$^\textrm{\scriptsize #1}$ #2 \\}

\newcommand{\corrauthora}[2]{#1$^{\textrm{\scriptsize #2}, \ddagger}$}
\newcommand{\corrauthorb}[2]{#1$^{\textrm{\scriptsize #2}, \mathsection}$}

\newcommand{\xGoogle}{\affiliation{1}{Google Research}}

\newcommand{\xNASA}{\affiliation{2}{Quantum Artificial Intelligence Laboratory, NASA Ames Research Center, Moffett Field, California 94035, USA}}

\newcommand{\xKBR}{\affiliation{3}{KBR, 601 Jefferson St., Houston, TX 77002, USA}}

\newcommand{\xUCONN}{\affiliation{4}{Department of Physics, University of Connecticut, Storrs, CT}}

\newcommand{\xNIST}{\affiliation{5}{National Institute of Standards and Technology (NIST), USA}}

\newcommand{\xUMass}{\affiliation{6}{Department of Electrical and Computer Engineering, University of Massachusetts, Amherst, MA}}

\newcommand{\xAU}{\affiliation{7}{Department of Electrical and Computer Engineering, Auburn University, Auburn, AL}}

\newcommand{\xCQCT}{\affiliation{8}{QSI, Faculty of Engineering and Information Technology, University of Technology Sydney, NSW, Australia}}

\newcommand{\xUCR}{\affiliation{9}{Department of Electrical and Computer Engineering, University of California, Riverside, CA}}

\newcommand{\xHU}{\affiliation{10}{Department of Chemistry and Chemical Biology, Harvard University, Cambridge, MA, USA}}

\newcommand{\xUOCA}{\affiliation{11}{Department of Physics and Astronomy, University of California, Riverside, CA}}

\newcommand{\Google}{1}
\newcommand{\NASA}{2}
\newcommand{\KBR}{3}
\newcommand{\UCONN}{4}
\newcommand{\NIST}{5}
\newcommand{\UMass}{6}
\newcommand{\AU}{7}
\newcommand{\CQCT}{8}
\newcommand{\UCR}{9}
\newcommand{\HU}{10}
\newcommand{\UOCA}{11}

\corrauthora{A. Morvan}{\Google},
\corrauthora{B. Villalonga}{\Google},
\corrauthora{X. Mi}{\Google},
\corrauthora{S. Mandr\`a}{\Google,\! \NASA,\! \KBR},
\author{A. Bengtsson}{\Google},
\author{P. V.~Klimov}{\Google},
\author{Z. Chen}{\Google},
\author{S. Hong}{\Google},
\author{C. Erickson}{\Google},
\author{I.~K.~Drozdov}{\Google,\! \UCONN},
\author{J. Chau}{\Google},
\author{G. Laun}{\Google},
\author{R. Movassagh}{\Google},
\author{A. Asfaw}{\Google},
% ...
\corrauthorb{K. Kechedzhi}{\Google},
\corrauthorb{Y. Chen}{\Google},
\corrauthorb{S. Boixo}{\Google},

\bigskip

\xGoogle
\xNASA
\xKBR
\xUCONN
\xNIST
\xUMass
\xAU
\xCQCT
\xUCR
\xHU
\xUOCA

{${}^\ddagger$ These authors contributed equally to this work.}\\
{${}^\mathsection$ Corresponding author: boixo@google.com}\\
{${}^\mathsection$ Corresponding author: bryanchen@google.com}\\
{${}^\mathsection$ Corresponding author: kostyantyn@google.com}

}
\end{flushleft}

\twocolumngrid

\break
\newpage

\bibliography{references}

\end{document}


\title{Supplement to Phase transition in Random Circuit Sampling}
\author{Google Quantum AI and Collaborators}

\maketitle
\tableofcontents

\appendix

\section{General RCS with XEB theory}
\label{app:rcs_distribution}

We show in this appendix that, under quite general conditions (see Eq.~\eqref{eq:chi_avg}), the effect of noise in XEB can be approximated as a global depolarizing channel. We use this to write an XEB estimator from any smooth and O(1) function $f(p_j)$ of the ideal probabilities $p_j$. 
 
It is non-trivial but true that the density of probabilities from a Haar random pure quantum state is uniform in the probability simplex~\cite{wootters1990random,caves_measures,petz2004asymptotics}
\begin{align}
  dP(p_1,\ldots,p_D) = (N-1)!\, dp_1 \cdots dp_D\;,~\label{eq:psimplex} 
\end{align}
where $D=2^n$ for $n$ qubits. The corresponding marginal distribution for any one probability $p_j$ is the Porter-Thomas (exponential or beta) distribution~\cite{boixo2018characterizing}. That is, for all $j$ we have
\begin{align}
dP(p_j) & = (D-1)(1-p_j)^{D-2} dp_j \label{eq:beta} \\ &\to D e^{-D p_j} dp_j \label{eq:PT}\;.
\end{align}
In the previous expression the bitstring index $j$ is fixed, and the distribution is over quantum states sampled uniformly in Hilbert space (Haar measure). 

One can sample a vector of the probabilities corresponding to a Haar random pure quantum state by sampling $D$ probabilities according to the distribution of Eq.~\eqref{eq:PT}, and then normalizing the result so that $\sum_j p_j = 1$~\cite{wootters1990random,caves_measures,zyczkowski2001induced}. Note that the sum of the independent $p_j$ is already $\sum_j p_j = 1 + \bigO(1/\sqrt D)$ before normalization. That is, for large $D$ the normalization introduces a small correlation between the previously independent $p_j$ that can be typically ignored. 

Approximate sampling of a random quantum circuit can be described by the probabilities
\begin{align}\label{eq:exp_mixture2}
    p^F_j = F p_j + (1-F) \Xi_j
\end{align}
where $F$ corresponds to the fidelity, $p_j$ is the ideal or simulated probability for the $j$th bitstring output of the quantum circuit, and $\Xi_j$ is a function over bitstrings corresponding to the effect of noise. In the quantum case, $\rho$ is the output of an experiment, $p^F_j = \bra j \rho \ket j$, $F = \bra \psi \rho \ket \psi$ where $\ket \psi$ is the ideal noiseless output, and $\Xi$ is defined by the equation $\rho = F \ket \psi\! \bra \psi + (1-F) \Xi$. Note that $\sum_j \Xi_j = 1$. For simplicity we sometimes denote $\Xi_j = 1/D$, the global depolarizing channel. 

In cross-entropy benchmarking (XEB) we use the expectation value of a random variable $f(p_j)$, which is defined as a function of the ideal probabilities $p_j$. That is, we associate the real value $f(p_j)$ to each sampled bitstring $\ket j$. We require $f(p_j)$ to be $\bigO(1)$ and $f$ smooth. For linear XEB $f(p_j) = D p_j -1$ and for log XEB $f(p_j) = \log(D p_j) + \text{Euler constant}$~\cite{arute_supremacy_2019}. In the following we assume that the output distribution is sufficiently close to the Porter-Thomas distribution, see Refs.~\cite{boixo2018characterizing,arute_supremacy_2019} and below.

The expectation value of $f(p_j)$ when sampling with noisy probabilities $p^F_j$ is 
\begin{align}
    \sum_j p^F_j f(p_j) = F \sum_j p_j f(p_j) + (1-F) \sum_j \Xi_j f(p_j)\;.\nonumber
\end{align}
The sum in the left hand side is an expectation value estimated with RCS sampling, within a statistical error $O(1/\sqrt k)$ where $k$ is the size of the sample. We explain below how to obtain the value of the two sums in the right hand size analytically for large circuits. Therefore solving for $F$ we obtain an estimator of the fidelity for any function $f(p_j)$ as specified above. 
 
Consider first the term $p_j f(p_j)$.  The expectation value over random circuits for fixed $j$ is
\begin{align}
    \aavg{p_j f(p_j)} = D\int_0^\infty  dp\, e^{- D p}\,  p f(p)\;,
\end{align}
where $\aavg{\cdot}$ is the average over random circuits.
Note that from the assumptions on $f$ above it also follows that $\aavg{p_j f(p_j)}$ is $\bigO(1/D)$.
Furthermore, the variance over random circuits for fixed $j$ is 
\begin{align}
    \var(p f(p)) \in \bigO\left( \frac 1 {D^2} \right)\;.
\end{align}
We saw above that the probabilities $p_j$ are almost independent. Treating the sum over $j$ as a sum of independent and identically distributed (i.i.d.) random variables, we have, by the central limit theorem,
\begin{align}
    \sum_j p_j f(p_j) = D^2\int_0^\infty  dp\, e^{- D p}\,  p f(p)+ \bigO\left( \frac 1 {\sqrt D} \right) \label{eq:ideal_pt}
\end{align}

Now we consider the term $\Xi_j f(p_j)$. We assume that, when averaged over random circuits for fixed $j$, the random variables $\Xi_j$ and $f(p_j)$ are independent. Therefore
\begin{align}
    \aavg{\Xi_j f(p_j)} = \aavg{\Xi_j} \aavg{f(p)}\;,
\end{align}
where
\begin{align}
    \aavg{f(p)} = D \int_0^\infty  dp\, e^{- D p}\,  f(p)\;.
\end{align}
We also assume that
\begin{align}
    \aavg{\Xi_j} &\in \bigO\left( \frac 1 D \right) \label{eq:chi_avg}.
\end{align}
Therefore
\begin{align}
    \var(\Xi_j f(p_j)) \in \bigO \left(\frac 1 {D^2} \right)\;.
\end{align}
Treating again the sum over $j$ as a sum of i.i.d. variables we obtain
\begin{align}
    \sum_j \Xi_j f(p_j) &= \aavg{f(p)} \sum_j \aavg{\Xi_j} +  \bigO\left( \frac 1 {\sqrt D} \right) \\
    &= \aavg{f(p)}  +  \bigO\left( \frac 1 {\sqrt D} \right) \label{app:chi_concentration}\;,
\end{align}
where we used $\sum_j \Xi_j =1$. We conclude that the averaged effect of noise $\Xi_j$ can be approximated as a totally depolarizing channel.  

For linear XEB we have $f(p) = Dp-1 $ and therefore
\begin{align}
 \sum_j p_j f(p_j) &= D^2\int_0^\infty  dp\, e^{- D p}\,  p(Dp-1) +  \bigO\left( \frac 1 {\sqrt D} \right) \nonumber \\ &= 1+  \bigO\left( \frac 1 {\sqrt D} \right) \\
 \sum_j \Xi_j f(p_j) &= D\int_0^\infty  dp\, e^{- D p}\,  (Dp-1) +  \bigO\left( \frac 1 {\sqrt D} \right) \nonumber \\ &= 0+  \bigO\left( \frac 1 {\sqrt D} \right)\;.
\end{align}
We obtain the same result for log XEB $f(p_j) = \log(D p_j) + \text{Euler constant}$~\cite{arute_supremacy_2019}. Therefore we have
\begin{align}
    F &\simeq \avg{Dp-1}_{\rm experiment} \\ 
    &\simeq \avg{\log(D p_j) + \text{Euler constant}}_{\rm experiment}\;.~\label{eq:XEB-lin-log}
\end{align} 

\begin{figure}[t!]
    \centering
    \includegraphics[width=0.85\columnwidth]{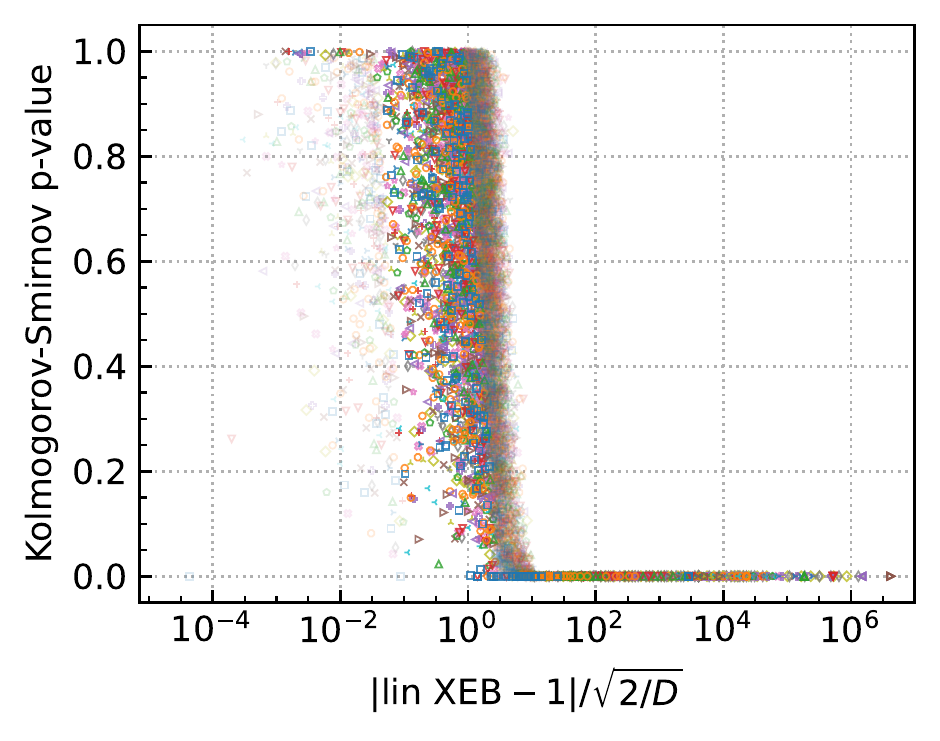} 
    \caption{The $y$-axis is the Kolmogorov-Smirnov $p-$value
between the probabilities $p_j$ at a given depth and the Porter-Thomas
distribution. The $x$ axis is the distance of linear XEB to the ideal value in units of standard deviation. 
    Each point corresponds to a different circuit size (with number of qubits ranging from $n=8$ to $n=25$) for a given fixed depth.
    Lighter points correspond to datapoints outside the $90\%$ two-sided confidence interval. For
    all the instances, the pattern ABCDCDAB is used.}
    \label{fig:xeb_scaling}
\end{figure}

We now check numerically at what depth the output distribution becomes Porter-Thomas. Figure~\ref{fig:xeb_scaling} shows that the probabilities $p_j$ truly follow a Porter-Thomas distribution
 (as measured by the Kolmogorov-Smirnov test) only if the linear XEB is exponentially close to its limit value. The scaling in the x axis comes from the variance (see also Ref.~\cite{arute_supremacy_2019})
\begin{equation}
    {\rm Var}\left(\textrm{lin XEB}\right) \simeq D^2\, {\rm Var}\left(p_j^2\right) =
    \frac{2}{D}. 
\end{equation}
Nevertheless, we find numerically and experimentally that XEB serves as an estimator of fidelity before this point, and closer to the transition point in Fig.~1a of the main text, as we don't require exponential $O(1/\sqrt{D})$ precision for this estimation.

\section{Device characterization and benchmarking}
    \subsection{Gate Optimization}\label{sec:gate_optimization}

\begin{figure}
    \centering
    \includegraphics[width=\columnwidth]{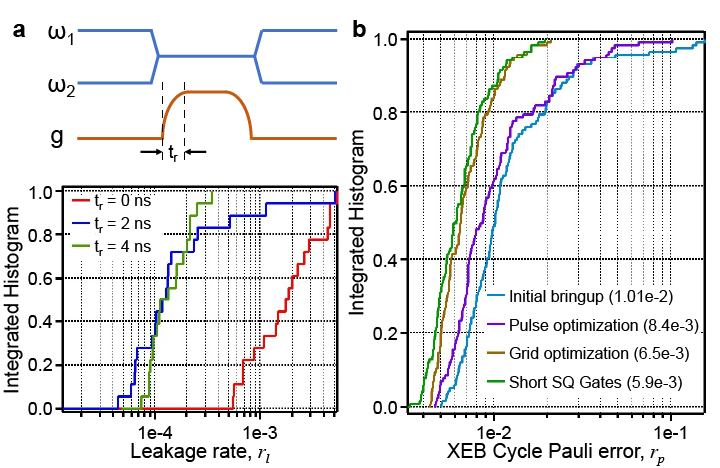} 
    \caption{Gate fidelity optimizations. (A) Upper panel: Schematic showing the flux pulses which detune the qubit frequencies ($\omega_1$ and $\omega_2$) and the inter-qubit coupling $g$ during the iSWAP-like gate. A cosine filter with a rise time $t_\text{r}$ is applied to the pulse on $g$. Lower panel: Integrated histograms of leakage per iSWAP-like gate, measured with three different values of $t_\text{r}$. Each histogram includes an identical set of 19 qubit pairs. (B) Integrated histograms of two-qubit Pauli error per cycle (which includes contributions from two single-qubit gates and one iSWAP-like gate) obtained from parallel XEB taken after different optimization steps indicated by the legend. Each histogram includes all qubit pairs on the quantum device. The median values of different histograms are quoted within the parentheses of the legend.}
    \label{fig:gate_opt}
\end{figure}

The gate fidelities of the quantum processor are carefully optimized through a series of steps. The first step involves shaping of the flux pulses used to realize the iSWAP-like gates, schematically shown in Fig.~\ref{fig:gate_opt}A. Here the computational states of two qubits, $\ket{10}$ and $\ket{01}$, are brought into resonance by pulsing the qubit frequencies $\omega_1$ and $\omega_2$ to nearly identical values. An inter-qubit coupling $g$ is then pulsed to a maximum value of $g_\text{max} \sim -13$ MHz over $t_\text{p} = 20$ ns to enable a complete population transfer from $\ket{10}$ to $\ket{01}$. 

An important error channel for such a two-qubit gate is the off-resonant oscillation between the $\ket{11}$ and $\ket{02}$ (as well as $\ket{20}$) states, which may result in appreciable leakage outside the computational space at the end of the pulses. One possible strategy for mitigating leakage is through simultaneous optimization of $g_\text{max}$ and $t_\text{p}$ such that the minima in leakage and iSWAP angle errors are synchronized  \cite{barends_iswap_2019}. However, due to the spread in qubit anharmonicities, such an optimization needs to be done for each individual qubit pair and is therefore a time-consuming process. An alternative method is to increase the rise time of the coupler pulse such that the transitions $\ket{11} \leftrightarrow \ket{02}$ and $\ket{11} \leftrightarrow \ket{20}$ are both adiabatic, thereby eliminating the need for synchronization. The leakage rates per iSWAP gate $r_l$, measured using a method adapted from Floquet calibration and applied to the two-exitation subspace \cite{Neill_Nature_2021}, are shown in the bottom panel of Fig.~\ref{fig:gate_opt}A. We observe that for short rise times in the coupler pulse ($t_\text{r} = 0$ ns), $r_l$ in excess of $10^{-3}$ is observed for most qubit pairs. The leakage rate is suppressed as $t_\text{r}$ is increased to 2 ns, although outlier qubit pairs with $r_l > 10^{-3}$ are still observed. For $t_\text{r} = 4$ ns, all pairs tested show $r_l < 4 \times 10^{-4}$. We therefore employ $t_\text{r} = 4$ ns for experiments described in this work.

The pulse shape optimization of the iSWAP-like gates has led to a reduction in two-qubit cycle Pauli errors $r_p$ in parallel two-qubit XEB from an initial median value of $1.01 \times 10^{-2}$ to $8.4 \times 10^{-3}$, as shown in Fig.~\ref{fig:gate_opt}B. In the same plot, we show two additional optimization steps that have further improved gate fidelities: By optimizing qubit frequency placements on the 2D grid~\cite{klimov2020snake} to mitigate cross-talk and coupling to two-level system (TLS) defects, we reduce $r_p$ to $6.5 \times 10^{-3}$. Finally, $r_p$ is reduced to only $5.9 \times 10^{-3}$ by shortening the execution time for the single-qubit gates from 25 ns to 18 ns.

\begin{figure}
    \centering
    \includegraphics[width=\columnwidth]{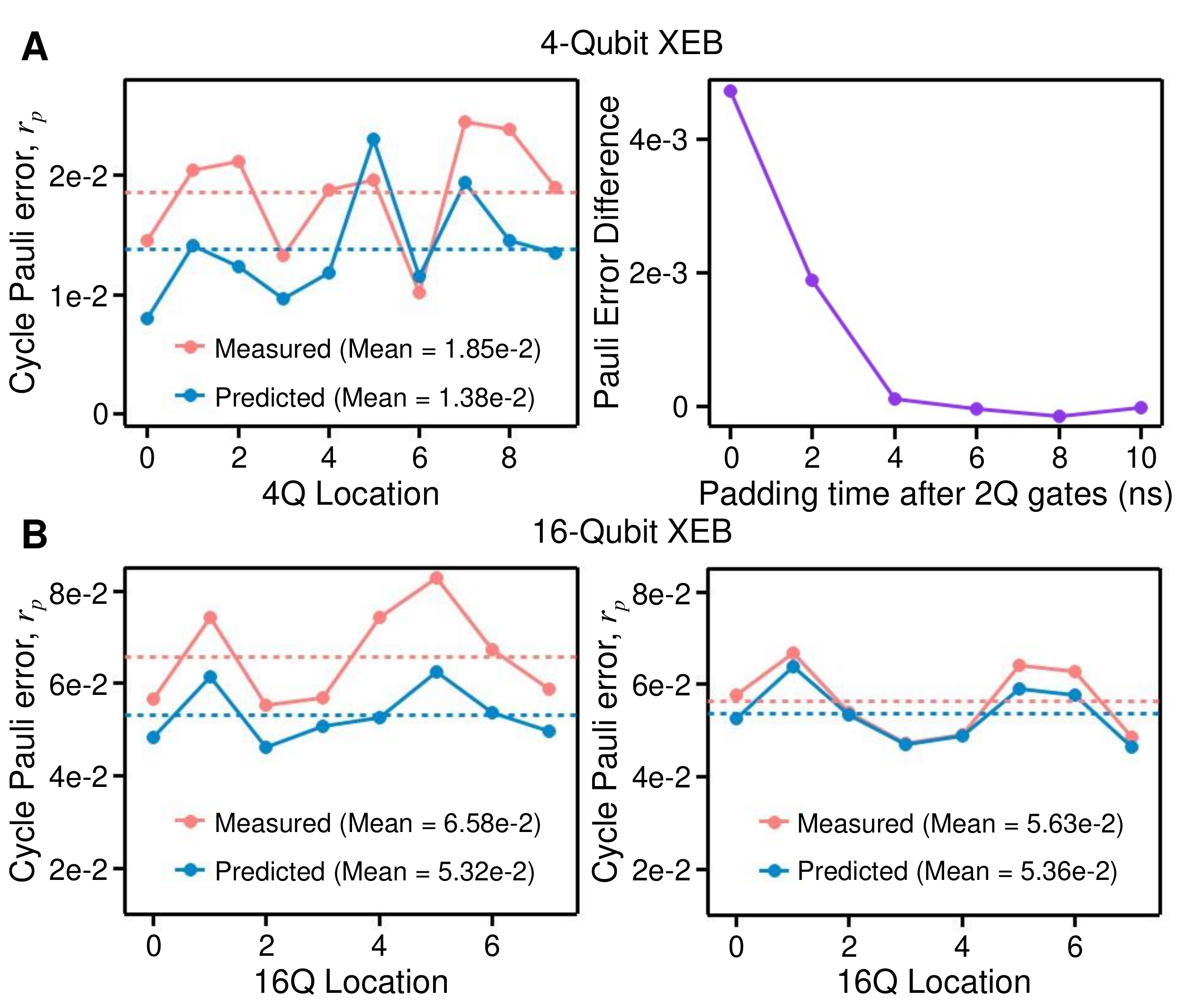} 
    \caption{Mitigating impact of z-tails. (A) Left panel: Comparison between the cycle Pauli error of a 4-qubit XEB experiment and the prediction from parallel 2-qubit XEB experiments. Horizontal axis corresponds to different 4-qubit choices. Dashed line indicates the mean values of the measured and predicted errors. Right panel: Mean difference between the measured and predicted 4Q XEB cycle errors as a function of padding times after the iSWAP-like gates. (B) Left panel: Comparison between the cycle Pauli error of a 16-qubit XEB experiment and the prediction from parallel 2-qubit XEB experiments. Horizontal axis corresponds to different 16-qubit choices. Dashed line indicates the mean values of the measured and predicted errors. Right panel: Same as the left panel but with median qubit detunings during the iSWAP-like gate reduced from 80 MHz to 40 MHz.}
    \label{fig:z_tails}
\end{figure}

After minimizing the cycle errors in two-qubit parallel XEB experiments, we benchmark performance of larger system sizes by performing a 4-qubit XEB experiment on ten different choices of 4 qubits across the quantum processor. We detect a substantial difference between the measured four-qubit cycle error and the predicted four-qubit cycle error based on two-qubit XEB measurements, as shown in the left panel of Fig.~\ref{fig:z_tails}A. The average 4-qubit cycle errors are over 30\% higher than predicted values. Through further characterizations, this discrepancy is understood to be arising from distortions in qubit flux pulses which lead to a slow settling of the qubit frequencies even after the pulses have nominally ended (a.k.a. ``z-tails''). To mitigate the impact of z-tails, we pad the moments between the two-qubit gates and single-qubit gates in the random circuits by an idling time. The right panel of Fig.~\ref{fig:z_tails}A shows the average difference between the measured and predicted four-qubit XEB cycle errors as a function of the padding time, where we observe that a padding time of 4 ns is sufficient to reduce the difference to nearly 0. This additional padding time has been applied to experiments described in this work.

Having reached agreements between two-qubit and four-qubit XEB experiments, we compare two-qubit parallel XEB predictions with 16-qubit XEB experiments. The initial result is shown in Fig.~\ref{fig:z_tails}B, where we again find that the measured 16-qubit XEB cycle errors are 24\% higher than predictions, even with padding between single- and two-qubit gates. To reduce this discrepancy, we have re-optimized the qubit frequency placements and reduced the detunings of the qubits during the iSWAP-like gates by a factor of two. The 16-qubit parallel XEB cycle errors measured after this qubit frequency re-optimization agrees closely with the predicted values from two-qubit parallel XEB measurements.
    \subsection{Benchmarking of gates and readout}
\label{supp:benchmarking}

In order to construct the error model for a random circuit, we use several experiments to predict the error rate of each element. The single qubit error is calibrated through Randomized Benchmarking using only $\pi/2$-pulses Fig. \ref{fig:supp_benchmarking}A. For the RB, we use 5 different depths logarithmically spaced up to a thousand Clifford, with 10 random Clifford circuit instances with 600 repetition per number of cycles and circuit. The two-qubit dressed error is measured through parallel XEB after optimizing for a phased-fSim model Fig. \ref{fig:supp_benchmarking}C. We used 20 random circuit instances, with 10 linearly spaced depths up to 150 cycles. The readout error is measured by preparing a random bitstring state and measuring the probability of wrong labeling of each qubit Fig. \ref{fig:supp_benchmarking}B. The error is averaged between the measurement error of the state $\ket{0}$ and $\ket{1}$. Finally, $T_1$ and $T_2$ echo, used for the idling on the edges of each patches, is measured through a standard population decay experiment and echo measurement Fig. \ref{fig:supp_benchmarking}D and E. respectively. Finally Fig. \ref{fig:supp_benchmarking}F shows the improvement over the results from \cite{arute_supremacy_2019} with the dashed line reporting the average fidelity achieved at the time. Every aspect of the experiment has improved, with a notable contribution from readout fidelity.

\begin{figure*}
    \centering
    \includegraphics[width=\textwidth]{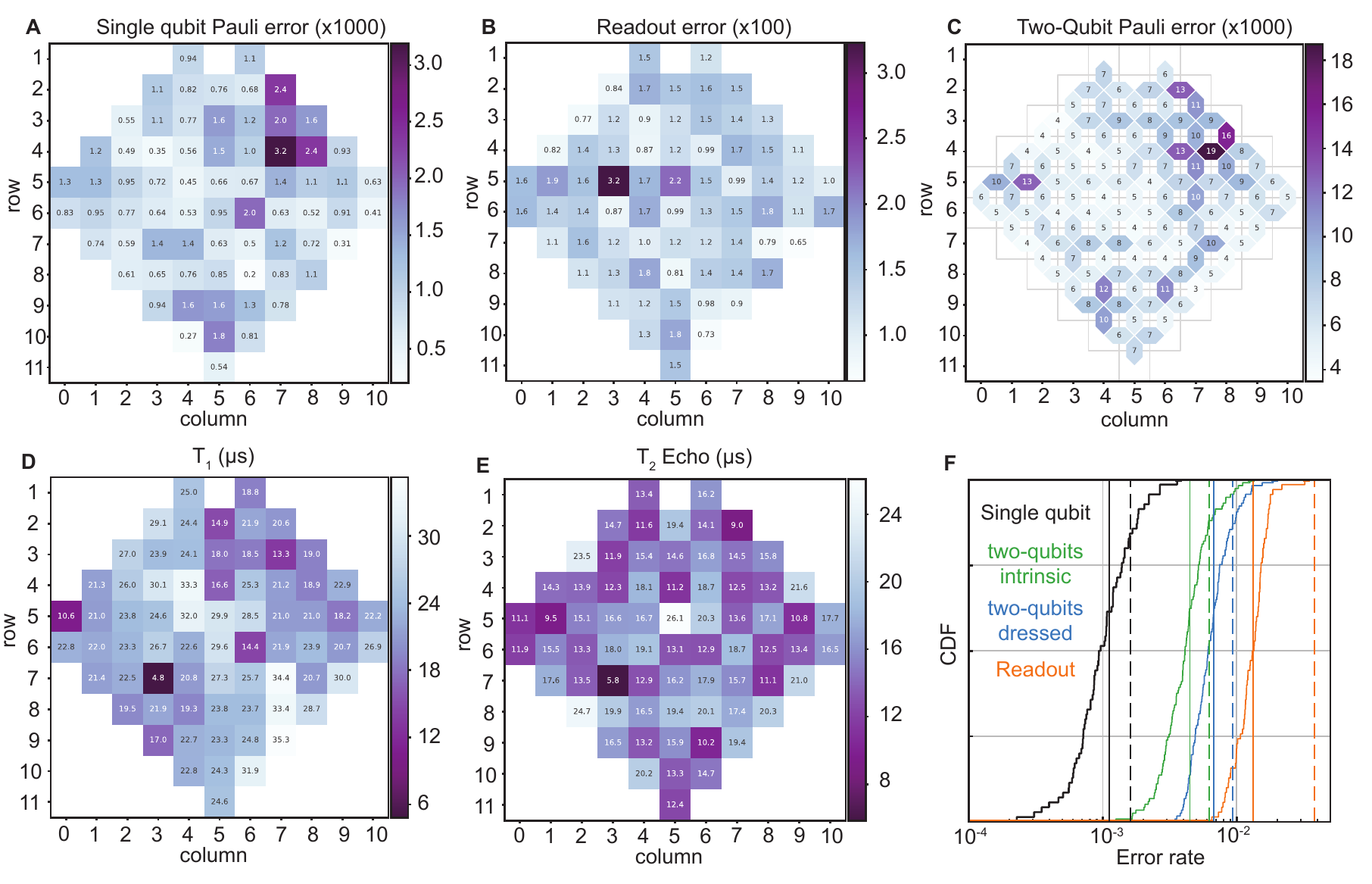} 
    \caption{\textbf{Benchmarking of the device for SYC-70}: Benchmarking of the random circuits elements. \textbf{A}: Single qubit Pauli error rate measured with Randomized Benchmarking. \textbf{B}: Readout error rate measured by preparing random bitstrings and averaging the errors over the bitstrings. \textbf{C}: Two qubit Pauli error rate measured with parallel 2-qubit XEB.
    \textbf{D} and \textbf{E}: $T_1$ and $T_2$ echo times. \textbf{F} CDF of the different errors, continuous vertical line is the average, and dashed line corresponds to the average from Ref.~\cite{arute_supremacy_2019}.}
    \label{fig:supp_benchmarking_syc70}
\end{figure*}

\begin{figure*}
    \centering
    \includegraphics[width=\textwidth]{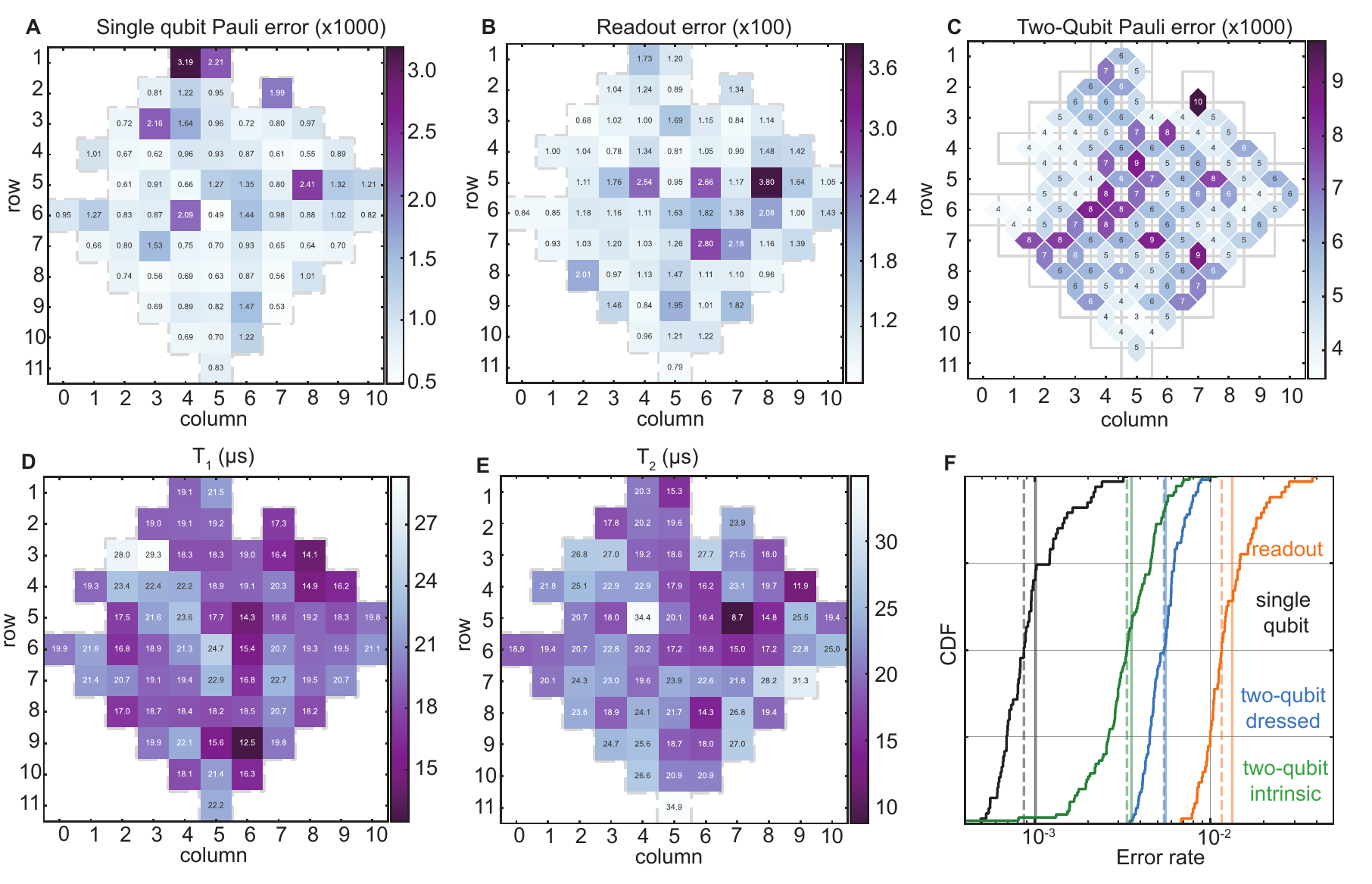} 
    \caption{\textbf{Benchmarking of the device for SYC-67}: Benchmarking of the random circuits elements. \textbf{A}: Single qubit Pauli error rate measured with Randomized Benchmarking. \textbf{B}: Readout error rate measured by preparing random bitstrings and averaging the errors over the bitstrings. \textbf{C}: Two qubit Pauli error rate measured with parallel 2-qubit XEB.
    \textbf{D} and \textbf{E}: $T_1$ and $T_2$ echo times. \textbf{F} CDF of the different errors, continuous vertical line is the average, and dashed line corresponds to median.}
    \label{fig:supp_benchmarking}
\end{figure*}

\begin{figure*}
    \centering
    \includegraphics{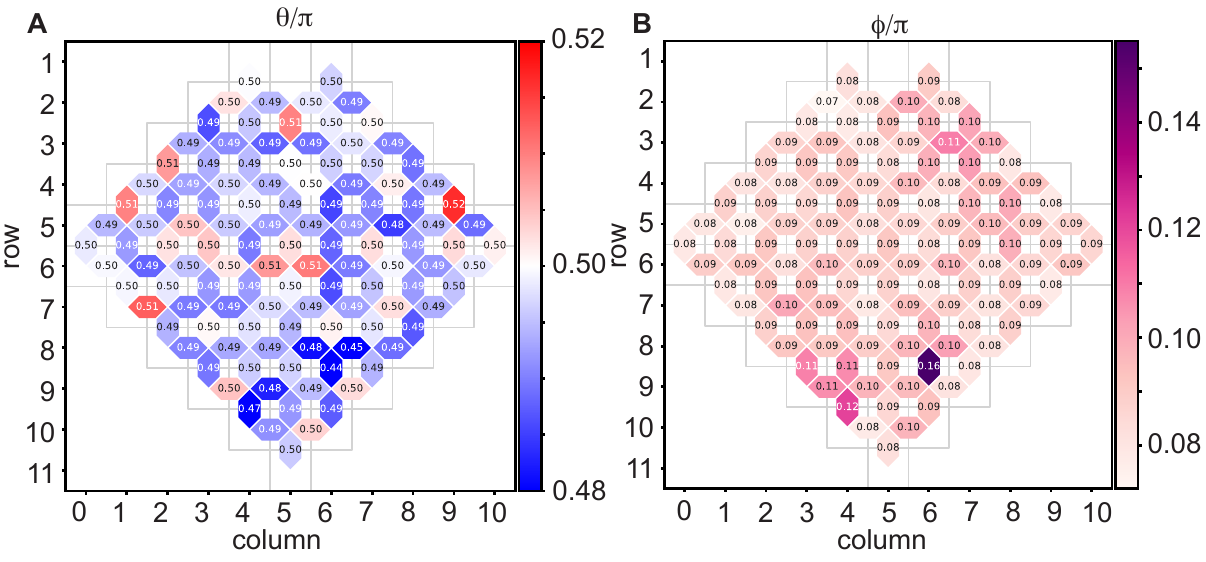} 
    \caption{\textbf{iSWAP-like characterization}: Measured angles of the iSWAP-like gate. On average, the angles are $\theta = 0.495(0.009) \times \pi$ and $\phi = 0.09(0.01) \times \pi$
    }
    \label{fig:supp_iswaplike_angles}
\end{figure*}

In Fig.~\ref{fig:supp_iswaplike_angles}, we report the angles of the iSWAP-like gate measured with parallel XEB. We note that the c-phase of the gate is now closer to $\pi/10$ compared to $\pi/6$ in \cite{arute_supremacy_2019}.

\section{Additional experimental data}
    \subsection{RCS experiment on a 70 qubits device: SYC-70}
\begin{figure*}
    \centering
    \includegraphics[width=\textwidth]{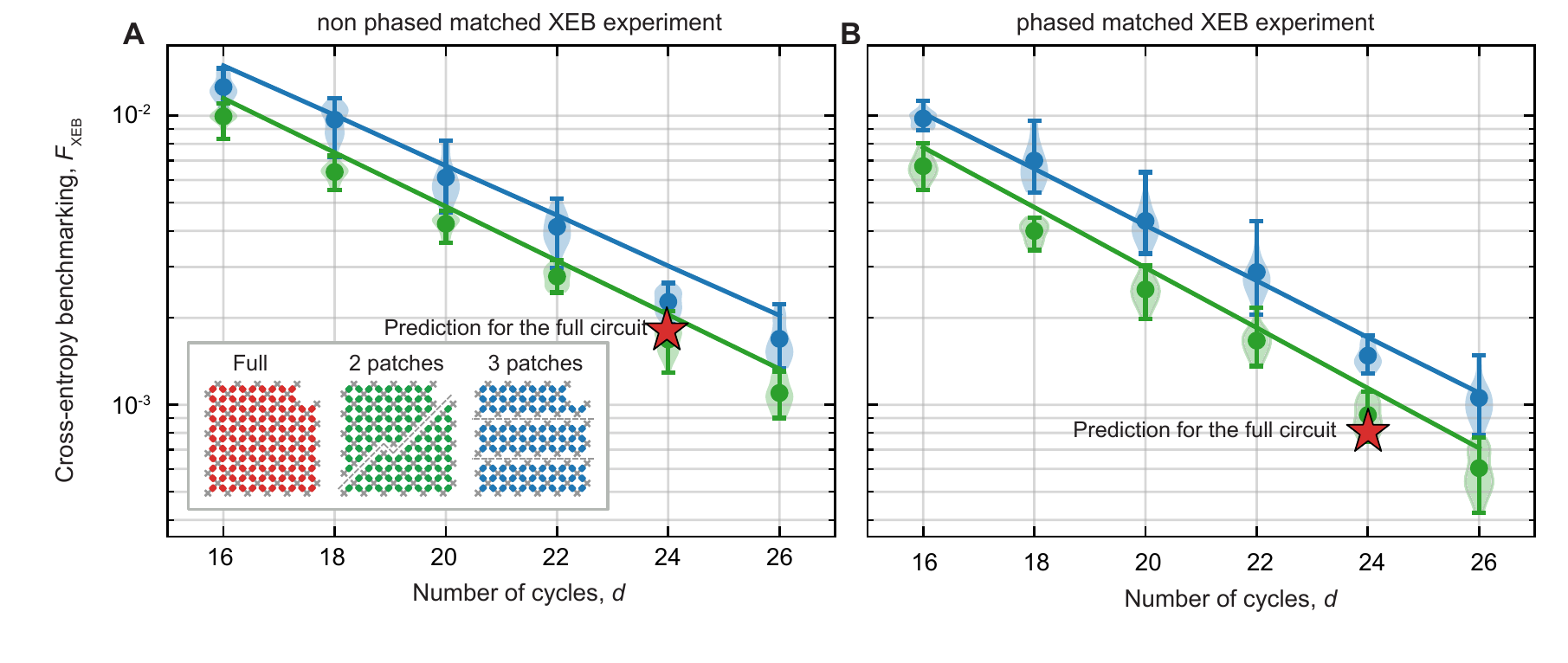} 
    \caption{\textbf{Random Circuit Sampling experiment SYC-70}: On a device with 70 qubits, we performed RCS experiments on 3 and 2 patches. Line indicate the predictions from the average error rate obtained by parallel XEB. \textbf{A} shows the data without phase matching (see text for more details) and \textbf{B} shows data with phase matching, similar to the SYC-67 experiment presented in the main text.  The difference between the two-patch and the three-patch fidelities is explained by the larger error rate of the two-qubit gates compared to the idling of the qubits for which two-qubit gates have been removed.
    }
    \label{fig:supp_phase_matching}
\end{figure*}

In this section, we present the RCS experiment on a 70-qubits device characterized in Fig.~\ref{fig:supp_benchmarking_syc70}. Figure~\ref{fig:supp_phase_matching} shows experiments without phase matching (similar to \cite{arute_supremacy_2019}) and with phase matching (similar to what is presented in the main text of this manuscript). In the Table I of the main text we report the estimated fidelity for the experiment without phase matching.

When performing a two-qubit gate, the actual unitary applied to the qubits differs from the ideal fSim by extra single-qubit Z-rotations from two sources: 1) the qubits are detuned during the gate, and 2) the qubit interaction Hamiltonian terms are not time-independent but rather oscillate due to the frequency difference between the qubits. The rotations arising from (1) do not depend on the time when the gate is applied, but the rotations arising from (2) do depend on this time, with a time-dependent phase $\gamma(t) = 2\pi(f_1 - f_0)t$.

When running quantum circuits, we typically implement Z-rotations as ``virtual'' gates by changing the phase of applied microwave pulses. This is equivalent to a circuit-level transformation where Z gates are pushed through the circuit by commuting them past other gates. The extra Z rotations associated with fSim gates can also be handled in this way by compiling them into the pulse sequence; in this case we say the fSim gates are ``phase-matched''. We can also ignore these extra Z rotations when compiling and then account for them in simulation by applying the appropriate time-dependent unitary for each gate instead of the ideal fSim unitary; in this case we say the fSim gate is not phase-matched. Note that Z-rotations only commute through fSim when $\theta$ is $0$ or $\pi/2$, that is, for c-phase-like or iSWAP-like gates. If the gate is not exactly iSWAP-like, then commuting Z-rotations through it for phase-matching introduces some error, which we can see in the slightly lower XEB fidelity when using phase-matched gates.

    \subsection{Adding noise}
\label{supp:add_noise}

\begin{figure}[t!]
    \centering
    \includegraphics[width=\columnwidth]{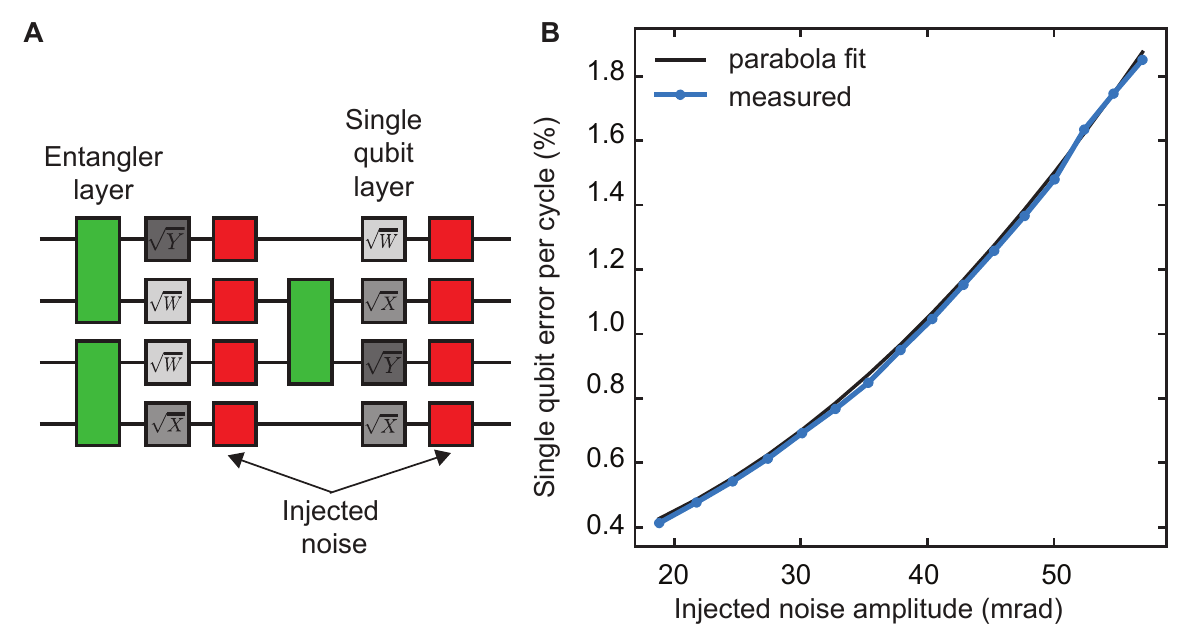} 
    \caption{\textbf{Controlling error rate.} \textbf{A}: Single qubit gates with random rotations are added after each single qubit layer. \textbf{B}: Single qubit error rate with the added random rotations measured with single qubit XEB. The error rate follows a parabolic law.}
    \label{fig:supp1}
\end{figure}

In order to probe the noise induced phase transition, we artificially increase the single qubit error rate by adding random rotations after each layer of single qubit gates in the circuit run on the hardware. The random single qubit gates are of the form:
\begin{equation}
    U = Z^{z} Z^{a}X^{x}Z^{-a}
\end{equation}
where $z$ and $x$ are sampled from a normal distribution centered on zero and with a standard deviation given by the injected noise amplitude. The axis $a$ is randomly sampled from a normal distribution centered on $-1$ with a standard deviation of $1$. In order to avoid correlated noise, the random gates are different from layer to layer in a single circuit and from circuit to circuit. These extra gates are not used in the classical simulation.  Figure~\ref{fig:supp1}~A shows the insertion of the random single qubit gates is done on each single qubit layer of a random circuit. In Figure \ref{fig:supp1} B we verify that adding these extra single qubit gates results in an average noise that scales as the square of the error angle, as expected.

    \subsection{Noise phase transition extended data}
\begin{figure*}
    \centering
    \includegraphics[width=\textwidth]{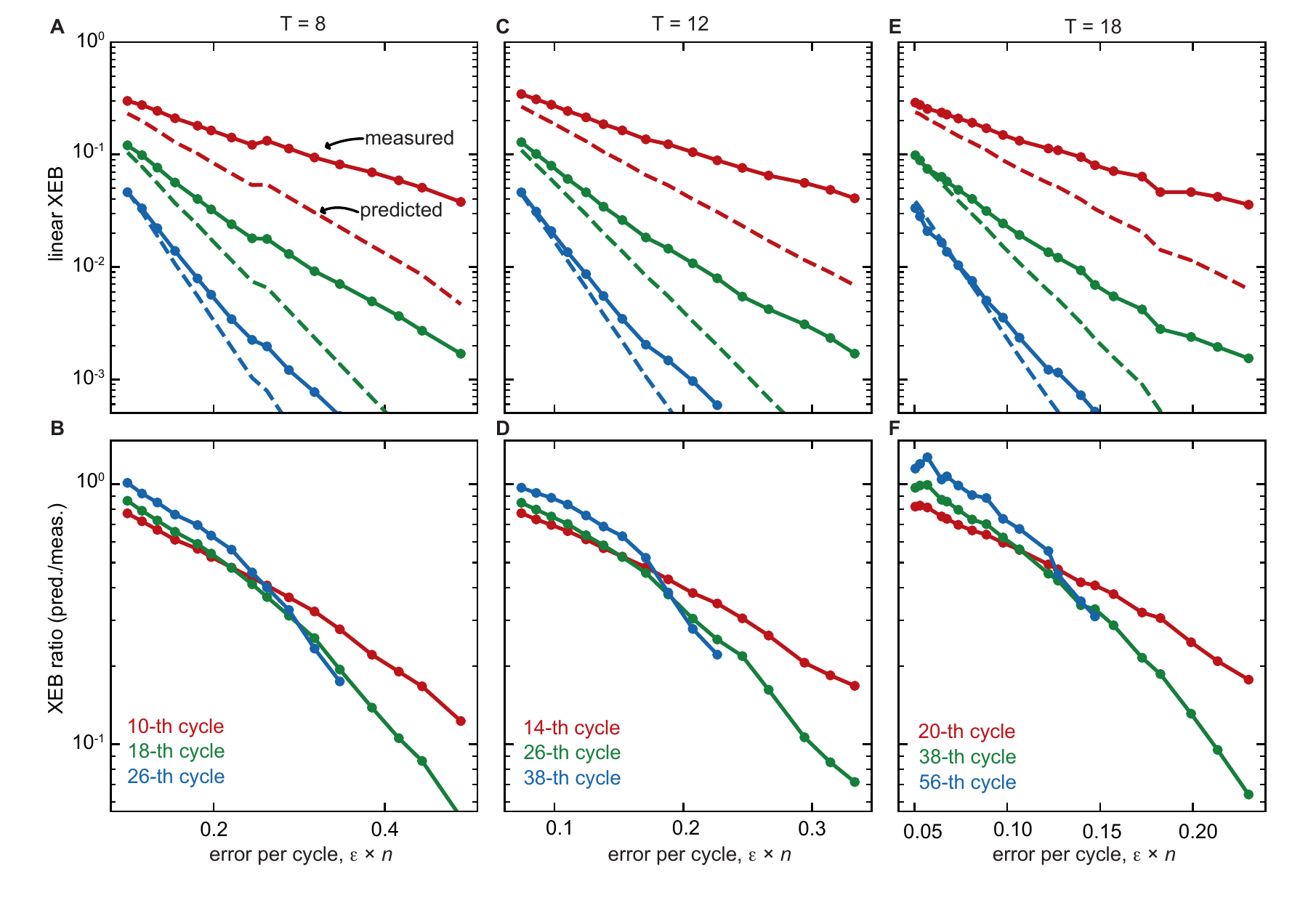} 
    \caption{\textbf{Weak-link model}: See main text Fig.~3 for more details. The first row shows the  measured XEB value as a function of the error per cycle. In the strong noise regime, the measured XEB value is far from the expected value, whereas in the weak noise regime, and sufficient depth, the measured value is correctly predicted by the component fidelity of the circuit. The second row shows the XEB ratio.
    }
    \label{fig:supp_weaklink_extended_data}
\end{figure*}

\begin{figure*}
    \centering
    \includegraphics{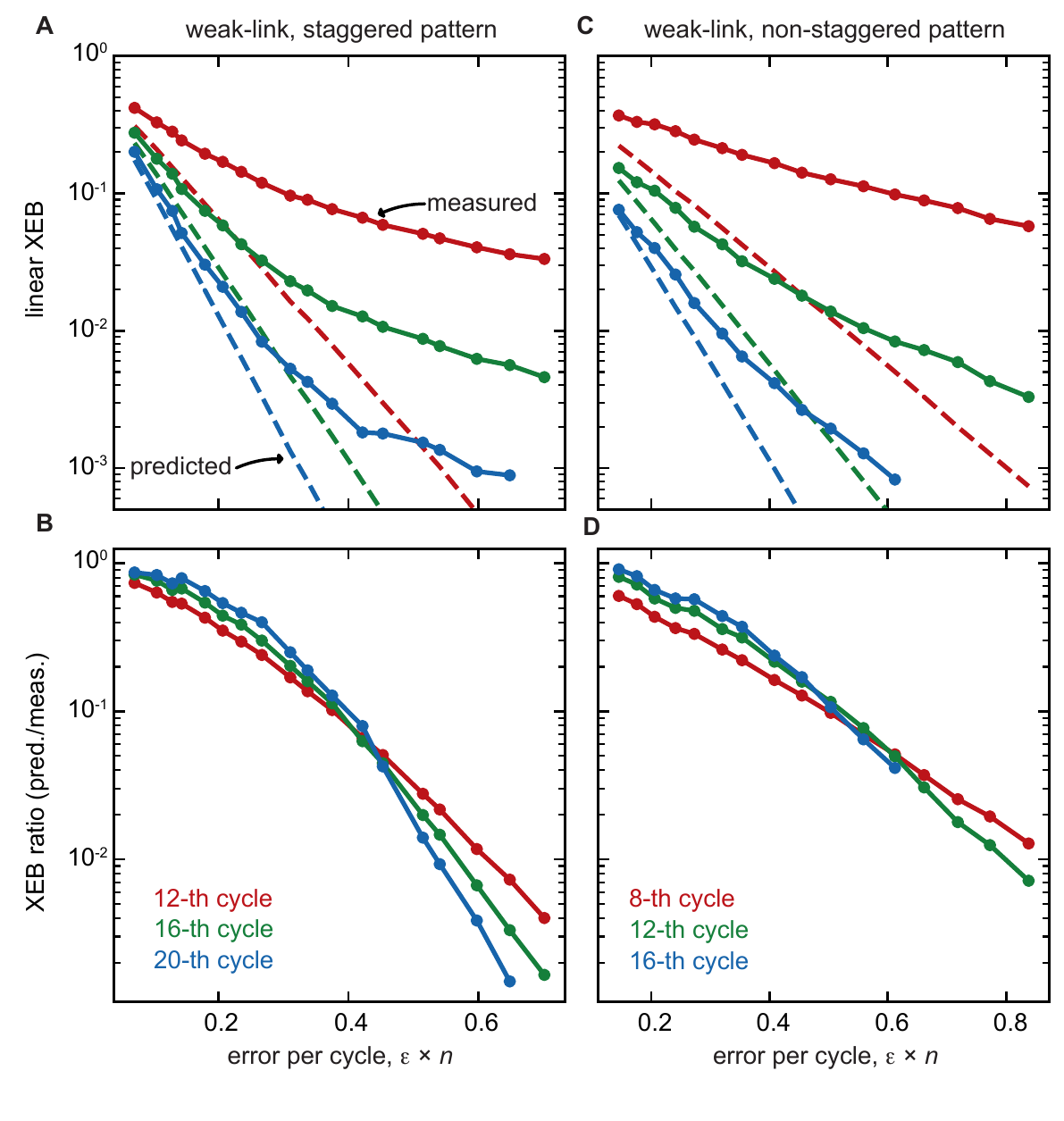} 
    \caption{\textbf{Noise phase transition in 2D}: The first row shows the measured XEB as a function of the error per cycle. In the strong noise regime, the measured XEB value is far from the expected value. See main text Fig.~3 for more details about the different patterns. For very strong noise, and large number of cycles the linear XEB value starts to be hard to measure and requires a large number of repetitions, making these measurements challenging.
    }
    \label{fig:supp_2d_extended_data}
\end{figure*}

In this appendix, we show the full dataset used for the characterization of the noise induced phase transitions identified in the main text. See Figs.~\ref{fig:supp_weaklink_extended_data} and \ref{fig:supp_2d_extended_data}.

\subsection{Weak-link model with local noise}
We performed an experiment corroborating the behaviour of the weak-link model under local noise, see main text and Sec.~\ref{sec:wl_model}. 
We rewrite Eq. (2) in main text splitting the contribution of the left and right fidelities as
\begin{equation}
    \text{linear XEB} = \lambda^{d/T}F_{\text{left}}^d + \lambda^{d/T}F_{\text{right}}^d + (F_{\text{left}} F_{\text{right}})^d\;.\label{eq:wl_splitted}
\end{equation}
We increases the noise only on the right side of the chain. We see that for low noise the last term dominates and the linear XEB decreases proportionally to the added error. However, for sufficiently large noise in the right side, the linear XEB becomes independent on the added noise. The reason is that the two last terms in Eq.~\eqref{eq:wl_splitted} become negligible compared to the first term.

\begin{figure}
    \centering
    \includegraphics[width=\columnwidth]{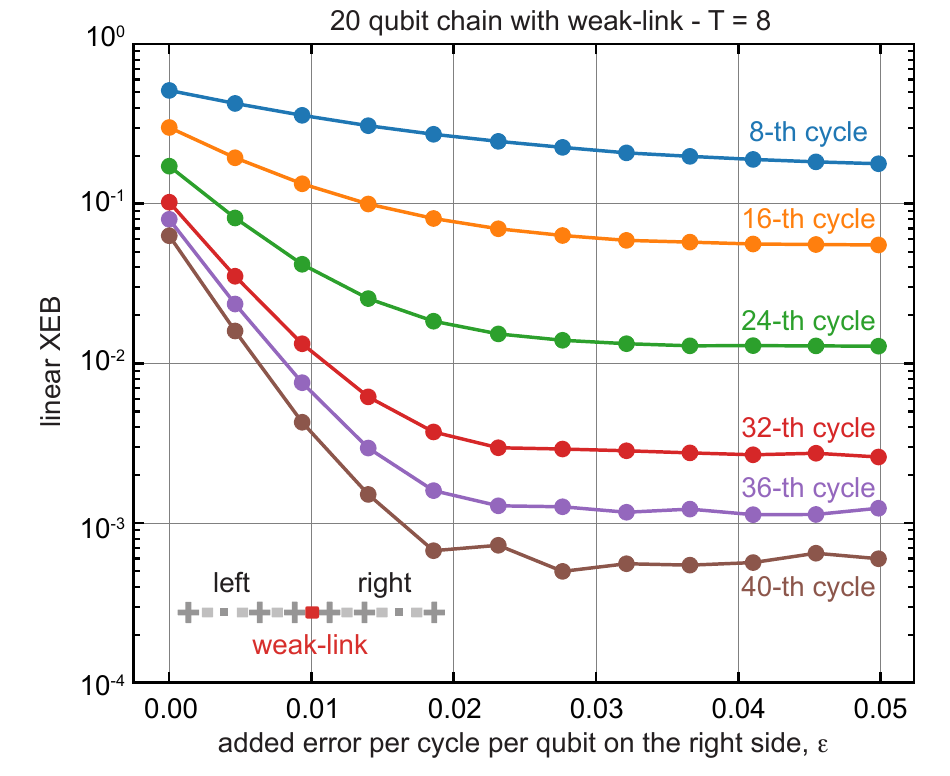}
    \caption{\textbf{Weak-link chain linear XEB with local noise}: We add extra noise only on one side of the weak-link chain. The linear XEB first decays following this added error. However, for strong enough noise, the linear XEB saturates for all depths. For this experiment we have taken a single dataset without extra error, and added coherent errors in the simulation of the circuit for the linear XEB calculation.
    }
    \label{fig:supp_half_chain}
\end{figure}

We  probe this behavior experimentally on a chain of 20 qubits with a weak-link applied with a period of $T=8$. In Fig. \ref{fig:supp_half_chain} we indeed observe that initially the fidelity decays linearly with the added noise. For very strong noise however the linear XEB plateaus at some value, indicating an insensitivity to added noise on the right side.

\section{Linear XEB via population dynamics}\label{app:popdyn}

\subsection{Population dynamics for the uniformly random single qubit gate ensemble}

The linear XEB over circuits may be written as 

\begin{gather}
    {\rm XEB}(d)
    =
    2^n C - 1, \\
    C = \sum _z 
    \bra{z}
    U  \rho_0 U^\dagger \ket{z}\bra{z} \mathcal{E}\left[ U \rho_0 U^\dagger \right]
   \ket{z},
\end{gather}
where $\mathcal{E}$ corresponds to a noisy evolution channel. We now explain how its average can be calculated via population dynamics~\cite{mi2021information,gao2021limitations}. 

Consider first a noise free evolution. Note that the average probability has the form of an out-of-time ordered correlator, $C = \sum _z \mathrm{Tr}\{\mathcal{O}_z \rho_0(d)\mathcal{O}_z\rho_0(d)\}$ where $\mathcal{O}_z = \ket{z}\bra{z}$, and $\ket{z} = \otimes_{i=1}^n \ket{z_i}, z_i =\{0,1\}$ is an $n$ qubit computational basis state.  
It can be described in terms of two copies of the evolution $ \rho_0(d) \otimes\rho_0(d)$.
After averaging over uniformly random (Haar) single qubit gates the dynamics in such doubled operator space is fully described in terms of two invariants: identity operator $\openone$ and $\mathcal{B} = (1/3)\sum_{\alpha=x,y,z} \sigma^\alpha\otimes\sigma^\alpha$, where $\sigma^\alpha$ are Pauli operators.

The average dynamics of a pair of identical operators $ \overline{ \mathcal{O}(d) \otimes\mathcal{O}(d)}$ in  the $n$ qubit system subject to a circuit consisting of cycles with two-qubit gates can be described by
a time dependent distribution \(P\left( \{v_i\},d\right)\) over an $n$ bit register $\{v_i\},
v_i \in \{0,1\}$ corresponding to $\left\{\openone_i, \mathcal{B}_i\right\}$, respectively. That is,
\begin{gather}
   \overline{ \mathcal{O}(d) \otimes\mathcal{O}(d)}
   =
   \sum_{ \{v_i\} } P(\{v_i\}, d) \bigotimes_i \left((1-v_i)\openone_i +\mathcal{B}_i v_i\right).
\end{gather}
For operators which satisfy $\mathcal{O}^2 = 1$ (true for Pauli operators) the coefficients are normalized probabilities $\sum_{\{v_i\}} P(\{v_i\}, d) =1$. 
Each two-qubit gate defines a Markov process with the update matrix,
\begin{gather}
P\left(\{v_i\},d+1\right)
=\sum _{v_j^{\prime }v_k^{\prime }} 
\Omega_{v_jv_k,v_j^{\prime }v_k^{\prime }} P\left(\{v_i^{\prime }\},d\right). \label{MarkovChain}
\end{gather}
where the indexes $j$ and $k$ correspond to the qubits involved in the corresponding two-qubit gate.

We can take the two-qubit gate to be approximately equal to an iSWAP,  
$U_{ij} = \exp\left( - i \tfrac{\pi}{4} (X_i X_j + Y_i Y_j) \right)$, for which the population dynamics update corresponds to
\begin{gather}
\hat{\Omega}^{(i,j)}=\left(
\begin{array}{cccc}
 1 & 0 & 0 & 0 \\
 0 & 0 & \frac{1}{3} & \frac{2}{9} \\
 0 & \frac{1}{3} & 0 & \frac{2}{9} \\
 0 & \frac{2}{3} & \frac{2}{3} & \frac{5}{9} \\
\end{array}
\right). \label{eq:Omega}
\end{gather}
Elements of the matrix $\hat{\Omega}$ correspond to the transition probabilities between different configurations induced by the application of a two-qubit gate. $\hat{\Omega}_{10,01}$ corresponds to $\mathcal{B}$ hopping from one qubit to another ($\Omega_{10,01} =\frac{1}{3}$ for iSWAP) whereas  $\Omega_{10,11}$ corresponds to creation of a new $\mathcal{B}$ ($\Omega_{10,11} =\frac{2}{3}$ for iSWAP).

The contribution of each configuration to XEB is determined by individual invariants, $\left(\openone_i, \mathcal{B}_i\right) \rightarrow \left(1, 1/3\right)$ as
\begin{gather}
     {\rm XEB} = 2^n \sum_{\{v_i\}} \frac{1}{3^{\sum v_i}} P(\{v_i\},d)-1.\label{eq:xeb_pd}
\end{gather}

To include the effects of noise the two-qubit gate  update rules need to be supplemented with the noise-induced decay rules at each two qubit gate~\cite{mi2021information},
\begin{gather}
 \openone_i  \openone_j\rightarrow  \openone_i  \openone_j, \\
\mathcal{B}_i \openone_j \rightarrow \exp(- \frac{16}{15}p_2) \mathcal{B}_i \openone_j, \\
\mathcal{B}_i \mathcal{B}_j \rightarrow \exp(- \frac{16}{15}p_2) \mathcal{B}_i  \mathcal{B}_j,
\end{gather}
where $p_2$ is the two-qubit depolarizing error. 

\subsection{Convergence of population dynamics to Porter-Thomas}\label{sec:pop_convergence}

The initial state for population dynamics is obtained by averaging the initial bitstring $\rho_0 = \prod_i (\openone_i +Z_i)/2$ over the first layer of single qubit gates. The result of this averaging is $\prod_i (\openone_i + \mathcal{B}_i)/4$. It can be interpreted as equal weight distribution $P(\{v_i\},0) = 1/2^n$ over all configurations $\{v_i\}$. 
After the first layer of one qubit gates XEB $= (4/3)^n$.

In a multi-qubit system a layer of gates corresponds to the evolution under $\hat{\Omega}^{(i,j)}$ applied to each pair of qubits subject to a gate of the layer. The circuit can be characterized by a transfer matrix $\hat{\mathcal{T}}$ that consists of a product of the layers that appears periodically, such that the whole circuit corresponds to $\hat{\mathcal{T}}^d$. 

There are two steady states of this Markov chain: (i) the vacuum $\{v_i=0\}$ for all $i$, (ii) the thermal state that corresponds to $P(\{v_i\})=\prod_i p(v_i)$, where $p(0)=1/4$ and $p(1)=3/4$. At long times in the noise free Porter-Thomas limit, $C = 2 / (2^n+1)$, and XEB $\approx 1$. Note that the vacuum configuration $\openone^{\otimes n}$ does not evolve, and in the presence of noise in the long depth limit the vacuum is the only remaining configuration. This produces the only non-vanishing contribution to $C$, resulting in XEB=0.

\subsection{Weak-link model analytical solution}\label{sec:wl_model}

In this section we provide details justifying Eq.~(2) of the main text.
We consider an example that can be analyzed analytically: a chain with a weak link connecting its two halves $A$ and $B$, that was introduced in the main text. At the weak link a two-qubit gate is applied only at depths $m T$ where $T$ is the number of layers applied to the rest of the chain. We describe the dynamics of XEB using the population dynamics formalism introduced in Ref.~\cite{mi2021information}.

We assume $T$ is long enough to establish the ``thermal" (or Porter-Thomas) state in each half of the chain independently. We introduce probabilities of four possible population dynamics configurations after time $T$: $g_{00}, g_{01}, g_{10}, g_{11}$ corresponding to both halves in the vacuum state, one half in the vacuum state and one in the thermal state and both halves in the thermal state. Initially all four configurations $g_{ij}$ give order one contributions to linear XEB, despite having exponentially different probability in the $\{v_i\}$ basis, due to the term $1/3^{\sum_{v_i}}$ in Eq.~\ref{eq:xeb_pd}. 

A single application of the weak link gate after $T$ cycles updates these probabilities as follows,
\begin{gather}
    g_{00} (d+T) = g_{00} (d)\;, \label{eq:weak-linkEqsg_00}\\
    g_{01} (d+T) = F^{T/2} \frac{1}{4} g_{01} (d)\;,\\
    g_{10} (d+T) = F^{T/2} \frac{1}{4} g_{10} (d)\;,\\
    g_{11} (d+T) \simeq F^{T} g_{11} (d)\;, \label{eq:weak-linkEqsg_11}
\end{gather} 
where as before $F$ is the fidelity per layer for the whole chain excluding the weak link, and the factor $1/4$ comes from the two-qubit iSWAP gate. In the last equation we drop the contributions of $g_{10}$ and $g_{01}$ to $g_{11}$ because  
it adds only an exponentially small contribution to XEB. This is because the initial thermal + vacuum state has exponentially smaller probability, as explained above. 
We therefore find
\begin{gather}
    {\rm XEB} (mT) = F^{mT} + 2\left( \frac{1}{4} F^{T/2}\right)^m.
\end{gather}
This gives a criteria for XEB to serve as a good fidelity estimate for the chain with weak link,
\begin{gather}
    F^T > \frac{1}{16}.
\end{gather}

\subsection{Numerical analysis of the phase transitions}

In this section we provide numerical simulations of XEB dynamics justifying the analysis of the data in the main text. Linear XEB is calculated numerically using the exact mapping on population dynamics introduced above, see Eq.~(\ref{eq:xeb_pd}). The time dependence of weights $P(\{v_i\},d)$ is computed by applying the transfer matrices corresponding to each two-qubit gate, Eq.~(\ref{MarkovChain}). This method requires memory that scales exponentially as $2^n$ because we evolve the full probability vector. At the same time it predicts the dynamics of the average linear XEB in the presence of noise and is quadratically more efficient than direct simulation of a noisy density matrix. Without loss of generality we simulate a simplified model of the noise including only single qubit noise applied to each qubit after each layer of two qubit gates. 

\begin{figure}
    \centering
    \includegraphics[scale=0.35]{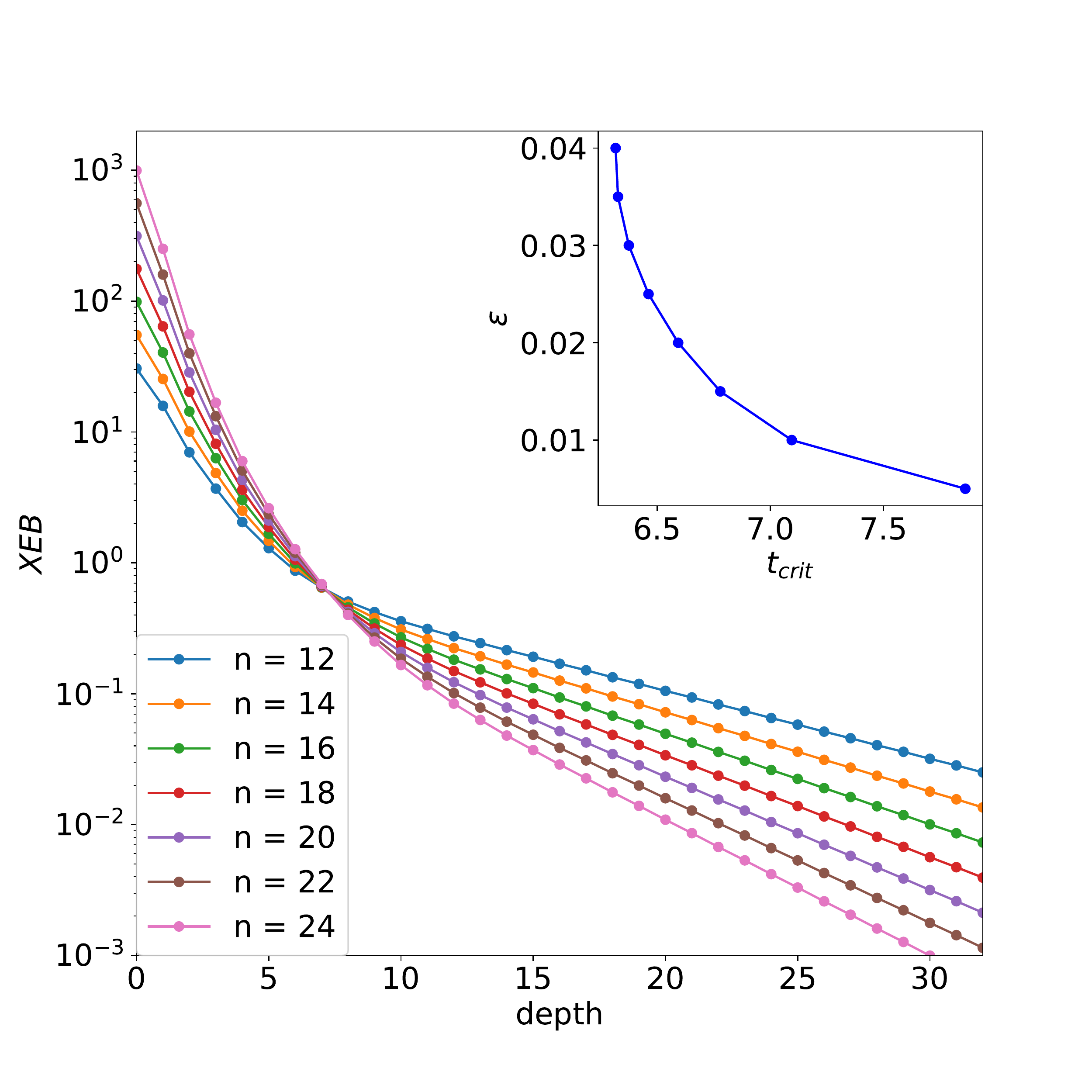}
    \caption{XEB as a function of depth for different size $n$ qubit chains. The error per qubit per unit time is $\epsilon = 0.01$. Inset shows the dependence of the critical depth on the error per qubit per unit time. }
    \label{fig:dyn_cross_1d}

\end{figure}

We first demonstrate the finite size critical scaling near the dynamical transition. Fig.~\ref{fig:dyn_cross_1d} is a numerical analog of Fig.~2a of the main text, showing the depth dependence of linear XEB for a fixed error rate per qubit $\epsilon =0.01$ and different size chains. The scaling of linear XEB with the system size changes from growth to decay at the transition point, whose value is approximately size independent. 
The transition point is $\tau \approx 1$ with corrections due to the error per cycle $\epsilon n \sim O(n^0)$. See Sec.~\ref{sec:phase_diagram} for an analytical description of this phase transition. 

\begin{figure}
    \centering
    \includegraphics[scale=0.35]{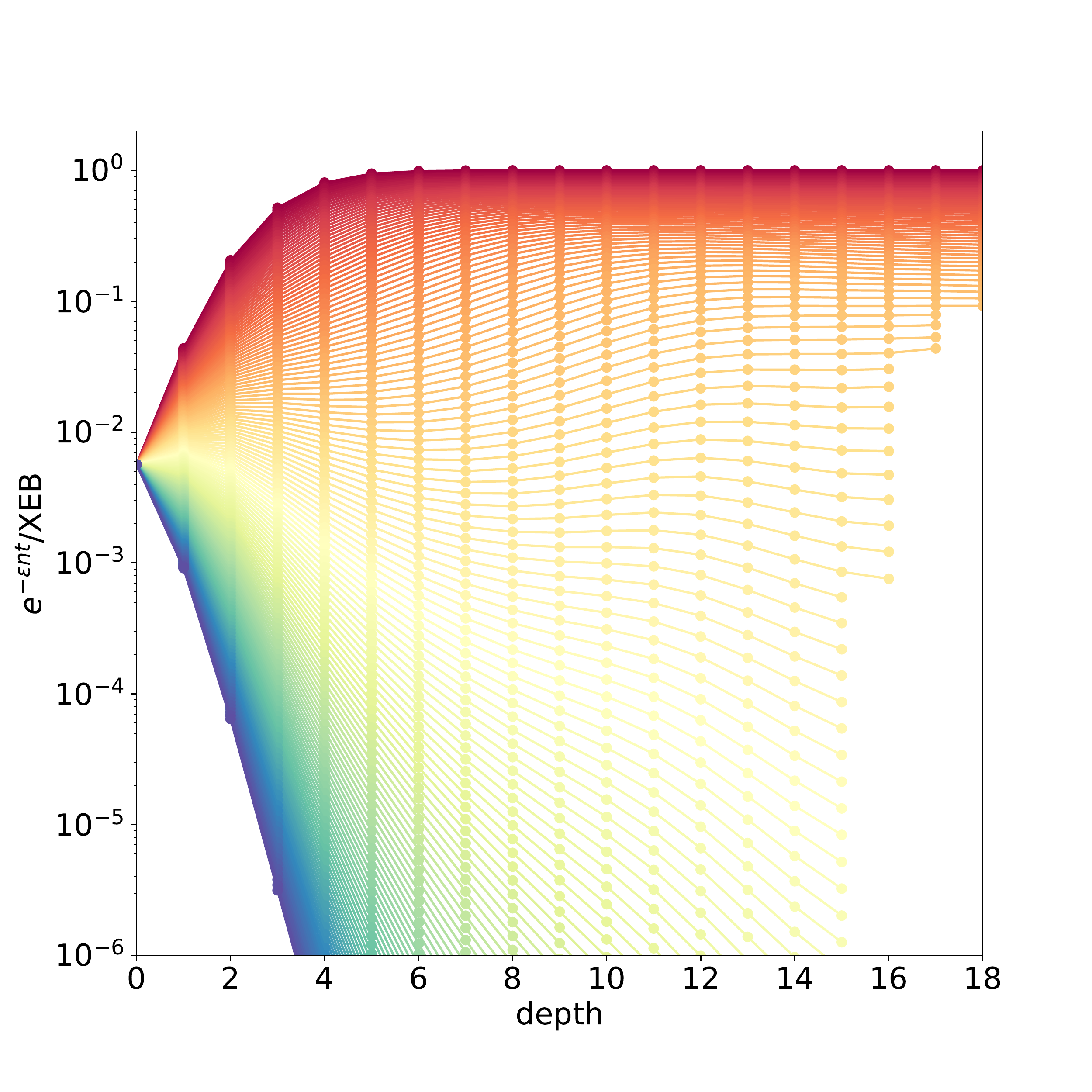} 
    \caption{ Order parameter of the noise induced phase transition as a function of depth for different levels of noise for $0\leq \epsilon n \leq 1.34$ (red to purple) on a $n=18$ qubit chain. 
    }
    \label{fig:nipt_op_v_d_1dchain}

\end{figure}

The noise induced phase transition (NIPT) is characterized by the change in the depth dependence of the order parameter $\Theta \equiv \exp(- \epsilon n d)/{\rm XEB}$. Fig.~\ref{fig:nipt_op_v_d_1dchain} shows the depth dependence of $\Theta$ for different values of the error per cycle $0\leq \epsilon n \leq 1.34$. At low error the order parameter converges to a constant, whereas in the presence of a sufficiently large error per cycle the order parameter converges to zero. 
See also Sec.~\ref{sec:phase_diagram} for an analytical description of this phase transition. 

\begin{figure}
    \centering
    \includegraphics[scale=0.35]{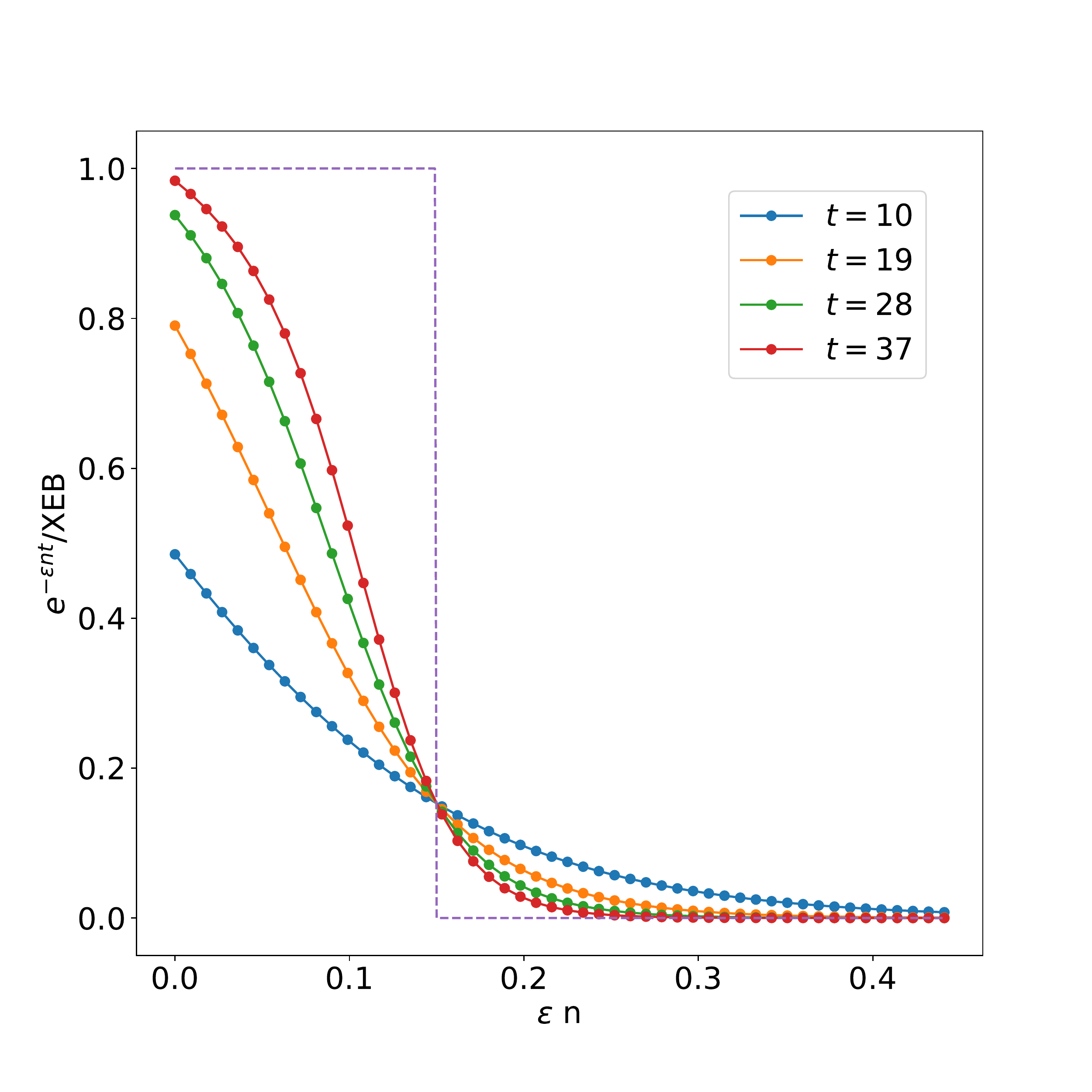} 
    \caption{ Order parameter of the noise induced phase transition as a function of error per unit time for different depths on a $n=18$ qubit chain. The link frequency is $1/T = 1/18$. The dashed line corresponds to the $d\rightarrow \infty$ limit. }
    \label{fig:nipt_op_1dwl}

\end{figure}

\begin{figure}
    \centering
    \includegraphics[scale=0.35]{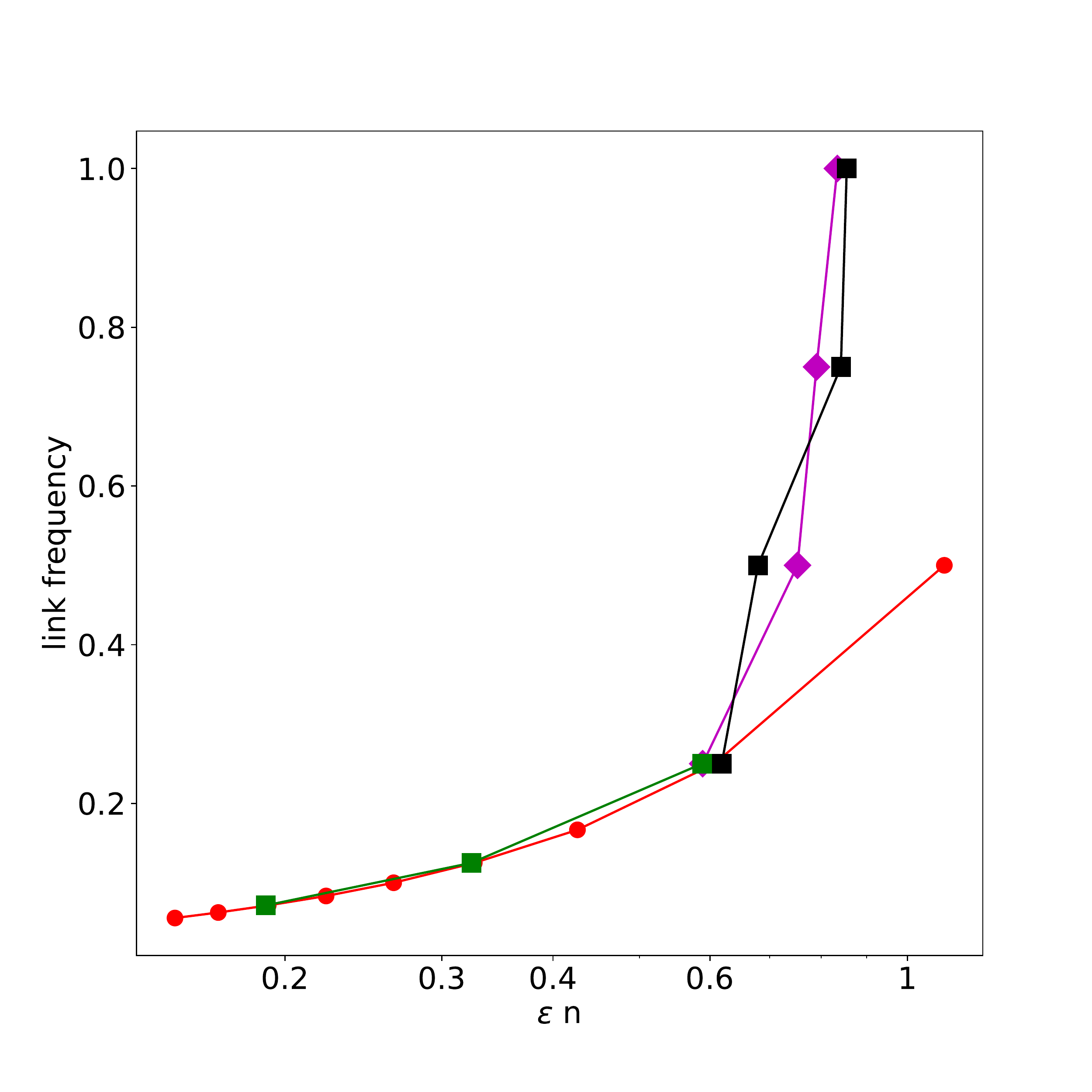} 
    \caption{Comparison of the transition point dependence on the weak link period $T$ for the discrete gate set used in the experiment and uniformly random ensembles of single qubit gates. Black squares and magenta rhombus correspond to $n=16$ ($4\times4$) ABCD pattern for the experimental discrete gate set and continuous ensemble, respectively. Green squares and red circles correspond to $n=16$ chain for the experimental and uniformly random gate set. }
    \label{fig:haar_v_sycamore}

\end{figure}

In the case of the weak link model introduced in the main text and in Sec.~\ref{sec:wl_model}, where the system is split into two parts with the gates entangling the two parts applied sufficiently rarely, there is a less data intensive procedure to identify the NIPT. This relies on the crossing point of the order parameter $\Theta $ as a function of $\epsilon n$ for different depths of the circuit (separated by a circuit period), as shown in Fig.~\ref{fig:nipt_op_1dwl}. This procedure was used to identify the transition experimentally in Fig.~3 of the main text, and works well for weak link frequency $1/T<1$ in 1D. In the absence of the weak link time dependence of the order parameter the method illustrated in Fig.~\ref{fig:nipt_op_v_d_1dchain} was used to identify the transition point. Numerical simulations for 2D circuits give similar results for the order parameter. The extracted transition points are summarized in the phase diagram presented in Fig.~3~G of the main text.

We also compare the transition point in linear XEB for a uniformly random ensemble of single qubit gates introduced above to the discrete single qubit gate set used in the experiment. The latter maps onto population dynamics in a space of three states per qubit and is more costly to implement, requiring memory $3^n$. Fig.~\ref{fig:haar_v_sycamore} shows a comparison of the NIPT location for 16 qubit systems of two different geometries: a chain and a $4\times4$ system. The role of the ensemble of single qubit gates does not appear to be significant for the NIPT point.

\section{XEB phase diagram}\label{sec:phase_diagram}

\subsection{XEB phase diagram in 1D}\label{sec:1d_phase_diagram}
The population dynamics explained in Sec.~\ref{app:popdyn} model the average dynamics for quantum circuits with a Markov chain. We focus on the case of uniformly random single qubit gate ensemble which results in a Markov chain with two states per site. In the absence of noise, this Markov chain has two steady configurations: the ``thermal'' or Porter-Thomas configuration; and the trivial ``vacuum'' configuration corresponding to the state normalization. After a short initial time, the state of a 1D circuit will be dominated by configurations with segments in the vacuum state and segments in the thermal configuration. We denote this configurations as  $g_{\sigma_1,...,\sigma_n}$ where $\sigma_k = \{0,1\}$ denotes the vacuum and thermal state respectively. 

This model is a generalization of the the weak-link model of Sec.~\ref{sec:wl_model}. Repeating the arguments of that section we obtain the following population dynamics update equation for multi-segment configurations
\begin{multline}
 g_{\sigma_1, ..., \sigma_n} (d+1) \\=  e^{-\epsilon \sum_i \sigma_i} 
 2^{- \sum_i (\sigma_i  - \sigma_{i+1})^2}
 g_{\sigma_1, ..., \sigma_n} (d)\;. \label{eq:multi-weak-link}
\end{multline}
The factor $e^{-\epsilon}$ account for the fidelity decay with noise strength $\epsilon$ for a segment in the $\sigma=1$ thermal state. The factor of $1/2$ between any two consecutive segments in different states comes from the application of an iSWAP, see Sec.~\ref{app:popdyn}, which happens every two circuit cycles in 1D. This equation reproduces Eqs.~\eqref{eq:weak-linkEqsg_00} to~\eqref{eq:weak-linkEqsg_11} of the weak-link model with $n=2$ and $T=2$. 

The initial state  corresponds to an equal population of all configurations, $g_{\sigma_1, ..., \sigma_n} = 2^{-n}$, as obtained by generalizing the initial state given in Sec.~\ref{sec:pop_convergence}. The linear XEB at depth $d$, which we denote by $\textrm{XEB}$ in this section, is given by 
\begin{equation}
\textrm{XEB} + 1 = 2^n \sum_{\{\sigma_i\}} g_{\sigma_1, ..., \sigma_n}(d)\;.
\end{equation}

Equation~\eqref{eq:multi-weak-link} can be solved for large $n$ by a transfer matrix method.  We will use the solution for a system of size $n$ and depth $d$ to express the solution of size $n+1$ at the same depth $d$.
Let's define
\begin{align}
  Z_n(\sigma,d) = 2^n \sum_{\{\sigma_i\}, i<n} g_{\sigma_1, \ldots, \sigma_{n-1},\sigma}(d)\;.
\end{align}
From Eq.~\eqref{eq:multi-weak-link} we obtain,
\begin{gather}
Z_{n+1}(\sigma,d) = e^{-\epsilon \sigma d}\sum_{\sigma_n = \{0, 1\}}
 2^{- (\sigma_n - \sigma )^2d}  Z_n(\sigma_n,d)\;. \label{eq:transfer-matrix}
\end{gather}
Equation~\eqref{eq:transfer-matrix} must be solved with the boundary condition,
\begin{gather}
 Z_1(\sigma, d) = e^{-\epsilon \sigma d}\;.\label{eq:init}
\end{gather}
The linear XEB is   
\begin{gather}
\textrm{XEB} + 1 =  \sum_{\sigma} Z_n(\sigma, t )\;.~\label{eq:XEBviaZ}
\end{gather}

We solve Eq.~\eqref{eq:transfer-matrix} in the large $n$ or continuous limit, approximating $Z_{n+1}(\sigma, d) \approx Z_{n}(d)+\partial_n Z_{n}(\sigma, d)$. Using the initial condition Eq.~\eqref{eq:init} and substituting the result into Eq.~\eqref{eq:XEBviaZ} we obtain,
\begin{gather}
\textrm{XEB}(n,d)+1 = 2e^{-\frac{\epsilon n d}{2}}\left(\cosh n \delta_d + \frac{2^{-d}}{\delta_d}\sinh n\delta_d \right),\\
\delta_d \equiv \sqrt{\frac{\epsilon^2 d^2}{4}+2^{-2d}}\;.\label{eq:xeb_analytic1D}
\end{gather}
Note that in the absence of noise $\epsilon\rightarrow 0$ we obtain
\begin{gather}
\label{eq:dalzel} \textrm{XEB}(n,d) + 1 = 2 e^{n 2^{-d}} \;.
\end{gather}
This result is consistent with Ref.~\cite{dalzell2022} which shows that 1D random circuits anticoncentrate in logarithmic depth. 

As explained in the main text, it is natural to introduce the scaling noise variable
\begin{gather}
f \equiv \epsilon  n\;. \label{eq:f-1d}
\end{gather}
To study the XEB phase diagram this variable is kept constant as $n\rightarrow \infty$. We also introduce the scaling depth variable 
\begin{gather}
\alpha = \frac{d}{ \log_2 n}. \label{eq:alpha-1d}
\end{gather}
Including these substitutions the expression for XEB takes the form
\begin{multline}
\textrm{XEB}(n;\alpha, f)+1 \\= 2e^{-\frac{f \alpha \log_2 n}{2}}\left(\cosh \Delta + \frac{n^{1-\alpha}}{\Delta}\sinh \Delta \right)\;,\label{eq:xeb_analytic_f}
\end{multline}
where
\begin{gather}
\Delta \equiv \sqrt{\frac{f^2\alpha^2 \log_2^2 n}{4}+n^{2(1-\alpha)}}\;.  \label{eq:1D_phase_diagram_delta}
\end{gather}

We can now explain the different phases of XEB by writing the Eq.~\eqref{eq:xeb_analytic_f} for fixed $f$ and $\alpha$ in the thermodynamic limit $n\rightarrow\infty$. At low depth $\alpha<1$, before anticoncentration, the second term under the square root in the Eq.~\eqref{eq:1D_phase_diagram_delta} for $\Delta$ dominates. We obtain
\begin{gather}
\textrm{XEB}(n;\alpha, f) = n^{-\frac{f\alpha}{\ln4}}\left(2e^{ n^{1-\alpha}}\right)-1\;.\label{eq:alessthenone}
\end{gather}
This result differs from the noise-free limit Eq.~\eqref{eq:dalzel} by a noise-dependent algebraic prefactor whose exponent  depends smoothly on noise.

After anticoncentration, $\alpha>1$, the first term in Eq.~\eqref{eq:1D_phase_diagram_delta} dominates in the thermodynamic limit $n \to \infty$. We obtain 
\begin{gather}
\textrm{XEB}(n;\alpha, f) = n^{-\frac{f \alpha}{\ln2}} + \frac{2n^{1-\alpha}}{f\alpha \log_2 n},~\label{eq:amorethenone}
\end{gather}
Clearly Eqs.~\eqref{eq:alessthenone} and \eqref{eq:amorethenone} cannot be merged to each other by an analytic function, which indicates that $\alpha = 1$ is a phase transition line. Indeed for $\alpha =1$ Eq.~\eqref{eq:alessthenone} is algebraic whereas Eq.~\eqref{eq:amorethenone} is logarithmic. This is in sharp contrast with the noise-free Eq.~\eqref{eq:dalzel} which does not have any singularity at $\alpha =1$. 

Equation~\eqref{eq:amorethenone} describes another phase transition defined by the noise induced phase transition line
\begin{gather}
f_c(\alpha) = \frac{\alpha-1}{\alpha} \ln2\;.\label{eq:crit-line}
\end{gather}
This line separates the weak and strong noise regimes. Note that the depth required in 1D for entanglement to spread across all the qubits is $\alpha \simeq n / \log_2 n$, and then $f_c \simeq \ln 2$. In the weak noise regime $f<f_c$ we obtain the XEB
\begin{gather}
\textrm{XEB}(n;\alpha, f) =
n^{-\frac{f \alpha}{\ln2}} = e^{-\epsilon n d}\;, \label{eq:weaknoise}
\end{gather}
which coincides with the circuit fidelity $F^{d}$. In the opposite regime $f>f_c$ we obtain the XEB
\begin{gather}
\textrm{XEB}(n;\alpha, f) =\frac{2n^{1-\alpha}}{f\alpha \log_2 n}\;, 
\end{gather}
which is much larger than the circuit fidelity.

\subsection{XEB phase diagram in 2D and higher dimensions}\label{sec:2d_phase_diagram}

The 1D update equation for multi-segment configuration Eq.~\eqref{eq:multi-weak-link} can be generalized to higher dimensions as
\begin{multline}
 g_{\sigma_1, ..., \sigma_n} (d+1)  \\ =  e^{- \epsilon \sum_i \sigma_i} 
 4^{- \frac{1}{2 \cdot \kappa}\sum_{ \langle ij \rangle} (\sigma_i  - \sigma_{j})^2}
 g_{\sigma_1, ..., \sigma_n} (d). \label{eq:multi-weak-link-2D}
\end{multline}
where $\langle i,j\rangle$ denotes the nearest neighbors on the D dimensional lattice and $\kappa$ is the number of neighbors per qubit. The factor of 1/4 in the second term  comes from the application of an iSWAP, as in Eq.~\eqref{eq:multi-weak-link} and Eqs.~\eqref{eq:weak-linkEqsg_00} to~\eqref{eq:weak-linkEqsg_11}. The exponent of 1/4 counts the number of segments in different states, devided by 2 because we count neighbors twice in the sum, and devided by $\kappa$ because an ISWAP is applied between two segments every $\kappa$ circuit cycles. The initial conditions
and the expression of linear cross-entropy is the same as in $1D$ case. 

The noise induced phase transition that we are interested in resides at $d\gg 1$. There we can expand the populations state around the two stationary configurations with all $\sigma_i=0$ (vacuum) or all $\sigma_i=1$ (thermal), using the so-called dilute flipped spin expansion. This gives the equation
\begin{multline}
  \textrm{XEB} + 1 \simeq \\\sum_k\frac{1}{k!}e^{-\epsilon kd} \left(\frac{n}{4^{d}} \right)^{k}
   + \sum_k\frac{1}{k!}e^{-\epsilon \frac{(n-k)d}{n}} \left(\frac{n}{4^{d}} \right)^{k} \;,\label{eq:dilute_expansion}
\end{multline}
where the first and second term in the right-hand-side correspond to the expansion around
$\sigma=0$ and $\sigma=1$ respectively, and the index $k$ describes the total number of the flipped spins. The usual combinatorial factor
$k!$ is needed to avoid the over-counting of same configurations. 
We can rewrite this equation as
\begin{multline}
   \textrm{XEB} + 1 \\=\exp\left[e^{-\epsilon d} \left(\frac{n}{4^d} \right)\right]
   + e^{- \epsilon n d}\exp\left[e^{\epsilon d} \left(\frac{n}{4^d} \right)\right]\;,
   \label{eq:2Dresult}
 \end{multline}
 which generalizes Eq.~\eqref{eq:xeb_analytic1D} to higher dimensions.

 Similarly to the one dimensional case, we introduce the scaling variables
 \begin{equation}
     \alpha=\frac{d}{\log_{4}n}\;,   \quad f=\epsilon n\;.\label{2Dscaling-variables}
 \end{equation}
In order the study the XEB phase diagram, we consider the thermodynamic limit $n \to \infty$ at fixed $\alpha$ and $f$. 
 We find
 \begin{align}
   \textrm{XEB} + 1 & = \exp\left[n^{1 - \alpha} e^{-\frac{f \alpha \log_4 n}{n}}\right] \nonumber \\ &\quad \quad + e^{ - f \alpha \log_4 n} \exp\left[n^{1 - \alpha} e^{\frac{f \alpha \log_4 n}{n}}\right] \\
   &\simeq \exp\left(n^{1-\alpha} \right)\left(1+
     e^{-f\alpha\log_4 n} \right) \\ 
    &= \exp\left(n^{1-\alpha} \right)\left(1+
     n^{-\frac {f\alpha}{\ln 4} } \right) \;.
   \label{eq:bulk_xeb}
\end{align}
We note that the first factor
\begin{align}
    \exp\left(n^{1-\alpha} \right) = e^{n 2^{-2d}}
\end{align}
describes the convergence to anticoncentration, in accordance to Eq.~\eqref{eq:dalzel} and Ref.~\cite{dalzell2022}. The second factor is
\begin{align}
    1+  n^{-\frac {f\alpha}{\ln 4} } = 1+e^{-\epsilon n d}\;.
\end{align}
 
In contrast with the one dimensional case, Eq.~\eqref{eq:bulk_xeb} does not have any singularity at finite $f$ and $\alpha=1$, and therefore does not exhibit a phase transition in the covergence to anticoncentration. We explain below how this phase transition appears due to a boundary effect.

The  noise induced phase transition line at $\alpha>1$  has a similar form to Eq.~\eqref{eq:crit-line}, 
\begin{equation}
f_c(\alpha) = \frac{\alpha-1}{\alpha}\ln 4\;,\label{eq:crit-line2}
\end{equation}
and terminates at $\alpha=1$.
The XEB value in the
 weak and strong noise regimes of this first order phase transition is similar to one dimensional case 
\begin{equation}
  \textrm{XEB}=\left\{
    \begin{array}{lr}
    n^{1-\alpha}, & f>f_c(\alpha)\\
      n^{-\tfrac{f\alpha} {\ln 4}} = e^{-\epsilon n d}, & f<f_c(\alpha)
  \end{array}
  \right.
\end{equation}
In the weak noise regime $f<f_c(\alpha)$ the value of the linear XEB $\textrm{XEB}$ is the circuit fidelity, as expected. We can trace this contribution to the thermal or Porter-Thomas state in the second term of Eq.~\eqref{eq:dilute_expansion}. In the strong noise regime $f>f_c(\alpha)$ the value of the linear XEB is dominated by local correlations above the vacuum which we can trace to the first term in Eq.~\eqref{eq:dilute_expansion}. This is the situation studied in Refs.~\cite{barak2020spoofing,gao2021limitations,aharonov2022polynomial}.

In the above analysis we neglected the existence of the boundary for a finite lattice. 
In the case of a regular lattice boundary qubits have different number of nearest neighbours from the bulk qubits. As we will see, boundary qubits dominate the convergence to the thermal state and the behavior of XEB is qualitatively different. 

We focus on the 2D case for simplicity. The presence of the boundary with $r$ nearest neighbours 
adds the following boundary contribution to the XEB Eq.~\eqref{eq:2Dresult}
\begin{align}
\exp\left( e^{- \epsilon  d} \frac{c_r \sqrt n}{4^{t\cdot r/4}} \right)\;,
\end{align}
where $c_r \sqrt{n}$ is the lattice perimeter and the exponent $r/4$ in the 1/4 factor accounts for the fact that in the boundary an iSWAP is applied every $r/4$ circuit cycles in 2D. In a regular 2D lattice each qubit in the boundary has $r=3$ nearest neighbours, while in the Sycamore device lattice the number of neighbours in the boundary is $r=2$.
In the thermodynamic limit $n \to \infty$ at fixed $\alpha$ and $f$ Eq.~\eqref{eq:bulk_xeb} gets modified to
\begin{equation}
   \textrm{XEB} + 1 \simeq  e^{n^{1-\alpha} +c_2 n^{\tfrac{1-\alpha}{2}}}\left(1+
     n^{-\frac {f\alpha}{\ln 4} }\right).
   \label{eq:2DS-boundary}
\end{equation}

Remarkably, the boundary introduces a phase transition at $\alpha = 1$ even without noise, in contrast to the one dimensional case. This phase transition arises from the competition between the two terms in the exponent of the prefactor, $n^{1-\alpha}$ and $n^{(1-\alpha)/2}$. Note that Ref.~\cite{dalzell2022}, which studies the convergence to anticoncentration, considers circuits without boundaries that do not exhibit this phase transition. 
The noise induced phase transition gets displaced to 
\begin{gather}
    f_c (\alpha) = \frac{\alpha - 1}{ \alpha} \ln2\;.
\end{gather}
Note that this is the same condition as the 1D case with boundary studied in the previous section.





\section{Noisy phase transition and spoofing}

The existence of the noise induced phase transition in the noisy random circuit dynamics discussed in the main text has important implications for the effectiveness of the so called spoofing algorithms~\cite{barak2020spoofing,gao2021limitations,aharonov2022polynomial}. Indeed, building up this theory, we give below a lower bound for the error per cycle below which spoofing algorithms can not match the experimental XEB. This boundary can also be interpreted as a first order phase transition. 

Spoofing algorithms aim to generate bitstrings from a distribution that maximizes the value of XEB without resorting to the exponential cost of full simulation of the quantum evolution, i.e. ideally with polynomial cost. These algorithms can be broadly split into two stages~\cite{gao2021limitations}: (i) approximation of the wave function, (ii) sampling of bitstrings to maximize XEB.
We analyze step (i) first.


\subsection{Spoofing in the weak-link model}

Consider the weak-link model of Eq.~(2) of the main text. The spoofing in this case corresponds to the elimination of entangling gates across the weak link. The omission of an iSWAP gate introduces a new two qubit operation into the population dynamics, corresponds to replacing $\hat{\Omega}^{(i,j)} \rightarrow \hat{\Omega}^{(i,j)}_{\rm omitted}$ in Eq.~(\ref{MarkovChain}),
\begin{gather}
\hat{\Omega}^{(i,j)}_{\rm omitted}=\left(
\begin{array}{cccc}
 1 & 0 & 0 & 0 \\
 0 & 0 & 0 & 0 \\
 0 & 0 & 0 & 0 \\
 0 & 0 & 0 & \frac{1}{3} \\
\end{array}
\right). \label{eq:OmegaOmitted}
\end{gather}
The system of equations for the dynamics of the weak-link model Sec.~\ref{sec:wl_model},  Eq.~\eqref{eq:weak-linkEqsg_11} is modified as follows,
\begin{gather}
    g_{00} (d+T) = g_{00} (d), \\
    g_{01} (d+T) = \frac{1}{4} g_{01} (d),\label{eq:weak-link_spoof_01} \\
    g_{10} (d+T) =  \frac{1}{4} g_{10} (d),\label{eq:weak-link_spoof_10} \\
    g_{11} (d+T) \simeq \frac{1}{4} g_{11} (d). \label{eq:weak-link_spoof_11}
\end{gather} 
Notice that $g_{11}$ is no longer constant even at $F=1$. The resulting linear XEB after $mT$ cycles reads
\begin{gather}
    \left\langle D p_{\text{sim}}(s) - 1\right\rangle_{\rm spoof} = 3\frac{1}{4^{m}}. \label{eq:wl-spoofing-xeb}
\end{gather}

We compare this to XEB of the noisy circuit with system fidelity $\approx F^{mT}$. Note that if the fidelity per cycle is smaller than $F^T_{cs} = 1/4$, the classical algorithm that omits the iSWAP gate in the weak link obtains a larger XEB value. This is different from the transition point of the corresponding noise induced phase transition, $F_c^T = 1/16$ from Eq.~(2) of the main text. This means that XEB is a good estimator of fidelity already for $F>F_c^T $, but nonetheless a spoofing algorithm could produce an XEB higher than experiment while $F \leq F^T_{cs}$. 



\subsection{Spoofing for general models}

The result of Eq.~\eqref{eq:wl-spoofing-xeb} can be extended naturally to general spoofing algorithms and models using the formalism of population dynamics, see Sec.~\ref{app:popdyn}. Spoofing algorithms can exploit two different contributions to cross entropy. The first is given by the small potential overlap that spoofing can still maintain with the ideal Porter-Thomas state, including some amount of global correlations. The ideal Porter-Thomas state is given by $g_{11}$ in the weak-link model above, and this contribution is modelled as removing two-qubit gates, as in Eqs.~\eqref{eq:OmegaOmitted} and~\eqref{eq:weak-link_spoof_11}. In the general case, we can break the full circuit into subsystems, each with a manageable number of qubits, typically $\sim n/2$~\cite{markov2018quantum,arute_supremacy_2019,gao2021limitations}. This is done by removing two-qubit gates along the cut that separates both subsystems. Each iSWAP gate removed lowers this contribution to XEB by a factor of 1/4~\cite{markov2018quantum,arute_supremacy_2019}. This contribution then scales as $1/4^{\nu d}$, where $\nu$ is the number of gates along the cut, and $d$ is the depth. Given the significant number of gates along any suitable cut~\cite{arute_supremacy_2019}, this contribution is subdominant compared to the second contribution which we study next.

The dominant contribution from spoofing algorithms to XEB is that they can potentially capture finite-depth local correlations outside the thermal or Porter-Thomas state~\cite{gao2021limitations,barak2020spoofing,aharonov2022polynomial}. In the weak-link model above, these finite depth local correlations are modelled by the  vacuum-``thermal'' and ``thermal''-vacuum states, $g_{01}$ and $g_{10}$. Their exponential decay for increasing depth is given in this model by Eqs.~\eqref{eq:weak-link_spoof_01} and~\eqref{eq:weak-link_spoof_10}.

In the general case, the thermal or Porter-Thomas state is the stationary eigenstate of a Markov chain modeling the dynamics when averaging over circuits, as explained in the  population dynamics Sec.~\ref{app:popdyn}. The magnitude of finite-depth local correlations outside of the Porter-Thomas state can be bounded by the difference between the noiseless XEB value at finite depth and the asymptotic value. This gives the equation
\begin{align}
    \left\langle D p_{\text{sim}}(s) - 1\right\rangle_{\rm spoof} \lesssim {\rm XEB} - C(n)\;,
\end{align}
where 
\begin{align}
    C(n) = {1-2^{-n} \over 1+2^{-n}} \simeq 1 
\end{align}
is the asymptotic value of XEB for given $n$.
The XEB value will converge exponentially with depth as
\begin{align}
    {\rm XEB} - C(n) \propto \lambda^{d}\;.
\end{align}




\begin{figure}
    \centering
    \includegraphics[width=\columnwidth]{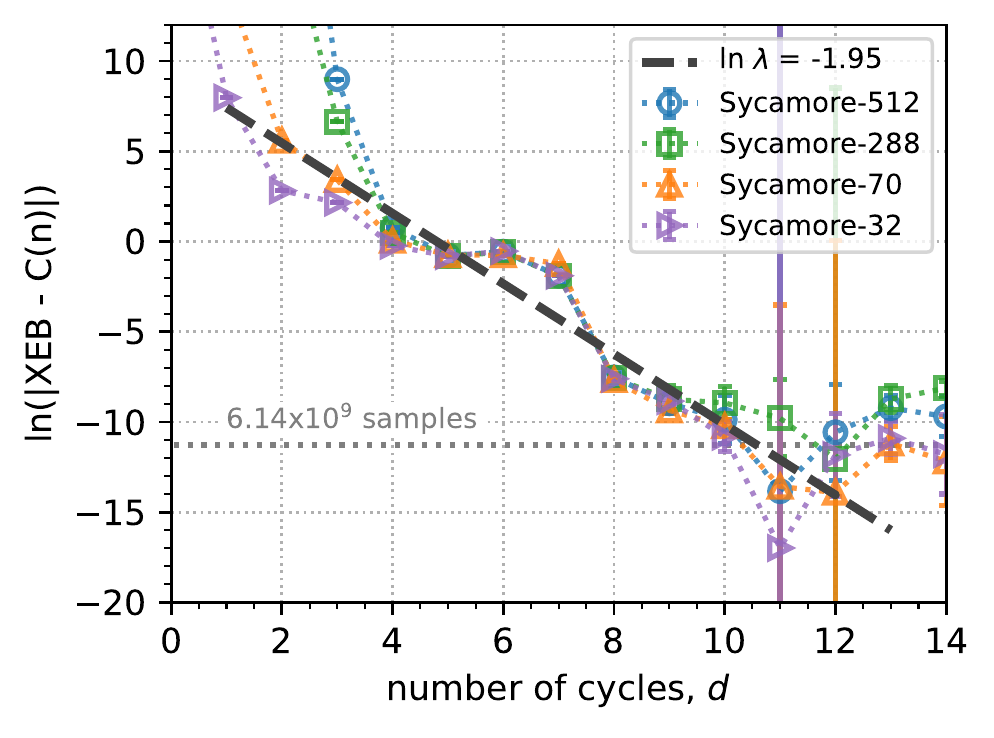} 
    \caption{Logarithm of ${\rm XEB} - C(n)$ as a function of the
    number of cycles $d$ for random noise-free Clifford circuits. Different colors correspond to
    different number of qubits.
    For all simulations, the Sycamore layout and the pattern ABCDCDAB is used. 
    Here $C(n)$ is the asymptotic
    value of the XEB for $n$ qubits. The horizontal dotted gray line
    corresponds to scale of the numerical error from the statistics resolution limit of $1/\sqrt M$,
    where $M = 6.14\times10^9$ is the number of samples used.}
    \label{fig:XEBclifford}
\end{figure}

We extract the decay rate $\lambda$ from numerical simulations, see Fig.~\ref{fig:XEBclifford}. 
We obtain a decay rate  $ \lambda \approx e^{-1.95}$. 
This is a much faster decay rate than the decay of fidelity observed in the experiments reported here. Note that we omit the non-Clifford gates for the numerics used Fig.~\ref{fig:XEBclifford} so we can scale up to 512 qubits. This can be done because it does not affect the average value of XEB for the ensemble of circuits, as shown in Sec.~\ref{app:cliff_purity}. The discrete gate set used here, together with the iSWAP gate, gives a faster convergence than other gate sets. In particular, we obtain a faster convergence than the estimation of Sec.~\ref{sec:2d_phase_diagram} which uses a uniformly random single qubit gate ensemble.

As we will see below, the post-processing stage (ii) of a spoofing algorithm can only affect a multiplicative factor, and not the exponential decay rate.

\subsection{Linear XEB amplification with post-processing}
A specific spoofing algorithm discussed in Ref.~\cite{gao2021limitations} approximates in step (i) the wave function by taking a product of two subsystem wave functions, each of size roughly half of the full system. This corresponds to omitting from the circuit all gates entangling the two subsystems. In the post-processing step (ii) the output bitstrings are chosen to maximize the approximate probabilities, instead of sampling the resulting approximate wave function. In Ref.~\cite{gao2021limitations} an average XEB corresponding to this procedure was estimated numerically. 

The post-processing stage (ii) in Ref.~\cite{gao2021limitations} is done sorting the probabilities within each subsystem, which incurs an exponential cost $\sim 2^{n/2}$. This is doable for relatively small subsystems. 
The number of distinct bitstrings to be produced has to match the number in the experiment $k\sim10^6 \ll D$, see Sec.~\ref{app:collisions}.

In what follows we show that the linear XEB is upper bounded by
\begin{multline}
    \left\langle D p_{\text{sim}}(s) - 1\right\rangle_{\rm spoof}  \lesssim  \\ \ln( D_L/k_L) \ln( D_R/k_R) \times \lambda^{d}, \label{eq:spoofing-xeb-stageii}
\end{multline}
where $D_L$ and $D_R$ are the Hilbert space dimensions for the left and right subsystems respectively, and $k_L$ and $k_R$ are number of samples from left and right subsystems. Note that $D = D_L D_R$ and $k = k_L k_R$. 
It is important to emphasize that the enhancement of XEB is at most logarithmic in the size of the Hilbert space of each subsystem (linear in the respective number of qubits). 
Contemplating equation~(\ref{eq:spoofing-xeb-stageii}) it may be tempting to split the Hilbert space into $r>2$ subsystems and expect an enhanced factor $\propto (\ln D^{1/r})^r$. However, while using more subsystems could in principle increase the post-processing multiplicative prefactor, it would result in a worse approximation because it ignores correlations between the subsystems. 

For the parameters of the experiment in Fig.~4 of the main text, $d=24, n =70$ and $\ln\lambda \simeq - 1.95$, the upper bound on spoofing XEB~(\ref{eq:spoofing-xeb-stageii}) is well below the value in the experiment. This indicates that spoofing cannot be successful. 

We now describe the derivation of the optimal the pre-exponential multiplicative factor in Eq.~(\ref{eq:spoofing-xeb-stageii}). Consider $M$ random numbers $w_i, i=\{1,...,M\}$ sampled from a joint probability density $\mathcal{P}(w_1, w_2, ..., w_M)$. Consider the sorted list of $w_i: w_1 > w_2 > ...$. The probability density for the $k$th term in the list, or $k$-th order statistic, is
\begin{widetext}
    \begin{align}
    \mathcal{P}_{k}(w) = 
  \frac{1}{(k-1)!(M-k)!}\sum_S \prod_{r=1}^{k-1}\int_w^\infty dw_r \prod_{q=k+1}^{M}\int_0^w  dw_q 
     \mathcal{P}(S(w_1, w_2, ..., w_M)),
\end{align}
\end{widetext}
where the sum is over all permutations of $w_i$. Identifying $w_j$ with a bitstring probability and $M$ with the dimension of the respective Hilbert space, the XEB is given by the following average,
\begin{align}
    \langle w \rangle &= \frac{1}{k} \sum_{k^\prime=1}^{k} \langle w \rangle_{k^\prime}, \\
     \langle w \rangle_{k^\prime} &=  \int_0^\infty dw w \mathcal{P}_{k^\prime}(w). \label{eq:w}
\end{align}

At the depths considered experimentally, each subsystem reaches the Porter-Thomas or exponential distribution. Then the corresponding $k$-th order statistic has probability density
\begin{multline}
     \mathcal{P}_{k}(w) =  \\ \frac{D!}{(k-1)!(D-k)!} D e^{-k D w} \left( 1 - e^{-D w} \right)^{D-k}.
\end{multline}
Substituting this into the expression for $\langle w \rangle$, Eq.~(\ref{eq:w}), we find for $k\gg1$,
\begin{gather}
    D\langle w \rangle = \ln (D/k).
\end{gather}
This expression is the origin of the logarithmic factors in Eq.~(\ref{eq:spoofing-xeb-stageii}). Note that is natural to expect this, because the probability to find a state on the tail of probability distribution decays exponentially.

This is the optimal enhancement from the stage (ii) of the spoofing algorithm if stage (i) is able to reproduce the  wave function  of each subsystem. Therefore we obtain
\begin{gather}
    \left\langle D p_{\text{sim}}(s) - 1\right\rangle_{\rm spoof}  \lesssim \lambda^{d} D 
    \frac{1}{k}
    \sum_{k^\prime,k^{\prime\prime}}\langle w \rangle_{k^\prime}^{L}\langle w \rangle_{k^{\prime\prime}}^{R},\label{eq:sspoof-xeb-split}
\end{gather}
where the sum over $k^\prime,k^{\prime\prime}$ includes $k$ terms in total and is defined to maximize the right hand side of Eq.~(\ref{eq:sspoof-xeb-split}). Using Eq.~(\ref{eq:sspoof-xeb-split}) we find that the maximum corresponds to a linear configuration such that $\ln D_L/k_L = \ln D_R/k_R$. 


\subsection{Logarithmic XEB amplification with post-processing}

It is worthwhile to consider the spoofing of the logarithmic XEB, see Eq.~(\ref{eq:XEB-lin-log}) for a formal definition. Logarithmic XEB is less sensitive to rare spikes of the wave function amplitude in the bistrsing basis. 

The first step of the XEB amplification in Ref.~\cite{gao2021limitations}, and in the previous section, is to devide the system in two subsections.
The ideal wave function is a superposition of products of left $\ket{\psi_{L,i}}$ and right $\ket{\psi_{R,i}}$ wave functions~\cite{markov2018quantum}, 
\begin{gather}
\ket{\psi} \simeq \frac{1}{\sqrt{\mathcal{N}}}\sum_{i} \ket{\psi_{L,i}} \ket{\psi_{R,i}}.\label{eq:wf-LR}
\end{gather}
Using the analysis similar to the previous subsection we obtain a bound on logarithmic XEB after stage (ii) of the spoofing algorithm
\begin{gather}
    \textrm{log XEB}_{\rm spoof} \simeq \ln\left[
    1 + \frac{1}{\mathcal{N}}\ln( D_L/k_L) \ln( D_R/k_R) \right].\label{eq:logXEB-stageii}
\end{gather}

Note that for large $\mathcal{N} \gg 1$ logarithmic XEB of the bitstrings produced by spoofing gives the same result as linear XEB. 



\section{Simulation of random circuit sampling using tensor network contraction}
\label{app:tn_simulation}

Tensor network contraction has been used extensively in the simulation of RCS over the last few years~\cite{markov2008simulating,boixo2017simulation,gray2021hyper,huang2020classical,kalachev2021multi,kalachev2021classical,pan2022solving,liu2022validating}.
Given a quantum circuit, it is straightforward to generate a tensor network whose contraction yields one or many of its output amplitudes.
In this tensor network, each one-qubit gate is expressed by a rank-2 tensor, each two-qubit gate is expressed by a rank-4 tensor, and the input state $\ket{0}^{\otimes n}$ is expressed by the tensor product of $n$ rank-1 tensors.

The time and memory complexities of the contraction of such a tensor network depend strongly on the order in which tensors are contracted.
Its time complexity has a lower bound related to the treewidth of the line graph of the tensor network.
Ref.~\cite{chen2018classical} introduced a method to alleviate the memory requirements for the contraction of a tensor network at the expense of a larger time complexity.
This method involves slicing (projecting) certain carefully chosen indices in the network to the different values in their support.
Each slice yields a tensor network that requires less memory to be contracted, although one has to contract a number of tensor networks that scales exponentially in the number of indices sliced.
Refs.~\cite{gray2021hyper,huang2020classical}
made substantial improvements in the optimization of contraction orderings and choice of slices.

These works focused on the contraction of an independent tensor network per output bitstring, which results in a simulation runtime that scales linearly in the number of bitstrings sampled.
Specifically, an algorithm to sample from the output distribution of a quantum circuit is as follows: 1) sample bitstrings uniformly at random; 2) calculate the ideal probabilities for these bitstrings; 3) perform rejection sampling to select a subset of these bitstrings as the output.  The frugal rejection sampling proposed in Ref.~\cite{markov2018quantum} requires computing only about 10 probabilities per output bitstring sampled. We can calculate the probabilities of $\gtrsim$10 very similar  bitstrings with just one contraction, and as we will only select one using rejection sampling and the probabilities are still uncorrelated, the result is the same as the algorithm above~\cite{villalonga2019flexible}. 

Ref.~\cite{pan2022solving} introduced a method to compute amplitudes of a large number of uncorrelated bitstrings with a much lower overhead than linear.
This implies the sparsification of the output of the tensor network as output tensors are being contracted: only those tensor entries that will lead to the computation of an amplitude of a bitstring in a pre-specified set are kept.
In addition, Ref.~\cite{pan2022solving} managed to make use of a special property of the fSim gate~\cite{arute_supremacy_2019} in order to propagate the slicing of certain indices to other ``related’’ indices at no extra cost.
These two advancements allowed Ref.~\cite{pan2022solving} to simulate sampling 1 million uncorrelated bitstrings from the largest circuit of Ref.~\cite{arute_supremacy_2019} in 15 hours using 512 NVIDIA Tesla V100 SXM3 GPUs with 32GB of RAM each.

Similarly, Ref.~\cite{kalachev2021classical} introduced dynamic programming techniques to reuse certain computations across the different instances of the slices taken over a tensor network.
This allowed them to simulate the largest circuit of Ref.~\cite{arute_supremacy_2019} in 14.5 days using 32 NVIDIA Tesla V100 GPUs with 16GB of RAM each.

In the present work we incorporate all of these advancements into a highly efficient simulated annealing optimizer in order to further reduce the time estimates for the simulation of RCS experiments.
Tensor network contraction orderings, choice of sliced indices, sparsification of the output of the circuit, and reuse of intermediate computation across slicing instances, are all taken into account simultaneously in the optimization.
This allows us to reduce the number of FLOPs required for the noisy simulated sampling of 1 million bitstrings from the hardest circuit of Ref.~\cite{arute_supremacy_2019} by about an order of magnitude compared to the requirements of Refs.~\cite{pan2022solving} and \cite{kalachev2021classical} when using a similar cluster with similar GPUs.
When running on a Google Cloud CPU with 12 TB of memory, we estimate FLOP counts about two orders of magnitude lower than in Refs.~\cite{pan2022solving} and~\cite{kalachev2021classical}.
We estimate a runtime of 2 days using a single CPU with a 20\% FLOP efficiency, similar to the efficiency found in Refs.~\cite{gray2021hyper,huang2020classical,pan2022solving}.
Table~I of the main text shows the runtime estimates for the simulation of the experiments of Refs.~\cite{arute_supremacy_2019,wu2021,zhu2022} and the present work when run on the Frontier supercomputer.

\section{Bounds to approximate tensor representations}
\label{app:mps}

The most promising numerical methods for more efficient approximations to random circuit sampling are based on approximate tensor representations~\cite{zhou2020limits, ayral2022density,gray2021hyper}. Let's write the state on $n$ qubits as
\begin{align}
    \ket{\psi} = \sum_{j_1,j_2,\ldots,j_n} \psi_{j_1,j_2,\ldots,j_n} \ket{j_1,j_2,\ldots,j_n}\;.
\end{align}
We study the simplest but illustrative case where the state is approximated with two tensors $M^{(1)}$ and $M^{(2)}$ as~\cite{ayral2022density}
\begin{align}\label{eq:mps2}
    \tilde \psi_{j_1,j_2,\ldots,j_n} = \sum_{\alpha = 1}^\chi M_{j_1,j_2,\ldots,j_l,\alpha}^{(1)}M_{\alpha,j_{l+1},j_{r+2},\ldots,j_n}^{(2)}\;.
\end{align}
The index $\alpha$ is called a virtual index and $\chi$ is called the bond dimension. The state is broken into qubits on the left $[1 \isep l]$ and the right $[r+1 \isep n]$, with the virtual index encoding the entanglement between these spaces. 
The size of the two sub-Hilbert spaces are respectively
$D_1 = 2^l$ and $D_2 = 2^{n-l}$.

Given a quantum state $\ket \psi$ and bond dimension $\chi$, the best approximation of the form ~\eqref{eq:mps2} can be found keeping the largest $\chi$ singular values (Schmidt coefficients) of a matrix with entries corresponding to the amplitudes of $\ket \psi$, and row and column dimensions corresponding to the left and right spaces. This gives the approximation
\begin{align}\label{eq:mps3}
\ket{\tilde \psi} = \frac 1 {\sqrt{\sum_{\alpha = 1}^\chi S_\alpha^2}}\sum_{\alpha = 1}^\chi S_\alpha \ket{l_\alpha}\ket{r_\alpha},
\end{align}
based on the Schmidt decomposition, where the Schmidt coefficients $S$ are ordered from large to small. This
representation is exact if the bond dimension is larger than the Schmidt rank. 

For approximate matrix-product state (MPS) simulations 
\cite{zhou2020limits,ayral2022density}, the initial state is 
a product state with Schmidt rank equal zero. Gates involving only qubits on the left (or
right) merely modify the tensor $M^{(1)}$ (or $M^{(2)}$) without affecting the Schmidt
rank. However, multi-qubit gates applied on qubits belonging to both left and right
partition, generally increase the Schmidt rank. The truncation of the quantum state
to a fixed bond dimension $\chi \ll 2^{\min(l, n-l)}$ eventually reduces the fidelity
to $F = |\!\braket{\psi}{\tilde\psi}\!|^2 = \sum_{\alpha=1}^\chi S_\alpha^2$.

We quantify the
fidelity for random Haar states in Sec.~\ref{app:fid_haar} and arbitrary states in Sec.~\ref{app:fid_any}. We also show that linear XEB remains a good estimator of fidelity in this case, see Sec.~\ref{app:mps_xeb}~\footnote{Ref.~\cite{ayral2022density} claims that using MPS methods the linear XEB scales as the square root of the fidelity, which contradicts our findings.
This is because a) at low number of cycles they compute XEB before the anti-concentration point and b) at high number of cycles they plot the absolute value of the linear XEB (instead of the linear XEB itself).}. We provide both numerical and analytical
bounds  on the bond dimension $\chi$ required to achieve a given fidelity for the simulation of RCS using MPS methods. We give a precise method to compute this lower bound, see Sec.~\ref{app:cliff_purity}. We also pinpoint the number of cycles of a sharp transition to the ``typical'' (quasi-maximum) entanglement in Sec.~\ref{app:reduced_purity} (see also Ref.~\cite{oliveira2007generic}).

In order to be of any practical use, $\chi$ must be much smaller than the Hilbert space dimension of the halves, with $\chi \lesssim 10^{3}$ for most realistic implementations. Fig.~\ref{fig:chi_bound} shows the required $\chi$ to reach a fidelity of $F \approx 10^{-4}$, as a function of the number of qubits and cycles. The reported memory footprint is the required memory to store two complex arrays of sizes $2^{n/2} \times \chi$. For 70 qubits and 24 cycles, the bond dimension $\chi$ required is of the order of $10^{7}$ (with 35 qubits in each half), which is well beyond the  capacity of Frontier.

\subsection{Fidelity for Haar random states}\label{app:fid_haar}
The output state  of a sufficiently deep quantum random circuit can often be approximated as a Haar random state. Therefore, we can explain the fidelity obtained with representation ~\eqref{eq:mps2} using the distribution of singular values of a complex matrix with Gaussian entries~\cite{petz2004asymptotics} and dimension $D_1 \times D_2$, where $D_1$ ($D_2$) is the dimension of the left (right) Hilbert space. We extend the study of Ref.~\cite{ayral2022density} to the case $D_1 \ne D_2$, and we assume without loss of generality $D_1 \le D_2$.

Let $S$ denote singular values and $s = \sqrt{D_1} S$ denote normalized singular values. The distribution of the normalized singular values follows the Mar\v{c}enko–Pastur distribution~\cite{marchenko1967distribution}
\begin{align}
 p_\lambda(s) =
 \frac{1}{\pi}\frac{\sqrt{(\lambda_+^2 - s^2)(s^2 - \lambda_-^2)}}{\lambda s},
\end{align}
with $D_1 \le D_2$, $\lambda = D_1/D_2$, $\lambda_\pm = (1 \pm \sqrt{\lambda})$ and $s\in[\lambda_-,\ \lambda_+]$.
For $\lambda = 1$, $p_\lambda(s)$ reduces to:
\begin{align}
p(s) = \frac{1}{\pi}\sqrt{4 - s^2}.
\end{align}
For a given bond dimension $\chi$ the fraction of singular values is $r = \chi/D_1$. Let us define the cumulative distribution of singular values
\begin{align}
C_\lambda(s^\prime) = \int_{s^\prime}^{\lambda_+} ds\,p_\lambda(s) 
\end{align}
Note that $r = C_\lambda(s^\prime)$ is the fraction of singular values up to a
given normalized singular value $s^\prime$, starting from the largest singular values.  
Therefore, the normalized singular
value corresponding to a given fraction $r$ is:
\begin{align}
s_\lambda(r) = C_\lambda^{-1}(r).
\end{align}
For $\lambda = 1$ one obtains~\cite{ayral2022density}
\begin{align}
s_{\lambda=1}(r) = 2\cos\left(\frac{1}{2}\mathcal{A}^{-1}(\pi r)\right)\;,
\end{align}
where $\mathcal{A}(\theta) = \theta - \sin\,\theta$.

The fidelity for a given fraction $r$ of singular values is given by
\begin{align}
\mathcal{F}_\lambda(r) = \int_{s(r)}^{\lambda_+} ds\,s^2 p_\lambda(s).
\end{align}
Recalling that
\begin{align}
\frac{ds_\lambda(r)}{dr} = \frac{dC_\lambda^{-1}(r)}{dr} =
  \frac{1}{\left.\frac{dC_\lambda}{ds}\right|_{s = s_\lambda(r)}} =
  -\frac{1}{p_\lambda(s_\lambda(r))},\nonumber
\end{align}
one also gets an alternative expression
\begin{multline}\label{eq:fidelity_as_s}
\mathcal{F}_\lambda(r^\prime) = -\int_{r^\prime}^0 dr \frac{1}{p_\lambda(s_\lambda(r))}
  s_\lambda^2(r)\, p_\lambda(s_\lambda(r)) \\= \int_0^{r^\prime} dr\,s_\lambda^2(r)\;.
  \end{multline}
For $\chi \ll D_1$ we have
\begin{align}\label{eq:small_chi_fi}
    \mathcal{F}_\lambda(\chi/D_1) \leq 
    \left.\frac{d\mathcal{F}_\lambda(r)}{dr}\right|_{r=0}\,\frac{\chi}{D_1} = 
    \lambda_+^2\,\frac{\chi}{D_1},
\end{align}

where we used the fact that $\mathcal{F}_\lambda$ is a monotonically increasing
and strictly concave function (that is, 
$d^2\mathcal{F}_\lambda/dr^2 = -2 s_\lambda(r)/p_\lambda(s_\lambda(r)) \leq 0$).
For $\lambda = 1$~\cite{ayral2022density}, 
the bound reduces to $\mathcal{F}_{\lambda=1} \leq \frac{4\chi}{\sqrt{D}}$.
This gives an upper bound for the fidelity of a single projection into the ansatz of
Eq.~\eqref{eq:mps2} once the ideal state is sufficiently entangled.

\subsection{Fidelity bound for arbitrary states}\label{app:fid_any}

\begin{figure}[t!]
    \centering
    \includegraphics[width=0.85\columnwidth]{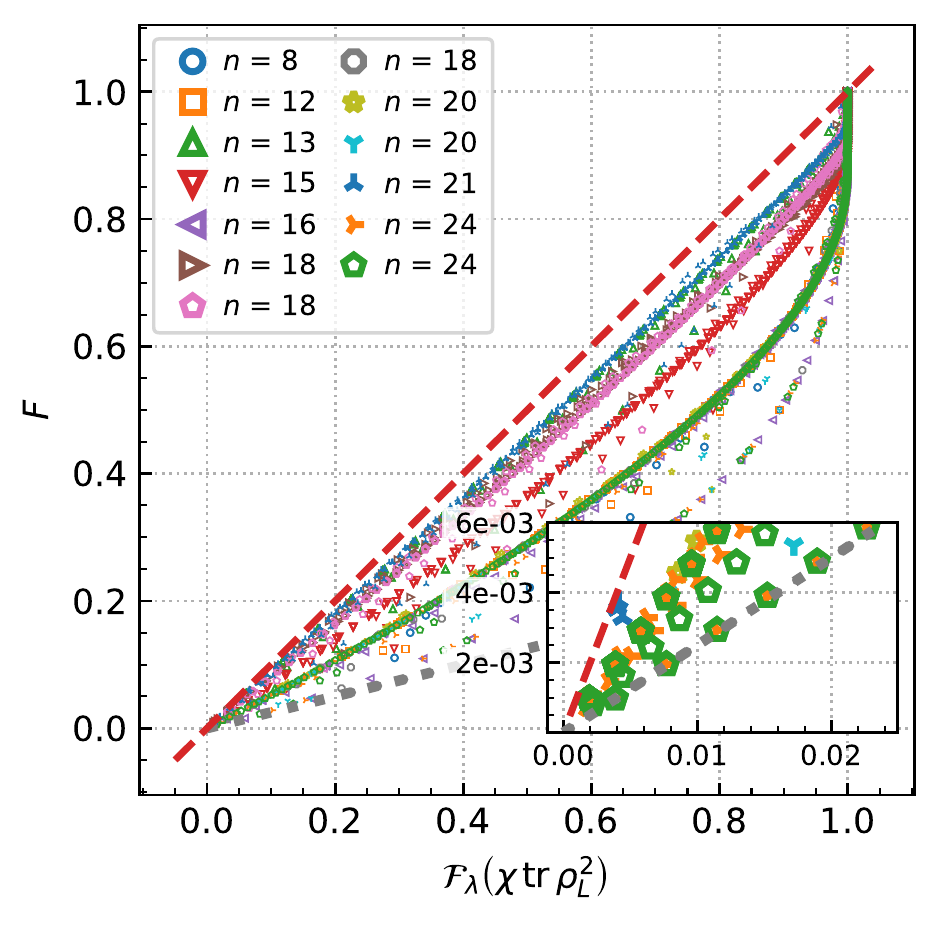} 
    \caption{Numerical fidelity bound. The plot compares the exact fidelity
    $F$ for different Sycamore layouts of different sizes, depths, and bond dimensions $\chi$ to the numerical upper-bound
    $\mathcal{F}_\lambda\left(\chi\, {\tr\!\rho_L^2}\right)$, with $\mathcal{F}_\lambda$ and $\tr\!\rho_L^2$ being respectively the fidelity in the Haar limit and the reduced purity. 
    The dotted-gray line corresponds to the bound $\mathcal{F}_\lambda\left(\chi\, {\tr\!\rho_L^2}\right) \leq 
    \lambda_+^2 \chi\, {\tr\!\rho_L^2} \leq 4\,\chi\, {\tr\!\rho_L^2}$. For all the instances, the pattern ABCDCDAB is used, and the qubits are partitioned in two equal halves using a diagonal cut.}
    \label{fig:fidelity_bound}
\end{figure}

The fidelity of a Schmidt-decomposed state as in Eq.~\eqref{eq:mps3} is
\begin{align}
    F = \sum_{\alpha=1}^{\chi} S_\alpha^2.
\end{align}
Using the Jensen's inequality, it follows that
\begin{align}
    \chi^2\left(\frac{1}{\chi}\sum_{\alpha=1}^{\chi} S_\alpha^2\right)^2 \leq
    \chi \sum_{\alpha=1}^{\chi} S_\alpha^4
\end{align}
and therefore
\begin{align}\label{eq:fidelity_bound_exact}
    F \leq \sqrt{\chi \sum_{\alpha=1}^{\chi} S_\alpha^4} \leq 
    \sqrt{\chi\, \tr \rho_L^2},
\end{align}
with $\tr \rho_L^2 = \sum_{\alpha=1}^{D_1} S_\alpha^4$ being the
reduced purity after tracing out the qubits on the right. Using exact numerics for
small systems (see Fig.~\ref{fig:fidelity_bound}), one can find a tighter bound
\begin{align}\label{eq:fidelity_bound_num}
    F \lesssim 
    \mathcal{F}_\lambda\left(
        \chi \tr \rho_L^2\right) \leq \lambda^2_+\,\chi
    \tr \rho_L^2 ,
\end{align}

with $\mathcal{F}_\lambda$ being the fidelity for Haar random states,
see Eq.~\eqref{eq:fidelity_as_s}, 
and the average is at fixed depth. 
Equations~\eqref{eq:fidelity_bound_exact} and~\eqref{eq:fidelity_bound_num}
gives an upper bound for the fidelity of a single projection for an
arbitrary quantum state in the form~\eqref{eq:mps2}.

We can compare the numerical bound Eq.~\eqref{eq:fidelity_bound_num} with the fidelity
for Haar random states Eq.~\eqref{eq:small_chi_fi} for \mbox{$D_1 = \sqrt D$}. 
The average purity for a Haar random state when $D_1 = \sqrt D$ is:
\begin{multline}\label{eq:purity_asym}
    \aavg{\tr \rho_L^2} = \sum_{\alpha=1}^{D_1} \aavg{S_\alpha^4}  = \frac 1 {D_1^2} \sum_{\alpha=1}^{D_1} \aavg{s_\alpha^4} \\
    = \frac 1 {D_1} \int_0^2 s^4 \frac 1 \pi \sqrt{4 - s^2} d s = \frac 2 {\sqrt D} \;.
\end{multline}
Therefore, for small $\chi / \sqrt{D}$, the numerical bound 
Eq.~\eqref{eq:fidelity_bound_num} is only twice larger than the correct fidelity in this case. 

\subsection{Open and close simulations using approximate tensor representations}\label{app:mps_open_close_sim}

As described in \cite{ayral2022density}, approximated tensor
representations can be used to sample bitstrings with a given
target fidelity $F$. More precisely, authors of \cite{ayral2022density}
present two different protocols: ``open'' and ``close'' simulations.

For open simulations at fixed bond dimension $\chi$,
the circuit $C$ is split in $k$ sub-circuits such that 
$C = C_k\ \cdots\, C_1 C_0$. Starting from the initial state
$\ket{\psi_0} = \ket{0}$, a new approximate state
$\ket{\tilde\psi_1}$ is obtained by truncating the state
$\ket{\psi_1} = C_1\ket{\psi_0}$ using its $\chi$ largest
singular values $\{S_{\alpha,1}\}$ only. The approximate
state $\ket{\tilde\psi_1}$ is then evolved to get
$\ket{\psi_2} = C_2\ket{\tilde\psi_1}$, which is again truncated
to its largest $\chi$ singular values to obtain
the approximate state $\ket{\tilde\psi_2}$.
Calling $f_k = |\braket{\psi_k}{\tilde\psi_k}|^2 = 
\sum_{\alpha=1}^\chi S_{\alpha,k}^2$ the partial
fidelity of the approximate state $\ket{\tilde\psi_k}$,
Ref.~\cite{ayral2022density} shows that, for random circuits,
the final fidelity can be expressed as the product of all the
partial fidelity, that is 
$F = |\braket{\tilde\psi_k}{C|\psi_0}|^2 = f_1 f_2 \cdots f_k$.

For a large number of qubits, computing $\ket{\psi_i} = C_i \ket{\psi_{i-1}}$ and finding its
singular values to get the approximate state $\ket{\tilde\psi_i}$ is
numerically intractable.
To overcome this limitation, authors of \cite{ayral2022density} introduce
a variational approach to compute $\ket{\tilde\psi_i}$ without
the need of the intermediate state $C_i \ket{\tilde\psi_{i-1}}$
and without explicitly computing its singular values. More precisely,
for any $C_i$, a new random approximate tensor 
$\ket{\tilde\phi_{i}}$ of bond dimension $\chi$ is used to compute
the objective:
\begin{equation}
    \tilde f_i(\ket{\tilde\phi_i}) =
    |\braket{\tilde\psi_{i-1}|C}{\tilde\phi_i}|^2.
\end{equation}
Recalling that both $\ket{\tilde\psi_{i-1}}$ and $\ket{\tilde\phi_i}$
are MPS of bond dimension $\chi$, computing
$\tilde f_i$ for sufficiently shallow $C_i$ and small $\chi$ is doable even for 
a large number of qubits \cite{ayral2022density}. Using the update strategy
proposed in \cite{ayral2022density}, one can find
$\ket{\tilde\psi_i}$ as
\begin{equation}
  \ket{\tilde\psi_i} = \underset{\ket{\tilde\phi_i}}{\rm argmax}
    \tilde f_i(\ket{\tilde\phi_i}).
\end{equation}

Because the quantum system becomes more and more entangled
by applying the sub-circuits $C_i$,
one expects that $f_{i+1} \leq f_{i}$, reaching the saturation value
of $\mathcal{F}_\lambda(\chi / D_1)$
when the quantum state reaches the random Haar limit.  
Close simulations at fixed bond dimension $\chi$ are useful to
sample bitstrings with an improved target fidelity than
the corresponding open simulations with the same bond dimension.
To start, the circuit is split in three parts $C = C_2 C_M C_1$.
Using the open simulation protocol, the approximate state
$\ket{\tilde\Psi_1}$ is computed by using $\ket{0}$ as
initial state and $C_1$ as circuit. Similarly, the approximate state
$\ket{\tilde\Psi_2(x)}$ is computed by using the open simulation
protocol with $\ket{x}$ as initial circuit and $C_2^\dagger$ as circuit.
Finally, the approximate amplitude $\tilde a_x$ is computed as:
\begin{equation}\label{eq:ampl_close_sim}
    \tilde a_x = \braket{\Psi_2(x)|C_M}{\Psi_1}.
\end{equation}

As discussed in \cite{ayral2022density}, $\tilde a_x$ can be seen
as amplitudes extracted from a quantum state with a fidelity
$F = F_1 \overline{F_2}$, with $F_1$ and $\overline{F_2}$
being the fidelity of $\ket{\tilde\Psi_1}$ and the average
fidelity of $\ket{\tilde\Psi_2(x)}$ respectively. Because both $\ket{0}$
and $\ket{x}$ are MPS with bond dimension $0$,
$F_1 \overline{F_2}$ might be larger than the fidelity one obtains with the open simulation protocol. However, unlike the open simulation,
multiple runs are needed to get the required number of bitstrings.

\begin{figure}[t!]
    \centering
    \includegraphics[width=0.85\columnwidth]{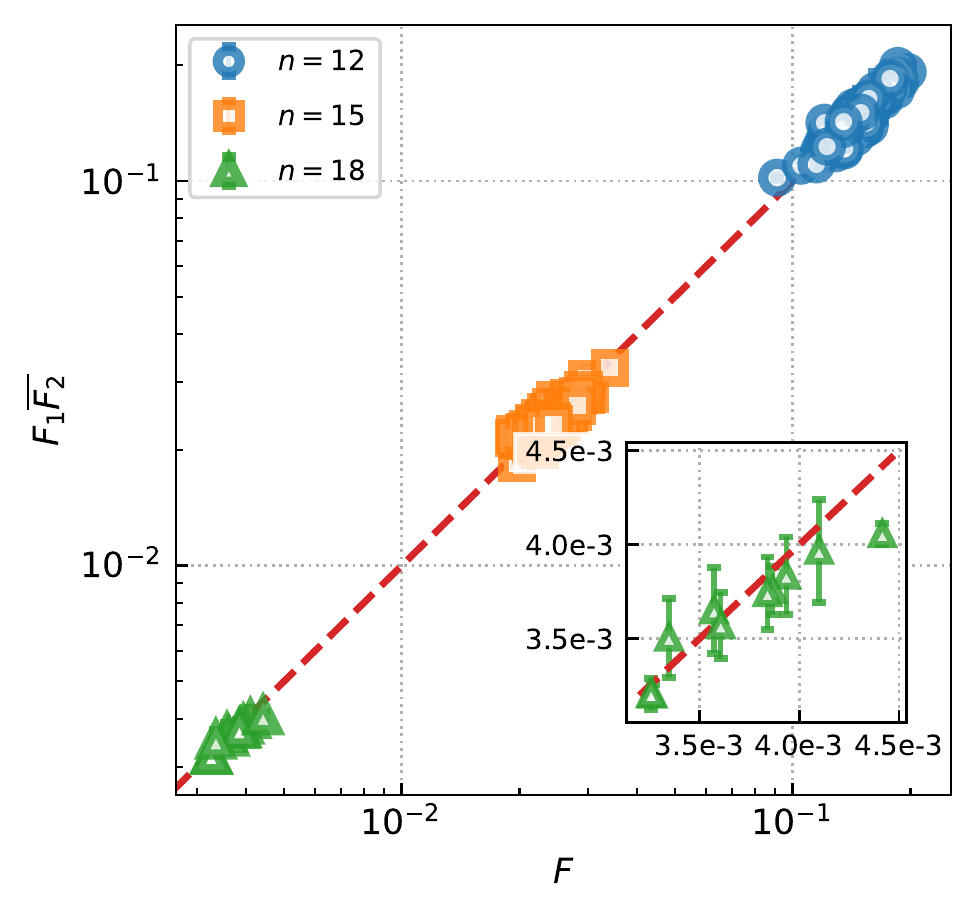} 
    \caption{Fidelity for close simulations $F_1 \overline{F_2}$ 
    (see App.~\ref{app:mps_open_close_sim}), compared to the exact fidelity $F$. The bond dimension is fixed to
    $\chi = 8$. Each point correspond to a different circuit, with the
    average being performed over bitstrings. Circuits are split in three
    parts of $|C_1|=8$, $|C_M|=4$ and $|C_2|=8$ cycles respectively. For all the instances, the pattern ABCDCDAB is used, and the qubits are partitioned in two equal halves using a diagonal cut.}
    \label{fig:fidelity_close_simulation}
\end{figure}

Fig.~\ref{fig:fidelity_close_simulation} shows the fidelity
of amplitudes sampled using the close simulation protocol at
fixed bond dimension \mbox{$\chi = 8$}, compared
to the exact fidelity. Each point correspond to a different
circuit, and circuits are split so that $C_1$ and $C_2$ 
contain $8$ cycles while $C_M$ contains $4$ cycles.
The average is performed over bitstrings. 

\subsection{XEB for approximate tensor representations}\label{app:mps_xeb}
We now explain why, at sufficiently large depth, XEB is still a good estimator of fidelity for approximate tensor representations. The optimal approximate tensor representation  Eq.~\eqref{eq:mps3} is based on the Schmidt decomposition, that is
\begin{align}\label{eq:mps4}
\ket{\tilde \psi} = \frac 1 {\sqrt F}\sum_{\alpha = 1}^\chi S_\alpha \ket{\nu_\alpha}\;,
\end{align}
where $\ket{\nu_\alpha} =  \ket{l_\alpha}\ket{r_\alpha}$ and $F = \sum_{\alpha = 1}^\chi S_\alpha^2$ is the fidelity. 

We first show that in the limit of large depth the left and right singular vectors $\{\ket{l_\alpha}\}$ and $\{\ket{r_\alpha}\}$ are Haar random states. Note that these are the singular vectors of a matrix $M$ with Gaussian entries (a Haar random matrix), see App.~\ref{app:fid_haar}. For any unitaries $U$ and $V$ we also have that $UMV^\dagger$ is a matrix with Gaussian entries. This implies that the distribution of singular vectors is invariant under unitary transformation, that is, the singular vectors are Haar random. Therefore, we approximate the singular vectors as having i.i.d Gaussian random real and imaginary parts. 

We first give an explanation for why XEB is a good estimator of fidelity in this case, followed by a formal proof. We can write
\begin{align}
    \ket{\psi} = \sqrt F \ket{\tilde \psi} + \sqrt{1-F} \ket{\perp}
\end{align}
where
\begin{align}
    \ket{\perp} = \frac 1 {\sqrt{1-F}} \sum_{\alpha = \chi+1}^{D_1} S_\alpha \ket{\nu_\alpha}\;.
\end{align}
In the case of linear XEB, $f(p_j) = D p_j$, we have
\begin{multline}
   D \sum_j \tilde p_j p_j = F \sum_j D \tilde p_j^2 + (1-F) D \sum_j \tilde p_j \perp_j \\+ \sqrt{F(1-F)} D \sum_j 2 {\rm Re} (\braket j {\tilde \psi}\braket{\perp}{j})\;,
\end{multline}
where $p_j$ are the ideal probabilities $p_j = |\braket j \psi|^2$ and $\tilde p_j = |\braket j {\tilde \psi}|^2$. 
For $D_1 \gg \chi \gg 1$ both $\ket{\tilde \psi}$ and $\ket{\perp}$ converge to independent Haar random states. Therefore 
\begin{align}
  D \sum_j \tilde p_j^2 & \simeq 2 \\
  D \sum_j \tilde p_j &\perp_j \simeq D^2 \aavg{ \tilde p_j \perp_j}\nonumber \\ & \simeq D^2 \aavg{ \tilde p_j} \aavg{\perp_j} \simeq 1\\
  2 D \sum_j {\rm Re} (\braket j {\tilde \psi}\braket{\perp}{j}) &\simeq 0\;,
\end{align}
where $\aavg \cdot$ denotes average over random states $\ket \psi$ (or circuits, see App.~\ref{app:rcs_distribution}). Therefore 
\begin{align}
   D \sum_j \tilde p_j p_j \simeq F\;,
\end{align}
as we wanted. 

\begin{figure}[t!]
    \centering
    \includegraphics[width=0.85\columnwidth]{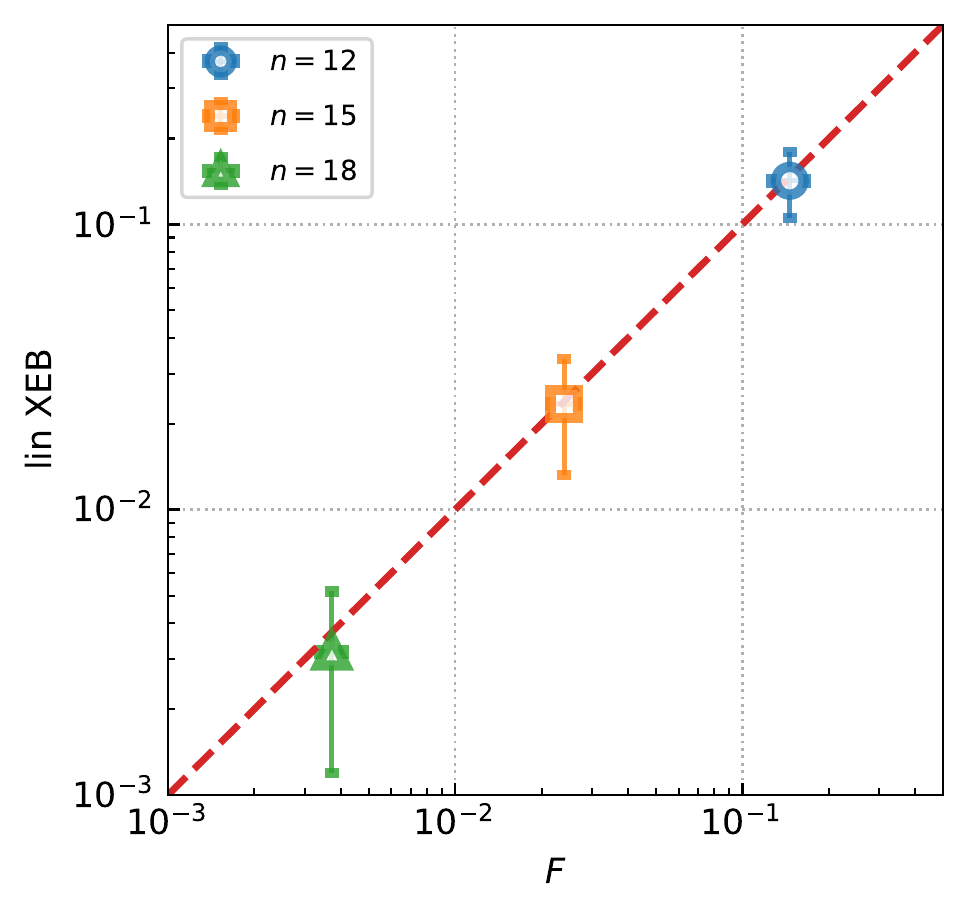} 
    \caption{The plot compares the exact fidelity $F$ to the linear XEB for close simulations with a fixed bond dimension of $\chi = 8$. Circuits
    for the close simulations are split in three parts of $|C_1|=8$, $|C_M|=4$ 
    and $|C_2|=8$ cycles respectively. The error is obtained by averaging over multiple 
    circuits. For all the instances, the pattern ABCDCDAB is used.}
    \label{fig:lin_xeb_close_simulation}
\end{figure}

We note that for very small $\chi$, such as $\chi = 1$, linear XEB 
overestimates the fidelity. Indeed, in this case we have
\begin{multline}
  D \sum_j \tilde p_j^2 = D \sum_{a,b}  |\braket a {l_\alpha}|^4
  |\braket b {r_\alpha}|^4 \simeq 4\;.
\end{multline}
Nevertheless, this case is uninteresting for simulations, 
as $F \simeq \lambda_+^2 / D_1$ corresponds to a very small fidelity.
Numerical results for small
quantum systems, Fig.~\ref{fig:lin_xeb_close_simulation}, 
confirm the correspondence between exact fidelity and the XEB
for approximate quantum states using close simulations.

We now give a detailed proof. Below till the end of this subsection, for the simplicity of notation, we will use $\mathbb{E}[\cdot]$ to denote the average over states $\langle\langle \cdot \rangle\rangle$.

We recall that for a $L\times L$ random $\beta-$Haar distributed
unitary $Q$ ($\beta=1$ for real and $\beta=2$ for complex and $\beta=4$ for quaternion unitaries)
with entries $q_{i,j}$ we have the following expectations (see Table IV in \cite{movassagh2010isotropic}): 
\begin{center}
\begin{tabular}{|c|c|}
\hline 
expectation values \tabularnewline
\hline 
\hline 
$\mathbb{E}[|q_{i,j}|^{4}]=\frac{\beta+2}{L(\beta L+2)}$\tabularnewline
\hline 
$\mathbb{E}[|q_{i,j}q_{i,k}|^{2}]=\frac{\beta}{L(\beta L+2)}$ \tabularnewline
\hline

\end{tabular}
\par\end{center}

We remind the reader that the exact and approximation outputs (from Eqs.~\eqref{eq:mps3} and~\eqref{eq:mps4}) are
\begin{eqnarray*}
|\psi\rangle & := & \sum_{\alpha=1}^{\sqrt{D}}S_{\alpha}|l_{\alpha}\rangle\otimes|r_{\alpha}\rangle:=\sum_{\alpha=1}^{\sqrt{D}}S_{\alpha}|\nu_{\alpha}\rangle\\
|\tilde{\psi}\rangle & := & \frac{1}{\sqrt{F}}\sum_{\alpha=1}^{\chi}S_{\alpha}|l_{\alpha}\rangle\otimes|r_{\alpha}\rangle:=\frac{1}{\sqrt{F}}\sum_{\alpha=1}^{\chi}S_{\alpha}|\nu_{\alpha}\rangle\,.
\end{eqnarray*}
Let $p_{j}:=|\langle j|\psi\rangle|^{2}$ and $\tilde{p}_{j}:=|\langle j|\tilde{\psi}\rangle|^{2}$. We have
$\langle j|\psi\rangle=\sum_{\alpha=1}^{\sqrt{D}}S_{\alpha}\langle j|\nu_{\alpha}\rangle$
\begin{eqnarray*}
p_{j} & = & \left|\sum_{\alpha=1}^{\sqrt{D}}S_{\alpha}\langle j|\nu_{\alpha}\rangle\right|^{2}=\sum_{\alpha,\beta=1}^{\sqrt{D}}S_{\alpha}S_{\beta}\;\langle j|\nu_{\alpha}\rangle\langle \nu_{\beta}|j\rangle\;,\\
\tilde{p}_{j} & = & \frac{1}{F}\left|\sum_{\alpha=1}^{\chi}S_{\alpha}\langle j|\nu_{\alpha}\rangle\right|^{2}=\frac{1}{F}\sum_{\alpha,\beta=1}^{\chi}S_{\alpha}S_{\beta}\;\langle j|\nu_{\alpha}\rangle\langle \nu_{\beta}|j\rangle\;.
\end{eqnarray*}
As stated above $\langle j|\nu_{\alpha}\rangle=\langle j|(|l_{\alpha}\rangle\otimes|r_{\alpha}\rangle)=v_{\alpha}w_{\alpha}$
is simply a product of two complex numbers, where $v_\alpha$ and $w_\alpha$ represent an entry of $|l_{\alpha}\rangle$ and $|r_{\alpha}\rangle$ respectively.
\begin{lem}
$D\mathbb{E}\left[\sum_{j}p_{j}\tilde{p}_{j}\right]-1\approx F+O(1/\sqrt{D})$,
where the expectation is over the random haar vectors $|l_{\alpha}\rangle$
and $|r_{\alpha}\rangle$ and singular values $S_{\alpha}$.
\end{lem}

\begin{proof}
We first compute $\mathbb{E}[p_{j}\tilde{p}_{j}]$. Treating singular
values as independent from the vectors we have
\begin{eqnarray}
\mathbb{E}[p_{j}\tilde{p}_{j}]  = & \frac{1}{F}\sum_{\alpha,\beta=1}^{\sqrt{D}}\sum_{c,d=1}^{\chi}\{\mathbb{E}[S_{\alpha}S_{\beta}S_{c}S_{d}]\times\nonumber\\&\quad\mathbb{E}[\langle j|\nu_{\alpha}\rangle\langle \nu_{\beta}|j\rangle\langle j|\nu_{c}\rangle\langle \nu_{d}|j\rangle]\}\nonumber \\
  = & \frac{1}{F}\sum_{\alpha,\beta=1}^{\sqrt{D}}\sum_{c,d=1}^{\chi}\mathbb{E}\{[S_{\alpha}S_{\beta}S_{c}S_{d}]\times\nonumber\\
  &\quad\mathbb{E}[v_{\alpha}w_{\alpha}\overline{v_{\beta}w_{\beta}}v_{c}w_{c}\overline{v_{d}w_{d}}]\}\;.\label{eq:Epp}
\end{eqnarray}
First of all the vectors $|l_{\alpha}\rangle$ and $|r_{\alpha}\rangle$
are independent and power of their entries vanish over the complex
field because of the invariance of Haar measure. Let
\begin{eqnarray}
g_{j,\chi,D}  := & \frac{1}{F}\sum_{\alpha,\beta=1}^{\sqrt{D}}\sum_{c,d=1}^{\chi}\{\mathbb{E}[S_{\alpha}S_{\beta}S_{c}S_{d}]\;\mathbb{E}[v_{\alpha}v_{c}\overline{v_{\beta}v_{d}}]\times\nonumber\\
&\qquad\mathbb{E}[w_{\alpha}w_{c}\overline{w_{\beta}w_{d}}]\}.\label{eq:g_j}
\end{eqnarray}
The non-zero contributions in the sum are three cases:
\[
\begin{array}{cc}
\text{case 1:} & \alpha=\beta\ne c=d\\
\text{case 2:} & \alpha=d\ne c=\beta\\
\text{case 3:} & \alpha=d=c=\beta
\end{array}
\]
Before we indulge in computing these cases  one by one, let us
focus on the expectation with respect to the \textit{entries }as the
entries of a Haar unitary have correlations. When $\alpha=\beta\ne c=d$
and $v_{\alpha}$ and $v_{c}$ do not belong to the same row of the
unitary matrix induced by singular value decomposition whose columns
are $|l_{\alpha}\rangle$, we have
\[
\mathbb{E}[|v_{\alpha}^{(1)}|^{2}|v_{c}^{(2)}|^{2}]=\frac{1}{D}\sqrt{D}(\sqrt{D}-1)\frac{1}{D}=\frac{1}{D}(1-\frac{1}{\sqrt{D}}).
\]
However, when $\alpha=\beta\ne c=d$ and $v_{\alpha}$ and $v_{c}$ come
from the \textit{same} row or column of the unitary matrix, then from the second row of the Table above we have
\begin{eqnarray}
\mathbb{E}[|v_{\alpha}^{(1)}|^{2}|v_{c}^{(1)}|^{2}]&=&\frac{1}{D}\sqrt{D}\frac{\beta}{\sqrt{D}\left(\beta\sqrt{D}+1\right)}\nonumber\\
&=&\frac{1}{D}\frac{\beta}{\left(\beta\sqrt{D}+1\right)}\nonumber\\
&=&\frac{1}{D\sqrt{D}}\left(1-\frac{1}{\beta\sqrt{D}}\right)\;.\label{eq:sameRow}
\end{eqnarray}
The latter is of lower order. We ignore the terms of lower
order below and proceed to calculate the three cases by assuming that
the entries do not come from the same row or column of the inducing random haar unitary
matrix. 

Recall from the asymptotic scaling of the purity Eq.~\eqref{eq:purity_asym} that $\sum_{c=1}^{\chi}\mathbb{E}[S_{c}^{4}]\le 2/\sqrt{D}$.

\textbf{Case 1}: $\alpha=\beta\ne c=d$. We have up to $O(1/\sqrt{D})$
from Eq.~\eqref{eq:g_j}
\begin{align}
g_{j,\chi,D} & =  \frac{1}{F}\sum_{\alpha=1,\alpha\ne c}^{\sqrt{D}}\sum_{c=1}^{\chi}\mathbb{E}[S_{\alpha}^{2}S_{c}^{2}]\;\mathbb{E}[|v_{\alpha}|^{2}|v_{c}|^{2}]\mathbb{\;E}[|w_{\alpha}|^{2}|w_{c}|^{2}]\nonumber \\
 & =  \frac{1}{FD^{2}}\sum_{\alpha=1,\alpha\ne c}^{\sqrt{D}}\sum_{c=1}^{\chi}\mathbb{E}[S_{\alpha}^{2}S_{c}^{2}]\nonumber \\
 & =  \frac{1}{FD^{2}}\left(\sum_{\alpha=1}^{\sqrt{D}}\sum_{c=1}^{\chi}\mathbb{E}[S_{\alpha}^{2}S_{c}^{2}]-\sum_{c=1}^{\chi}\mathbb{E}[S_{c}^{4}]\right)\nonumber \\
  & =  \frac{1}{FD^{2}}\left(F\sum_{\alpha=1}^{\sqrt{D}}\mathbb{E}[S_{\alpha}^{2}]-\sum_{c=1}^{\chi}\mathbb{E}[S_{c}^{4}]\right)\nonumber \\
  & =  \frac{1}{FD^{2}}\left(F\sum_{\alpha=1}^{\sqrt{D}}\mathbb{E}[S_{\alpha}^{2}]-O(1/\sqrt{D})\right)\nonumber \\
 & =  \frac{1}{D^{2}}\left(1-O(1/\sqrt{D})\right) .\label{eq:case1}
\end{align}

\textbf{Case 2}: $\alpha=d\ne c=\beta$. We have up to $O(1/\sqrt{D})$
from Eq.~\eqref{eq:g_j}
\begin{align}
g_{j,\chi,D} & =  \frac{1}{F}\sum_{\alpha=1,\alpha\ne c}^{\chi}\sum_{c=1}^{\chi}\mathbb{E}[S_{\alpha}^{2}S_{c}^{2}]\;\mathbb{E}[|v_{\alpha}|^{2}|v_{c}|^{2}]\;\mathbb{E}[|w_{\alpha}|^{2}|w_{c}|^{2}]\nonumber \\
 & =  \frac{1}{FD^{2}}\sum_{\alpha=1,\alpha\ne c}^{\chi}\sum_{c=1}^{\chi}\mathbb{E}[S_{\alpha}^{2} S_{c}^{2}]\nonumber \\
  & =  \frac{1}{FD^{2}}\left(\sum_{\alpha=1}^{\chi}\sum_{c=1}^{\chi}\mathbb{E}[S_{\alpha}^{2} S_{c}^{2}]-\sum_{c=1}^\chi \mathbb{E}[S^4_c]\right)\nonumber \\
    & =  \frac{1}{FD^{2}}\left(\sum_{\alpha=1}^{\chi}\sum_{c=1}^{\chi}\mathbb{E}[S_{\alpha}^{2} S_{c}^{2}]-O(1/\sqrt{D})\right)\nonumber \\
        & =  \frac{1}{FD^{2}}\left(F\sum_{\alpha=1}^{\chi} \mathbb{E}[S_{c}^{2}]-O(1/\sqrt{D})\right)\nonumber \\
 & =  \frac{1}{D^{2}}\left(\sum_{\alpha=1}^{\chi}\mathbb{E}[S_{\alpha}^{2}]-O(1/\sqrt{D})\right)\nonumber \\
 & =  \frac{F}{D^2}\left(1-O(1/\sqrt{D})\right)\nonumber
 .\label{eq:case2}
\end{align}

\textbf{Case 3}: $\alpha=\beta=c=d$. From Eq.~\eqref{eq:g_j} and the first row of the Table above, we have 
\begin{eqnarray}
g_{j,\chi,D} & = & \frac{1}{F}\sum_{c=1}^{\chi}\mathbb{E}[S_{c}^{4}]\;\mathbb{E}[|v_{c}|^{4}]\mathbb{E}[|w_{c}|^{4}]\nonumber \\
 & = & \frac{1}{F}\left(\frac{\beta+2}{2\sqrt{D}(\sqrt{D}+1)}\right)^{2}\sum_{c=1}^{\chi}\mathbb{E}[S_{c}^{4}]\nonumber \\
 & \approx & \frac{4}{D^{2}}\frac{\sum_{c=1}^{\chi}\mathbb{E}[S_{c}^{4}]}{F}\left(1-\frac{2}{D^{1/2}}\right)\nonumber\\
 &=& O(D^{-5/2}).\label{eq:case3}
\end{eqnarray}

We can now compute the desired result. First of all from Eq.~\eqref{eq:Epp}
and the above 3 cases we have
\[
\mathbb{E}\sum_{j}\tilde{p}_{j}p_{j}=\frac{1}{D}+\frac{F}{D}+O(D^{-3/2})\,.
\]
Therefore,
\begin{equation}
D\mathbb{E}\left[\sum_{j}\tilde{p}_{j}p_{j}\right]-1=F+O(1/\sqrt{D}),\label{eq:final}
\end{equation}
which proves our result. 
\end{proof}

We now consider the special case of $\chi=1$ where the approximate state $|\tilde{\psi}\rangle$ is taken to be a product state.
\begin{cor}
When $\chi=1$,  then $D\mathbb{E}\left[\sum_{j}\tilde{p}_{j}p_{j}\right]-1=3F+O(F/\sqrt{D})$.
\end{cor}
\begin{proof}
In this case 
\begin{eqnarray}
\mathbb{E}[p_{j}\tilde{p}_{j}] & = &\sum_{\alpha,\beta=1}^{\sqrt{D}}\mathbb{E}[S_{\alpha}S_{\beta}]\;\mathbb{E}[v_{\alpha}w_{\alpha}\overline{v_{\beta}w_{\beta}}|v_{1}|^2 |w_{1}|^2]\qquad\label{eq:EppChi1} 
\end{eqnarray}
The only non-zero contributions come from $\alpha=\beta=1$ and $\alpha,\beta\ge 2$. We have 
\begin{eqnarray}
\mathbb{E}[p_{j}\tilde{p}_{j}] & = &\mathbb{E}[S^2_{1}]\;\mathbb{E}[|v_{1}|^4 |w_{1}|^4]\nonumber \\
&+&\sum_{\alpha,\beta=2}^{\sqrt{D}}\mathbb{E}[S_{\alpha}S_{\beta}]\;\mathbb{E}[v_{\alpha}w_{\alpha}\overline{v_{\beta}w_{\beta}}|v_{1}|^2|w_{1}|^2]\nonumber\\
& = & F\; \mathbb{E}[|v_1|^4] \mathbb{E}[|w_1|^4]\nonumber\\
&+& \sum_{\alpha=2}^{\sqrt{D}}\mathbb{E}[S^2_{\alpha}]\;\mathbb{E}[|v_{\alpha}|^2]\mathbb{E}[|w_{\alpha}|^2]\mathbb{E}[|v_{1}|^2]\mathbb{E}[ |w_{1}|^2]\nonumber\\
& = & F\; \mathbb{E}[|v_1|^4] \mathbb{E}[|w_1|^4] + \frac{1}{D^2}(1-F)\nonumber\\
& = & F\;\left(\frac{\beta+2}{2\sqrt{D}(\sqrt{D}+1)}\right)^{2} + \frac{1}{D^2}(1-F)\nonumber\\
& = & F\;\frac{4}{D^2}\frac{1}{\left(1+1/\sqrt{D}\right)^{2}} + \frac{1}{D^2}(1-F)\nonumber\\
& = & F\;\frac{4}{D^2}\left(1-2/\sqrt{D}\right)+ \frac{1}{D^2}(1-F)\nonumber\\
& = & \frac{3F}{D^2}+ \frac{1}{D^2}+O(F\;D^{-5/2}).\nonumber
\end{eqnarray}
We conclude the desired final result
\[
D\sum_j\mathbb{E}[p_{j}\tilde{p}_{j}] = 3F+ 1+O(F\;D^{-1/2}).
\]
\end{proof}

\subsection{Quantifying entanglement with Clifford circuits}\label{app:cliff_purity}

We derived an analytical and numerical bound on the fidelity of the tensor product approximation, Eqs.~\eqref{eq:fidelity_bound_exact} and \eqref{eq:fidelity_bound_num}, which depend on the reduced purity. We now study the reduced purity growth rate in phase matched circuits (see SM \ref{app:phase_match}). The average Pauli error between the fsim gates used in the experiment and the Clifford gate ${\rm iSWAP}^{-1}$ is $\sim 1\%$. For the purposes of quantifying reduced purity growth we therefore approximate fsim gates with ${\rm iSWAP}^{-1}$. The one-qubit gates are $Z^p X^{1/2} Z^{-p}$ with $p \in \{-1, -1/4, -1/2, \ldots, 3/4\}$~\footnote{We study the circuit ensemble consisting of random choices of one-qubit gates $Z^p X^{1/2} Z^{-p}$ with $p \in \{-1, -1/4, -1/2, \ldots, 3/4\}$ and the two-qubit gate ${\rm iSWAP}^{-1}$. We obtain the same ensemble if we use ${\rm iSWAP}$ instead of ${\rm iSWAP}^{-1}$. This follows from ${\rm iSWAP}^{-1} = {\rm iSWAP} \cdot (Z \otimes Z) = (Z \otimes Z) \cdot {\rm iSWAP} $ and the fact that the set $\{Z Z^p X^{1/2} Z^{-p}\}$ is the same as $\{Z^p X^{1/2} Z^{-p} Z\}$. Therefore we can move Z gates between layers as we transform ${\rm iSWAP}^{-1}$'s to ${\rm iSWAP}$'s.}. We can study a related ensemble of Clifford circuits by reducing the parameter $p$ of the one-qubit gates to the set $p \in \{-1, -1/2, 0, 1/2\}$.  Note that the reduced purity produced by Clifford circuits is efficient to calculate numerically~\cite{audenaert2005entanglement}.

Consider the average purity of the reduced state $\aavg{\tr \rho_L^2}$,  where $\rho_L$ is the partial trace of $\ket \psi$ on the left qubits. We now show that layer by layer this average is the same for the random circuits and the corresponding Clifford circuits of the previous paragraph. One intuition why this might be true is that Clifford circuits are a two-design. Nevertheless, we are interesting in the growth rate, not the average over Clifford circuits. Therefore, we use a different technique. 

First note that $\tr \rho_L^2$ can be written as 
\begin{align}
    \sum_{z_L} \bra{z_L} \sum_{z_R} \bra{z_R} \rho \ket{z_R} \ket{z_L} \otimes \bra{z_L} \sum_{z_R} \bra{z_R} \rho \ket{z_R} \ket{z_L}\;,\nonumber
\end{align}
where $\{z_L\}$ ($\{z_R\}$) is a basis for the left (right) patch. Therefore, this quantity can be calculated from two replicas of the output state of the circuit of interest $\rho \otimes \rho$. As shown in SM~\ref{app:popdyn} and Ref.~\cite{mi2021information}, for the circuits of interest, the average of an observable with two replicas can be calculated with a Markov chain describing the evolution of the density matrix in the basis of Pauli strings. If we denote the basis of normalized Pauli strings as $\{s\}$ we can write any operator as $\rho = \sum_s \tr( \rho s) s$. It follows that 
\begin{align}
    U \rho\, U^\dagger = \sum_s \tr(\rho s) \sum_{s'} \tr(s' U s\, U^\dagger) s'\;.
\end{align}
Using this relation repeatedly we reduce the evolution of a circuit to the evolution of gates over Pauli strings. Furthermore, the average over random circuits is composed of averages over the corresponding set $\{g\}$ of random gates. In the case of the evolution of two replicas the elementary step is
\begin{align}
    \E_g&[\sigma^\alpha \otimes \sigma^{\alpha'}] = \\ &\sum_{\beta,\beta'} \frac 1 {|g|} \sum_g \tr(\sigma^\beta \otimes \sigma^{\beta'} g \otimes g\, \sigma^\alpha \otimes \sigma^{\alpha'}\, g^\dagger \otimes g^\dagger ) \,\sigma^\beta \otimes \sigma^{\beta'}\nonumber
\end{align}
where the tensor product is between the two replicas, and each gate is the same for both replicas. 
The initial state of each qubit with two replicas is 
\begin{align}
    \ket 0 \bra 0 \otimes \ket 0 \bra 0 = \frac {1 + Z} 2 \otimes \frac {1 + Z} 2\;.
\end{align}
We obtain, both the Clifford and non-Clifford single-qubit gates above,
\begin{align}
    \E[II] &= II \\
    \E[IZ] &= \E[ZI] = 0 \\
    \E[ZZ] &= \frac 1 2 ( XX + YY) \equiv \perp
\end{align}
where we also define the symbol $\perp$. For shorthand, we also introduce the notation $\mathbb{Z} = ZZ$ and $\mathbb{I} = II$. We also obtain, again for both the Clifford and non-Clifford gates,
\begin{align}
    \E[\perp] = \frac 1 2 (\mathbb{Z} + \perp)\;.
\end{align}
Finally the transformation from applying ${\rm iSWAP}^{-1}$ to two qubits in each replica gives
\begin{align}
    \mathbb{I}\mathbb{Z} &\to \mathbb{Z} \mathbb{I} \\
    \mathbb{I}\perp &\to \perp \mathbb{Z} \\
    \mathbb{Z}\mathbb{Z} &\to \mathbb{Z} \mathbb{Z} \\
    \perp \perp &\to \perp \perp \\
    \perp \mathbb{Z} &\to \mathbb{I} \perp
\end{align}
Therefore the average purity of the reduced state $\aavg{\rho_L}$ is the same for both ensembles, as the transformations for the initial state of interest is the same for the average of Clifford and non-Clifford gates. 

Entanglement is typically measured with the von Neumann entropy $-\rho_L \tr \rho_L$. Using Jensen's inequality this can be bounded with the R\'enyi entropy 
\begin{align}
    -\tr \rho_L \log_2 \rho_L \ge - \log_2  \tr \rho_L^2\;.
\end{align}
The average R\'enyi entropy can be bounded with purity of the reduced states using Jensen's inequality as
\begin{align}
   -  \aavg{\log_2  \tr \rho_L^2} \ge -\log_2 \aavg{\tr \rho_L^2}\;. 
\end{align}

\begin{figure}[t!]
    \centering
    \includegraphics[width=0.85\columnwidth]{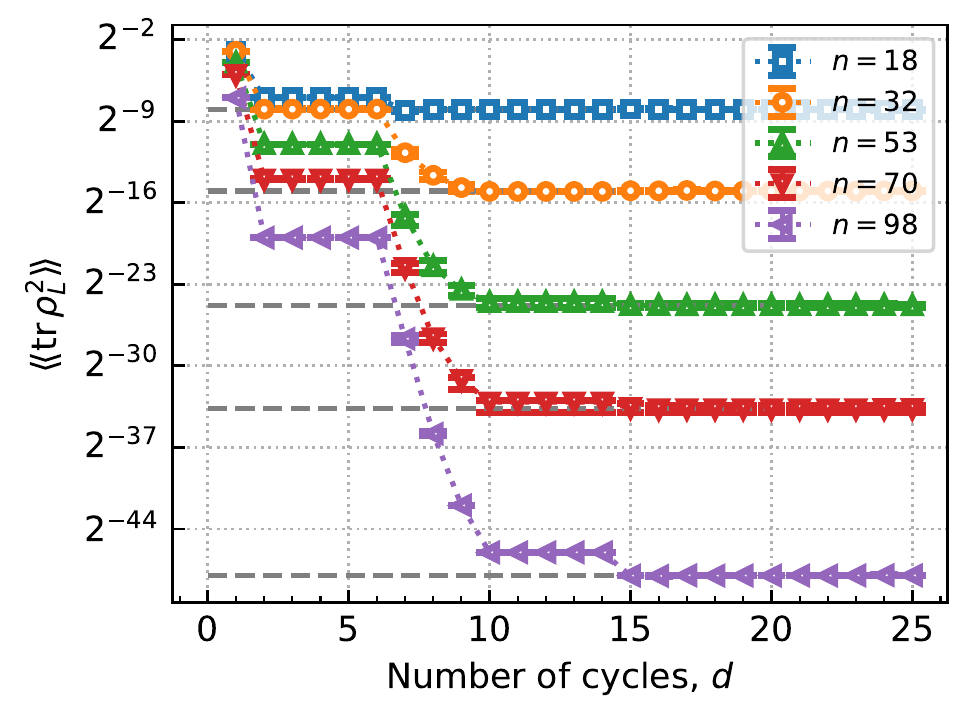}\\
    \includegraphics[width=0.85\columnwidth]{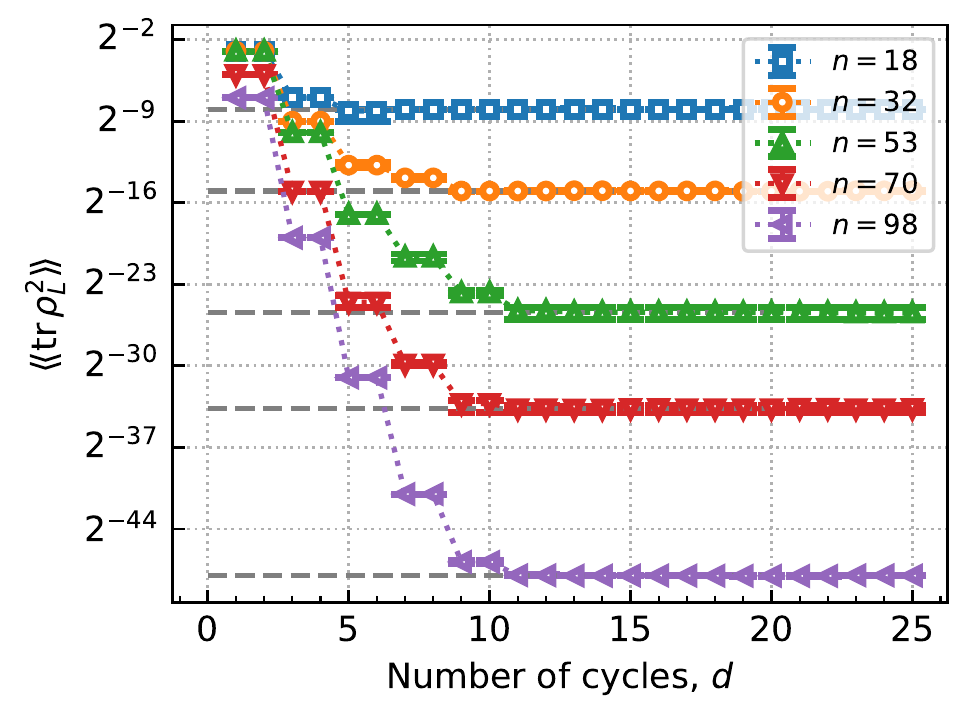} 
    \caption{The plots show the reduced purity as a function of the number of cycles, for different circuit sizes and cuts (diagonal cut for the top figure, and vertical cut for the bottom figure), using Clifford gates only. The dashed lines correspond to the reduced purity $\overline{\rm pur}$ in the random Haar limit.
    see Eq.~\eqref{eq:purity_limit}. For all the instances, the pattern ABCDCDAB is used.
    }
    \label{fig:red_purity}
\end{figure}

Figure~\ref{fig:red_purity} shows the reduced purity, as a function of the number of cycles,
for different circuit sizes and cuts. Dashed lines are the reduced purity limit values, see Eq.~\eqref{eq:purity_limit}. 
For Sycamore-70 (this work), the reduced purity is close to its limit value
at depth $10$.

\subsection{Reduced purity and distribution of singular values}\label{app:reduced_purity}
We will now show numerically that the reduced purity is a good witness for the distribution of singular values, which undergoes through a sharp transition to its limiting value. We first need to understand what is the expected reduced purity and the corresponding standard deviation for a Haar random state. In Sec.~\ref{app:fid_haar} we gave the distribution of the normalized singular values $s = \sqrt{D_1} S$, where $D_1$ is the small Hilbert space dimension (between the halves in which the state is being divided). The expected purity is 
\begin{align}\label{eq:purity_limit}
   \overline{\rm pur} = \Big\langle\!\Big\langle \sum_{\alpha=1}^{D_1} S_\alpha^4\Big\rangle \! \Big\rangle \simeq \frac 1 {D_1^2}\Big\langle\!\Big\langle \sum_{\alpha=1}^{D_1}  s_\alpha^4 \Big\rangle\!  \Big\rangle = \frac 1 {D_1} \aavg{ s_\alpha^4 }\;.
 \end{align}
The variance is
\begin{align}
   \var({\rm pur}) &= \var \left( \sum_{\alpha=1}^{D_1} S_\alpha^4 \right) \simeq \frac 1 {D_1^4} \sum_{\alpha=1}^{D_1} \var( s_\alpha^4 ) \nonumber \\ &= \frac 1 {D_1^3} \var ( s_\alpha^4 )\;.
\end{align}
Therefore, we can say that the reduced purity has converged to its limiting value at a given depth when 
\begin{align}
    \frac {\tr\rho_L^2 - \overline{\rm pur}}{\sqrt{\var({\rm pur})}} \in O(1)
\end{align}
Note that the variance decreases exponentially. 

\begin{figure}[t!]
    \centering
    \includegraphics[width=0.85\columnwidth]{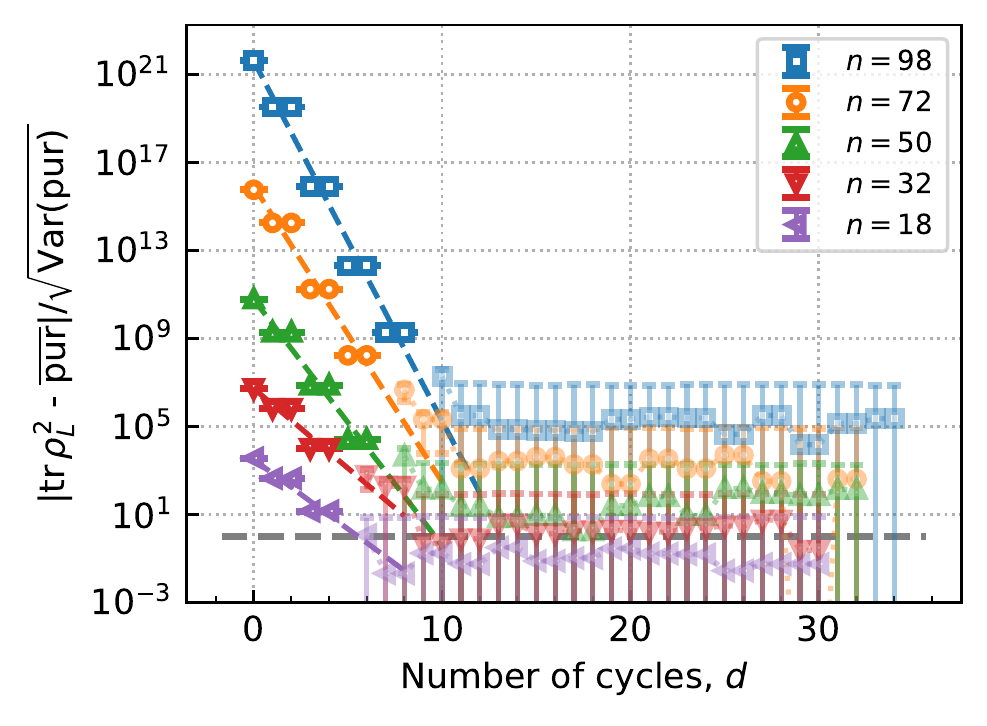}\\
    \includegraphics[width=0.85\columnwidth]{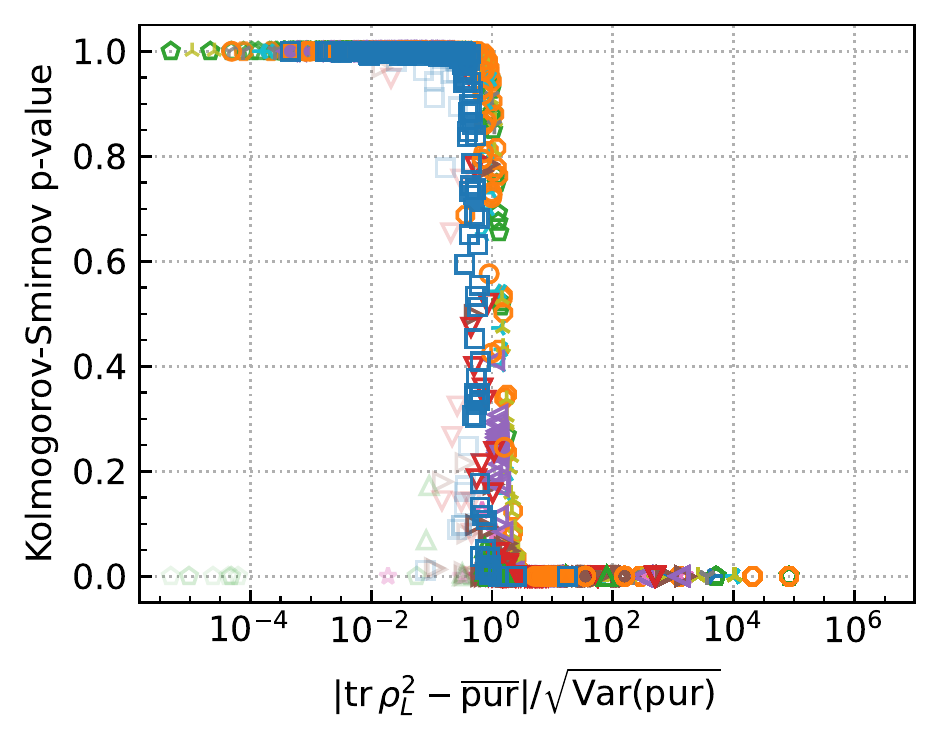} 
    \caption{(Top) Distance of the reduced purity from its limit value in units of the standard 
    deviation as a function of depths, using Clifford gates only. The dashed line corresponds to $1$.
    (Bottom) Kolmogorov-Smirnov $p-$value between the singular values $S_\alpha$ and 
    the distribution of singular values for large depth, Eq.~\eqref{eq:fidelity_as_s}, as 
    a function of the distance between the reduced purity and its limit value in units
    of standard deviation. Each point corresponds to a different circuits (with the number of qubits ranging from $n=8$ and $n=24$) at 
    given fixed depth. Lighter points correspond to datapoints outside the $90\%$ two sided confidence interval. For all the instances, the pattern ABCDCDAB is used, and qubits are partitioned in two equal halves using a diagonal cut.}
    \label{fig:red_purity_rd_limit}
\end{figure}

Figure~\ref{fig:red_purity_rd_limit}-Top shows the distance of the reduced
purity in units of the standard deviation as a function of the number of cycles. As expected,
the depth for which the reduced purity is exponentially close to its limit value
increases with the system size. Figure~\ref{fig:red_purity_rd_limit}-Bottom shows instead
how the Kolmogorov-Smirnov $p-$value between the singular values $S_\alpha$ and 
the Haar random distribution of singular values Eq.~\eqref{eq:fidelity_as_s}
transitions when the purity reaches its limiting value in units of standard deviation. 
As one can see, there is a sharp transition so that only once the
reduced purity is appropriately close to its limit value, the distribution of
$S_\alpha$ truly follows the distribution of singular values in the random
Haar limit.

\subsection{Bounding the approximate tensor representation performance for close simulations}
Using Eqs.~\eqref{eq:fidelity_bound_exact} and~\eqref{eq:fidelity_bound_num}, 
it is possible to lower bound the required $\chi$ to achieve a target
fidelity $F$ for close simulations (see App.~\ref{app:mps_open_close_sim}). 
For our bounds, we split circuits $C$
of $m$ cycles in three parts of $|C_1| = m-2$, $|C_M| = 4$ and $|C_2| = m - 2$
cycles respectively. Moreover, we assume that it is possible to compute
$\ket{\tilde\Psi_1}$ and $\ket{\tilde\Psi_2(x)}$ with a single truncation
each. While being unrealistic for any practical purpose, it allows us
to find analytical and semi-analytical bounds since every realistic 
simulation would require more than one truncation.

\begin{figure}[t!]
    \centering
    \includegraphics[width=0.85\columnwidth]{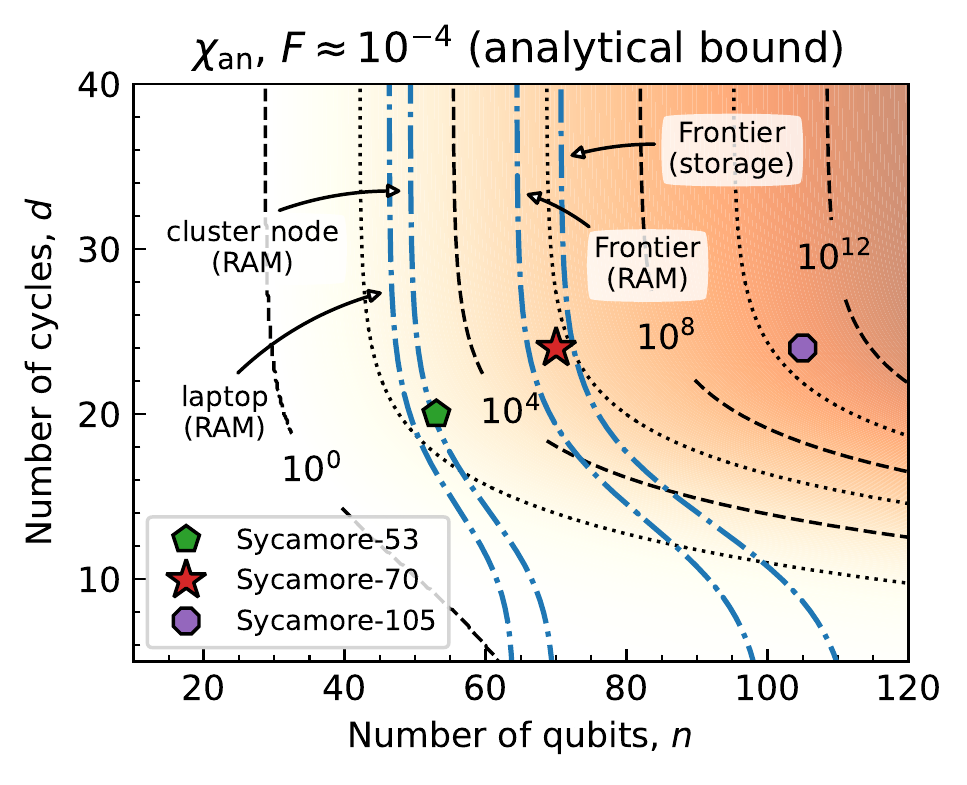}\\
    \includegraphics[width=0.85\columnwidth]{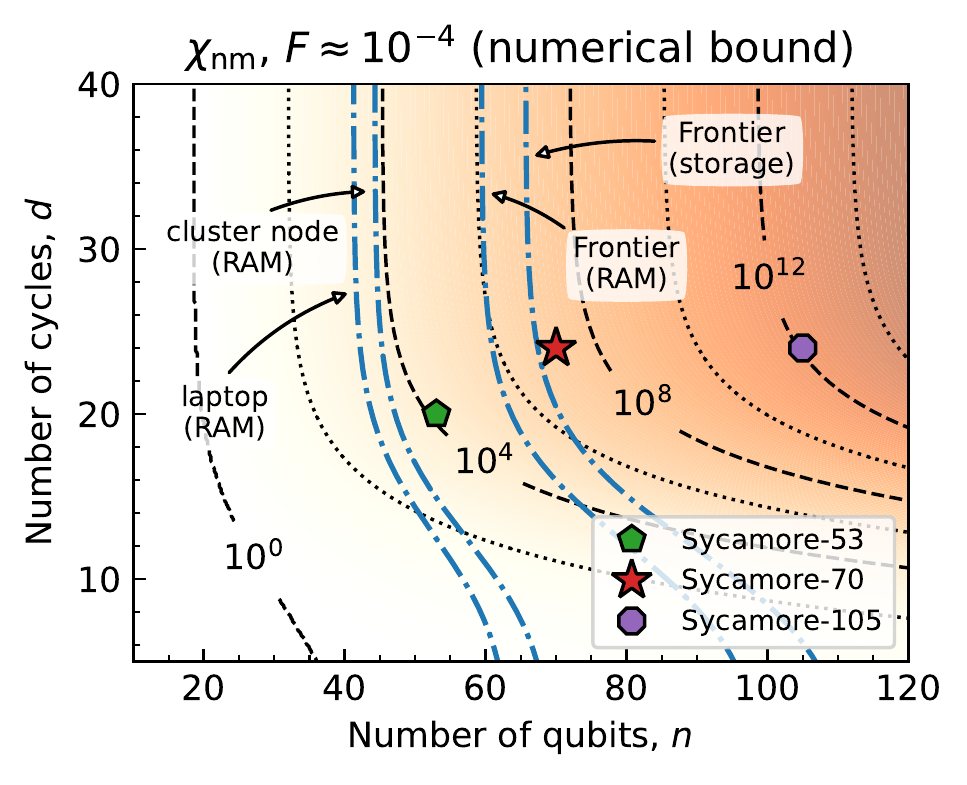} 
    \caption{Analytical (Top) and numerical (Bottom) upper bounds for the required bond dimension to achieve a target fidelity of $F = 10^{-4}$, by varying the number of qubits
    and depths. The memory footprint (blue dashed-and-dotted lines) is computed as the amount of memory required
    to store two \texttt{complex32}  tensors of dimension $2^{n/2}\times \chi$.
    Laptop (RAM) = 32GB, cluster node (RAM) = 256GB, Frontier (RAM) = 9.2PB and
    Frontier (storage) = 700PB. The density map is obtained by averaging
    circuits with patterns ABCDCDAB/BADCDCBA and with diagonal/vertical
    cuts.
    }
    \label{fig:chi_bound}
\end{figure}

Recalling that the final fidelity of a close simulation is 
$F = F_1 \overline{F_2}$, with $F_1$ and $\overline{F_2}$ being the fidelity of
$\ket{\tilde\Psi_1}$ and the average fidelity $\ket{\tilde\Psi_2(x)}$ respectively,
it is possible
to get an estimate of the optimal bond dimension $\chi$ for a given target fidelity $F$
as:
\begin{subequations}\label{eq:chi_bounds}
\begin{align}
    \chi_{\rm an} &= \frac{F}{\aavg{\tr \rho_L^2}},\\
    \chi_{\rm nm} &= \frac{\mathcal{F}_\lambda^{-1}(\sqrt{F})}{
    \aavg{\tr \rho_L^2}} \geq \frac{\sqrt{F}}{
    \lambda_+^2 \aavg{\tr \rho_L^2}},
\end{align}
\end{subequations}
with $\lambda_+^2 \leq 2$ being the largest singular value, and $\chi_{\rm an}$ and 
$\chi_{\rm nm}$ being respectively the estimate for the bond dimension
using either the analytical or the numerical bound. 
It is important to stress that the upper bounds provided by Eqs.~\eqref{eq:chi_bounds} are valid
for arbitrary depths and bond dimensions $\chi$, even if the quantum state has not yet reached
the Porter-Thomas limit.
For small target fidelity $F$, the ratio
between $\chi_{\rm an}$ and $\chi_{\rm nm}$ becomes:
\begin{equation}
    \frac{\chi_{\rm an}}{\chi_{\rm nm}} \approx \lambda_+^2 \sqrt{F}.
\end{equation}
For a cut that split the qubits in two equal halves 
(\mbox{$\lambda_+ = 2$}), and for a target
fidelity of $F = 10^{-4}$, one gets $\chi_{\rm nm} \approx 25\, \chi_{\rm an}$, that is
the numerical bound is only $25$ times larger than the analytical bound.
This is consistent with what we observe in Fig.~\ref{fig:chi_bound}. Note the required FLOPs scale as $O(2^n \chi)$ if we represent the state with two equal size tensors.

\section{Client-certified randomness generation with RCS}
\label{app:crng}

Randomness is a valuable resource with many applications and is a key resource in much of modern cryptography. In classical physics, the outcome of any experiment can in principle be determined from the initial conditions, so there is no such thing as true randomness. On the other hand, quantum physical systems exhibit true randomness.
The outcome of certain quantum processes is inherently random, meaning that no amount of prior information is sufficient to predict the outcome.

Client-certified randomness generation has been proposed as a potential application of RCS~\cite{aaronson2018,bassirian2021certified,aaronson2023certified}. The proposed protocol works as follows. The client generates challenge random quantum circuits which she sends to
an untrusted server that operates a quantum processor.
The server responds to each challenge by sending a requested number of bitstrings within a given short amount of time.
When the server is honest, it produces the bitstrings
by sampling from the circuit using the quantum processor, so the bitstrings contain entropy due to
the inherent randomness of quantum measurements. The client can then pass the raw bitstrings through
a randomness extractor
to obtain random bits of higher quality, in the sense of being closer to the uniform distribution.

The client is able to gain confidence that the returned sample of bitstrings is consistent with executing the challenge circuits by performing statistical tests, such as XEB. For cryptographic certification of randomness, these tests need to account for the possibility that the quantum operator is adversarial. Such an adversary will try to construct a set of bitstrings that pass
the statistical tests despite having low or no entropy. 
Challenge circuits must be executed with fidelity greater than an agreed value $F$. In principle, the client allots a sampling time sufficiently shorter than the necessary time to simulate the same sampling (number of bitstrings and fidelity) by all known classical algorithms using reasonable computing resources. This way, the client can gain confidence that the bitstrings were indeed obtained using a quantum computer, and therefore are a source of quantum randomness. Unfortunately, in this protocol, the client needs to perform classical simulations if she wishes to certify the quantum randomness using XEB. However, the client can do this simulation a posteriori, running classical computations for a much longer time. Furthermore, the client can in principle use a large number of challenge circuits and select only a subset of them for verification. This way the client can force an adversarial server to perform expensive simulations on a large number of circuits while only expending computing resources verifying a much smaller subset. 

There exist a tension between the need for practical verification, which incurs an exponential cost in this proposal, and the requirement that an adversary could not pass the same test deterministically. Furthermore, "spoofing" is typically a factor of $F$ cheaper than the verification~\cite{markov2018quantum}. The 70 qubit circuits presented in this work are currently too big to be verified with XEB. At the same time, the computational cost of classical algorithms keeps improving (see SM~\ref{app:tn_simulation}), as well as the performance of implementations in specialized hardware~\cite{morningstar2022simulation}. In this work we do not resolve this tension, and we leave open the problem of finding an efficient verification protocol, perhaps along the line of cryptographically secure proposals~\cite{brakerski2018cryptographic,mahadev2022efficient}, or more near-term obfuscation techniques~\cite{yung2020anti}. We nevertheless study how this protocol could work if this problem is resolved or if a client is willing to expend sufficient compute resources to gain enough confidence against a potential deterministic adversarial server. 

We summarize how our study of certified randomness compares to the original proposal introduced in Refs.~\cite{aaronson2018,aaronson2023certified}. The original reference is more rigorous, under the assumption that the total system fidelity remains constant as the number of qubits scales, which requires quantum error correction~\footnote{The parameter $b$ in the ``Linear Cross-Entropy Benchmarking'' definition of Problem 1 in Ref.~\cite{aaronson2023certified} corresponds to our XEB + 1. Note that this parameter $b$ is assumed to be constant as the number of qubits $n$ scales.}. Our proposal is less rigorous but more practical. First, we argue that the amount of randomness per circuit should scale as the number of samples times the system fidelity, not as the number of qubits as in Refs.~\cite{aaronson2018,aaronson2023certified}. Second, we introduce new statistical tests to detect adversarial sampling with lower randomness. Third, we implement an optimized (quantum-proof) Trevisan’s randomness extractor~\cite{aaronson2023certified,trevisan2001extractors}, which is important for a potential practical application given that the required input sizes are several orders of magnitude larger than those considered by previous implementations. We also propose a faster but less rigorous randomness extractor.

\subsection{Entropy estimation}

The output of the protocol is produced by applying a randomness extractor to the output of the quantum computer; see Sec.~\ref{sec:extractor}. A randomness extractor takes an input from the source with a given min-entropy, together with a uniformly random seed, and it outputs a near-uniformly random bitstring of length proportional to the input min-entropy. The min-entropy of a random variable $X$ is defined as minus the log of the maximum probability:
\begin{align}
    \textrm{ min-entropy} = - \max_x \log_2\left(\Pr[X=x]\right).
\end{align}
Below, we give bounds for the quantum min-entropy (see also Ref.~\cite{BP2020:NotesInterrRandQuantCircs}).

\subsubsection{Entropy estimation for an honest server}
The experimental output of a noisy quantum random circuit can be described by (see SM~\ref{app:rcs_distribution})
\begin{align}
F \ket \psi\!\! \bra \psi + (1-F) \Xi\;,
\end{align}
where $F$ is the experimental fidelity (probability of no error), $\ket \psi$ is the ideal output of the quantum circuit, and $\Xi$ has trace one and is the result of errors. Measurement of this state can be interpreted as measuring the ideal quantum state $\ket \psi$ with probability $F$, and measuring the operator $\Xi$ with probability $1-F$. This is depicted in the following diagram:
\begin{equation}
\begin{tikzpicture}[scale=2, baseline=(current  bounding  box.center)]
\tikzstyle{atomic} = [draw, thick,minimum width=1.5cm,minimum height=1.5cm]
\node (i) at (0,1) [atomic] {$\ket \psi$};
\node (n) at (0,0) [atomic] {$\Xi$};
\node (s) at (1.5,.5) [atomic] {Sampler};
\draw[-latex,thick] (s) edge node[above] {$F$} (i);
\draw[-latex,thick] (s) edge node[above] {$1-F$}(n);
\end{tikzpicture}
\end{equation}

For the purposes of quantum randomness generation, we take an adversarial approach with respect to the noise operator $\Xi$ and consider it to be deterministic. 
The reason is that we are purely interested in the quantum entropy that is generated experimentally, and not in using potential ``noise'' or ``errors'' in the experiment as a source of entropy. Arguably, if we were willing to accept an entropy source based on ``noise'', there are simpler setups that do not require the use of a quantum processor. Furthermore, the potential entropy coming from the noise cannot be certified. Therefore, we model the sampling as depicted in the next diagram:
\begin{equation}
\begin{tikzpicture}[scale=2, baseline=(current  bounding  box.center)]
\tikzstyle{atomic} = [draw, thick,minimum width=1.5cm,minimum height=1.5cm]
\node (i) at (0,1) [atomic] {$\ket \psi$};
\node (n) at (0,0) [atomic] {Deterministic};
\node (s) at (1.5,.5) [atomic] {Sampler};
\draw[-latex,thick] (s) edge node[above] {$F$} (i);
\draw[-latex,thick] (s) edge node[above] {$1-F$}(n);
\end{tikzpicture}
\label{fig:entropy_model}
\end{equation}

Given the adversarial model above, the bitstring with the highest probability is the deterministic noise with probability $1-F$. Therefore for a sample of size $k$ the min-entropy is
\begin{align}
    \textrm{min-entropy} = -\log_2\left((1-F)^k \right) \approx kF
\end{align}

It is possible to obtain a tighter bound by ignoring the very unlikely event that all outputs of the experiment correspond to the ``noise'' term, which we are treating as deterministic. The approach is to use the $\varepsilon$-min-entropy or smooth min-entropy, that is, we bound the min-entropy ignoring events with cumulative probability smaller than some suitable small $\varepsilon$.  

In the simplified model given by (\ref{fig:entropy_model}) within a sample of size $k$ the expected number of bitstrings obtained from the ideal output $\ket{\psi}$  is $kF$ with the variance $kF(1-F)$. We wish to bound the probability of obtaining sufficiently many bitstrings from the ideal output. We now estimate this probability for obtaining a sequence of bitstrings from the ideal distribution up to a cumulative probability of $1-\varepsilon$.
Assuming that we are sampling from a quantum processor with experimental fidelity $F$, then we can choose a constant $c_1$ and upper-bound the probability of obtaining at least
\begin{align}
    q:=kF-c_1\sqrt{kF(1-F)}
\end{align}
bitstrings from the {\it ideal} distribution, where $c_1$ is the number of standard deviations below the mean. Since the sample size is large enough (order of millions) we can use gaussianity where by $c_1=5$ implies that $\epsilon=1.5 \times 10^{-12}$ rendering the probability of success  $1-1.5 \times 10^{-12}$. 

Suppose that the distribution of the ideal output probabilities $p(s) = | \braket s \psi |^2$ follows the Porter-Thomas distribution with the probability density function
\begin{align}
    f(x) = e^{-x},
\end{align}
where $x = Dp(s)$ is the ideal bitstring probability scaled by the Hilbert space dimension $D = 2^n$.
The average of minus log of the probability of one bitstring from the Porter-Thomas distribution is (see Eq.~\eqref{eq:ideal_pt})
\begin{align}
    -D^2\int_0^\infty \log(p) e^{- D p}  p\, dp = \log(D)-1 + \gamma\;,
\end{align}
where $\gamma$ is Euler's constant. Then minus the log of the probability of a sequence of $q$ independent ideal bitstrings is, by central limit theorem, a normal distribution with average $q(\log(D)-1 + \gamma)$. The variance is upper-bounded by $q \pi^2/6$. Similar as above, we can choose a constant $c_2$ such that with high probability the $\varepsilon$-min-entropy is $q(\log(D)-1+\gamma)-c_ 2 \sqrt{q \pi^2/6}$. Putting it all together, we obtain the following lower-bound for the $\varepsilon$-min-entropy 
\begin{align}
    q(\log(D)-1 + \gamma) - c_ 2 \sqrt{\frac{q \pi^2 }{6}}.
    \label{eq:bare_me}
\end{align}

\subsubsection{Correction to the min-entropy}
\label{sec:entropy-estimation:corrections}

The bound of Eq.~\eqref{eq:bare_me} represents the pure quantum min-entropy obtained from sampling a random quantum circuit. We now consider the situation in which a client has only black-box access (say via the cloud) to the quantum processor held by a server. We discuss deviations from the scenario of the previous section due to potential adversarial actions of the server, while still assuming that the server calls a quantum processor to obtain the output bitstrings. 

In the previous section we bounded the number $q$ of bitstrings obtained in a sample of size $k$ using a quantum processor of fidelity $F$. An adversarial quantum server might be able to oversample $sq$ bitstrings with $s \ge 1$ in the allotted time from the ideal quantum state before returning $q$ bistrings. The server can also rearrange the bistrings  in any predetermined way before returning them to the client. These operations lower the min-entropy and do not necessarily affect statistical tests such as the cross entropy. In order to bound the min-entropy of this multiset oversampling, we consider first a simplified model where the server samples from a uniform distribution of size $D$ instead of the ideal quantum state. 

We now give an expression for the probability of obtaining a given multiset $S$ of size $q$ when sampling $sq$ times from the uniform distribution of $D$ values. We can assume $q \ll D$ and that all the values in the set $S$ are  distinct. Let $A_i$ denote the set of all sequences missing value $i$. We have
\begin{align}
    P(S) = 1 - {\left|\bigcup_{i \in S} A_i \right| \over D^{sq}}\;.
\end{align}
Note that
\begin{align}
    \left|\bigcap_{i \in I} A_i \right| = (D-|I|)^{sq}\;.
\end{align}
Therefore, by the inclusion-exclusion principle, we have
\begin{align}
    \left|\bigcup_{i \in S} A_i \right| = \sum_{j=1}^q (-1)^{j-1} \binom q j (D-j)^{sq}\;,
\end{align}
and 
\begin{align}
    P(S) &= 1-\sum_{j=1}^q (-1)^{j-1} \binom q j\left({D-j \over D}\right)^{sq}\;.
\end{align}
Although this expression is exact, note that all the terms have the same order in $1/D$, so it does not result in a compact estimate of the probability in the case of interest, $D \to \infty$.

We can also write different upper and lower bounds for $P(S)$. Let $\alpha$ denote a set of $q$ indexes and $B_\alpha$ denote the set of words with the $q$ given values in positions $\alpha$. We first have
\begin{align}
    P(S) \le \frac{\sum_\alpha |B_\alpha|} {D^{sq}}\;.
\end{align}
There are $q!$ ways in which the $q$ values can appear in the $\alpha$ positions, and $D^{sq-q}$ possible choices for the other $sq-q$ positions. Therefore
\begin{align}
    |B_\alpha| = q! D^{sq-q}\;.
\end{align}
There are are $sq$ chose $q$ ways to choose $\alpha$. Therefore
\begin{align}
  P(S) \le \binom {sq} q \frac {q!} {D^q} = \frac {(sq)_q} {D^q}\;,
\end{align}
where 
\begin{align}
    (sq)_q = \Pi_{j=0}^{q-1}(sq -j)
\end{align}
This gives the following bound for the min-entropy
\begin{align}
  \textrm{ min-entropy} \ge q \log D - \log(sq)_q\;.
\end{align}
We can obtain a related lower bound by considering the sets $C_\alpha$ including words with the $q$ values of interest in positions $\alpha$, and none of those values anywhere else. Then
\begin{align}
    P(S) &\ge \frac{\sum_\alpha |C_\alpha|} {D^{sq}} \\ &= \binom {sq} q {q! (D-q)^{sq-q} \over D^{sq}} \\ &= \frac {(sq)_q} {D^q} \left(1- \frac q D \right)^{q(s-1)}\;.
\end{align}
Therefore, in the limit of $D \to \infty$, we have 
\begin{align}
    P(S) & = \frac {(sq)_q} {D^q} \left(1- O\left( \frac {q^2 s} D \right)\right) \\ 
    \textrm{ min-entropy} &= q \log D - \log(sq)_q + O\left( \frac {q^2 s} D \right) \\
    &\simeq q \log D - q \log sq + q\;.
\end{align}

We have seen that multiset sampling can lower the min-entropy by a factor $\log((sq)_q)$. Applying this to the honest server min-entropy bound, Eq.~\eqref{eq:bare_me}, gives a bound for the multiset sampling min-entropy
\begin{align}\label{eq:entropy-rejection-sampling-bound}
    q(\log(D)-1 + \gamma) - c_ 2 \sqrt{\frac{q \pi^2 }{6}} -\log((sq)_q)\;.
\end{align}

\subsection{Repeated bitstrings}\label{app:collisions}
In the previous section we ignored the possibility of repeated bitstrings in the adversarial server sampling. We study this now. We denote the total sampling budget of the adversarial server ($sq$ in the previous section) by $\beta$. We will see that the client can require the server to return unique bitstrings, and this has little effect in the linear XEB as long as $\beta \ll D$. An adversarial server can also postselect to bistrings that appear at least twice to artificially boost the nominal ``fidelity" as measured by XEB. We will see that this effect is negligible as long as $s \ll \sqrt {D/q}$.

\subsubsection{Probabilities for repeated bitstrings}
The probability that a bitstring $j$ appears exactly $c$ times is
\begin{align}
  p_j'(c) = \binom \beta c p_j^c (1-p_j)^{\beta-c}\;.
\end{align}

For large enough sampling budget $\beta$, there may occur collisions, i.e., repeated strings.
    We can calculate the expected number $M$ of strings appearing 
with each multiplicity $c$, and the corresponding ideal probability value $A$.
    In the following, we derive closed formulas up to first-order approximation, 
confirming the formulas \eqref{eq:m_simple}, 
 \eqref{eq:Exp-QC-val-multiplicity-for-fid-1} conjectured in Ref.~\cite[App.~D]{BP2020:NotesInterrRandQuantCircs}.

\begin{lemma}\label{lem:m}
 Assuming $D+\beta \gg c$, the expected number of bitstrings that appear exactly $c$ times is
\begin{align}
  M_{\beta, c} = \binom \beta c {D^{1-c}c! \over \left(1 +\frac  \beta D\right)^{c+1}}\left(1+ O\left(\sqrt{(2c)! \over  D}\right)\right)\;.\label{eq:m_lemma}
\end{align}
\end{lemma}
\begin{proof}
The expected number of bitstrings that appear exactly $c$ times is
\begin{align}
  M_{\beta, c} = \sum_j p_j'(c)\;.
\end{align}

First note that 
\begin{align}
  M_{\beta,c} &=  \sum_j p_j'(c) \\ &= D \aavg{p'(c)} + O\left(\sqrt {D \,{\rm Var}(p'(c)}\right)\;,
\end{align}
where, as in App.~\ref{app:rcs_distribution}, we use the approximation that for large $D$ the probabilities $p'_j(c)$ are i.i.d.

We can write
\begin{align}
  \aavg{p'(c)}  = \binom \beta c I(\beta,c)\;,
\end{align}
where $I(\beta,c)$ is the expectation value of $p^c (1-p)^{\beta-c}$. This can be calculated as
\begin{align}
  I(\beta,c) &= \int_0^1 p^c (1-p)^{\beta-c}  dF(p) \\
  &=\int_0^1 p^c (1-p)^{\beta-c} (D-1)(1-p)^{D-2} d p \\
  &= (D-1) c! {(D + \beta -c -2)! \over (D + \beta -1)!} \\
  &= (D-1) c! {1 \over (D +\beta -1)_{c+1}}\;,
\end{align}
where the last expression uses a falling factorial in the denominator. 
Therefore
\begin{align}
  \aavg{p'(c)} = \binom \beta c {(D-1) c!  \over (D +\beta -1)_{c+1}}\;,  
\end{align}
We are interested in the value for large $D$, so we can use the approximation
\begin{align}
  \aavg{p'(c)} \simeq \binom \beta c {(D - 1) c! \over \left(D + \beta -1 - \frac c 2\right)^{c+1}}\;.
\end{align}
This approximation is valid for
\begin{align}
  D + \beta \gg c\;,
\end{align}
which is always the case in the regime of parameters we are interesting in. 

We can also estimate the variance
\begin{align}
 & {\var(p'(c)) \over \binom \beta c ^2} \\&= {(D - 1) (2c)! \over \left(D + 2 \beta -1 - c \right)^{2c+1}}  \nonumber \\&\qquad\qquad- \left( {(D - 1) c! \over \left(D + \beta -1 - \frac c 2\right)^{c+1}}\right)^2 \\ &= \nonumber  D^{-2c} {(1 - \frac 1 D) (2c)! \over \left(1 + 2 \frac \beta D - \frac 1 D -\frac  c D \right)^{2c+1}} \\ &\qquad\qquad -D^{-2c}  \left( {(1 - \frac 1 D) c! \over \left(1 + \frac \beta D -\frac 1 D - \frac c {2 D}\right)^{c+1}}\right)^2 \;.
\end{align}
Ignoring small terms we get
\begin{align}
  \var(p'(c)) =  \binom \beta c ^2 D^{-2c} \left((2c)! - (c!)^2 +O(\beta/D)\right)\;.
\end{align}
Keeping only the dominant term $(2c)!$ completes the proof of the lemma.

Note that we also ignore terms $O(1/D)$ and $O(c/D)$ for consistency with the fluctuations from the variance. 
\end{proof}

In Eq.~\eqref{eq:m_lemma} we can use $\beta \gg c$ to write
\begin{align}
  \binom \beta c \simeq {\left(\beta - \frac c 2\right)^c \over c!}\;.
\end{align}
Plugging this back and ignoring again terms $O(1/D)$ and $O(c/D)$ we get
\begin{align}
  M_{b,c} \simeq D {b^c \over (1+b)^{c+1}} \label{eq:m_simple}\;,
\end{align}
where $b = \beta/D$. 

We are also interested in the expected value of the simulated or ideal probability for the bistrings that are obtained exactly $c$ times in a $\beta$-sample. 
\begin{lemma}
  The expected value of the ideal probability for the bistrings that are obtained exactly $c$ times is
  \begin{align}
    A_{\beta,c} = \frac 1 D {c+1 \over 1 +b}\;.
  \end{align}
\end{lemma}
\begin{proof}
The probabilities of bitstrings conditioned on appearing exactly $c$ times are proportional to $p_j'(c)$, normalized so that their sum is $1$. That is, the conditional probabilities are $p_j'(c)/M_{b,c}$. Therefore, the expected value of the simulated probability conditioned on appearing $c$ times has the expression
\begin{align}
  A_{\beta,c} = \frac 1 {M_{\beta,c}} \sum_j p_j'(c) \,p_j\;. \label{eq:abc}
\end{align}

Using the same methodology as in Lemma~\ref{lem:m} we have
\begin{align}
  A_{\beta,c} &= {\int_0^1 p^{c+1} (1-p)^{\beta-c} dF(p) \over \int_0^1 p^c (1-p)^{\beta-c} dF (p)} \\
              &= {I(\beta+1,c+1 \over I(\beta,c)} \\
              &= {(c+1)! \over c!}{ (D +\beta -1)_{c+1} \over (D +\beta)_{c+2}} \\
  &= \frac 1 D {c+1 \over 1 +b}\;.\label{eq:Exp-QC-val-multiplicity-for-fid-1}
\end{align}

\end{proof}

The expected number of unique bitsrings in a $\beta$-sample follows from Eq.~\eqref{eq:m_simple} and is given by the expression
\begin{align}
    M_\beta = \sum_{c=1}^\infty M_{\beta,c} = D {\beta \over D + \beta}\;.
\end{align}

\subsubsection{Linear cross-entropy with unique bitstrings}
Following the same logic as in Eq.~\eqref{eq:abc}, we can calculate the expected value of the linear cross entropy when an honest server returns unique bitstrings. The probabilities of bitstrings conditioned on appearing at least one time are proportional to
\begin{align}
    p''_j(c) = \sum_{c =1}^\infty p'_j(c)\;,
\end{align}
normalized so that their sum is 1. The corresponding linear cross entropy is 
\begin{align}
    D \sum_j {p''_j(c) \over M_\beta} p_j-1 &= \frac D {M_\beta} \sum_{c =1}^\infty A_{\beta,c} M_{\beta,c} -1\\ 
    &= {2+ b \over 1 + b} -1 = {1 \over 1 +b}\;.
\end{align}
The expectation value of the linear cross entropy is 1 when allowing repeated bitstrings if sampling from a Haar random quantum state. Therefore, the perturbation to the linear cross entropy when requiring unique bitstrings can be ignored when $b\ll 1$ or, equivalently, $\beta \ll D$. 
Note that sampling with less fidelity results in a lower frequency of collisions. 

\subsubsection{Adversarial postselection of repetitions}
Consider now the situation where an adversarial server is asked to return $k$ unique bitstrings, but the server secretly oversamples many more bitstrings to take advantage of collisions. That is, the server can postselect bitstrings that appear at least twice, and therefore, in the ideal case of fidelity 1, have a higher expectation value for the simulated probability $D A_{\beta,2} \sim 3$, instead of the usual average simulated probability $\avg{D p} =2$. In this way, the server could pass the linear cross entropy test returning a smaller number of quantum generated bitstrings, that is, a sample with similar estimated fidelity but less quantum entropy. 

Next we bound how many bitstrings can be oversampled with still a negligible effect in the estimated fidelity from linear cross entropy. In order to cover the case of an adversarial server with non-ideal fidelity $\phi < 1$, we consider an idealized model where errors are heralded. That is, we treat sampling $s k$ bitstrings with fidelity $\phi$ as sampling $\beta = s q$ bitstrings with fidelity 1, where $q = \phi k$. The contribution to the linear cross entropy for bitstrings that appear $c=2$ times is
\begin{align}
    \frac D q M_{\beta,2} A_{\beta,2} &= \frac D q {\beta^2 \over D^2} 3 (1 + O(\beta/D)) \\
    &= 2 {s^2 q \over D}3 (1 + O(\beta/D))\;.
\end{align} 
This effect is negligible for $s \ll \sqrt{D/q}$.

\subsection{Additional statistical tests}
We explained in the main text the conditions under which XEB is an estimator of fidelity, which is the main test for experimental RCS (see also SM~\ref{app:rcs_distribution}). Ref.~\cite{arute_supremacy_2019} also introduced the idea of checking the consistency between log and linear XEB, and a Kolmogorov-Smirnov test for the simulated probabilities of the experimental bitstrings. We now introduce two additional statistical tests which might be useful in an adversarial setting such as client-certified randomness generation.  

\subsubsection{Hamming distance filter}
\label{sec:hd_filter}
In order to sample from the output distribution of a quantum circuit one can use an independent tensor contraction per output bitstring using frugal rejection sampling (see SM~\ref{app:tn_simulation} and Refs.~\cite{markov2018quantum,villalonga2019flexible}). This results in a simulation runtime that scales linearly in the number of bitstrings sampled. Ref.~\cite{pan2022solving} introduced a method to compute amplitudes of a large number of uncorrelated bitstrings with a much lower overhead than linear.

An adversarial server using tensor network contractions could avoid the remaining overhead from Ref.~\cite{pan2022solving} using less tensor network contractions to calculate the probabilities of many bitstring with small Hamming distance between them, although this does not perform RCS (see for instance Ref.~\cite{pan2021simulating}).
We now give a Hamming distance filter test which detects this pseudo-sampling.

We can approximate a Porter-Thomas sampling as a uniform sampling of bitstrings for the purpose of analyzing the Hamming distance between unique sampled bitstrings.
We denote the distance between bitstrings $j$ and $k$ as $h_{jk}$.
For fixed $j$, the distribution of the Hamming distance to other bitstrings is binomial with $n$ the number of qubits and $p = 1/2$.
As an example we can consider $n = 70$ and Hamming distance 15.
The probability of $h_{jk} \le 15$ is
\begin{align}
  p_h = \frac 1 {2^n} \sum_{c=0}^{15} \binom {70} c =  8.26 \cdot 10^{-7}\;.
\end{align}

The experimental readout measurement error has a bias which we can take into account. Let $e_{01}$ bet the probability of measuring state $0$ when the quantum state is $1$ and $e_{10}$ the probability of measuring state $1$ when the quantum state is $0$.
This gives a bias $b= e_{01} - e_{10}$.
The probability of sampling a 1 on a qubit is, on average, $p_b = (1 - b) / 2$, while the probability of sampling a 0 is $1 - p_b$. 
The probability of obtaining a given Hamming distance between two bitstrings is therefore given by a binomial distribution with the slightly biased value of $p_b$, 
which is slightly higher than in the unbiased case.

For a given sample $S$ with $k$ bitstrings, we can eliminate sufficient bitstrings so that there are no pairs of bitstrings within Hamming distance less than some bound, such as 15.
One way to do this is to process the bitstrings one by one in the sample $S$.
For each bitstring, we eliminate all the other bitstrings at Hamming distance 15 or less.
With this method, we keep more than half of the bitstrings if $k = 10^6$.
Note that each random ordering of bitstrings results in a different sub-sample.
Therefore, this is equivalent to implementing bootstrapping in the initial sample.
That is, we can repeat this sub-sampling a large number of times calculating the XEB fidelity estimator each time.
The average of the XEB of all the sub-samples corresponds, in the honest case, to the sample average.
We can do this for larger Hamming distances also.

In conclusion, with some small computational cost we can prevent a potential attack using a tensor network algorithm to calculate probabilities of sets bitstrings with small Hamming distance between them.

\subsubsection{Statistical test of large probabilities}

The value of the XEB fidelity estimator is higher if, instead of sampling, an adversarial server outputs the bitstrings with the highest simulated probabilities. This might be detected already by tests introduced in Ref.~\cite{arute_supremacy_2019}, such comparing linear and log XEB, or the Kolmogorov-Smirnov test. Here we give another option, namely  using a truncated XEB fidelity estimator which ignores the bitstrings with simulated probability beyond some threshold $t$.

Consider the XEB estimator based on the function (see SM \ref{app:rcs_distribution})
\begin{align}
  f_t(p_j) :=  D \,p_j \,\openone_{p_j \le t/D} \label{eq:tobs}
\end{align}
where $p_j$ is the simulated or ideal probability for bitstring $j$, $D=2^n$ is the Hilbert space dimension, and $\openone_{p_j \le t}$ is an indicator function with value $1$ if $p_j \le t$ and value $0$ in other case. As explained in SM~\ref{app:rcs_distribution} we can model the sampling probabilities of a quantum processor with fidelity $F$ as
\begin{align}
  p^F_j = F p_j + {1-F \over D}\;.
\end{align}
The expectation value of the sampling with function \eqref{eq:tobs} is
\begin{align}
  \sum_j  p^F_j \,D\,p_j \openone_{p_j \le t/D}\;.
\end{align}
Assuming that the simulated probabilities are distributed according to the Porter-Thomas or exponential distribution we have
\begin{align}
  {\rm tXEB} &=\sum_j  p^F_j \,D\,p_j \openone_{p_j \le t/D} \nonumber\\
          &= D \aavg{p^F_j \,D \,p_j \openone_{p_j \le t/D}}  \\
          &= D^2 \int_0^t (F x + 1 - F) x e^{-x} dx \\
  &= F + 1 - e^{-t} \left(1+t+(1+t+t^2)F\right)\;. 
\end{align}

As with any other XEB fidelity estimator, we can estimate the value tXEB sampling bitstrings from an experimental implementation. This gives the tXEB fidelity estimator
\begin{align}
  F = {{\rm tXEB} -1 + e^{-t} (1+t) \over 1 - e^{-t}(1+t+t^2)}\;.
\end{align}
Note that we are interested in the case $t \gtrsim 2$, which makes the denominator positive. 

In order to calculate the variance of this estimators, we need the expectation value of the square of tXEB. This is
\begin{align}
  \sum_j  &p^F_j \,\left(D\,p_j \openone_{p_j \le t/D}\right)^2\nonumber\\
  &= D^2 \int_0^t (F x + 1 - F) x^2 e^{-x} dx\;.
\end{align}
This integral can be calculated analytically to obtain an expression for the variance of the corresponding estimator of fidelity. For simplicity, we only give the variance in the limit $F \to 0$, which is
\begin{align}
  {\rm VAR}({\rm tXEB}) \simeq {1 - e^{-t} t^2 - e^{-2t}(1+2t+t^2) \over \left( 1 - e^{-t} (1+t+t^2)\right)^2}\;.\label{eq:txeb_std}
\end{align}
Table~\ref{tab:truncated_ce_variance} shows the variance for different values of the truncation parameter. We see that in an experimental sample we can ignore all bitstrings with ideal probabilities $\ge 4/D$ without affecting much the variance of the corresponding tXEB estimator. This gives another statistical test sensitive to a potential adversary which postselects bitstrings with unusually large ideal probabilities. 

\begin{table}[h!]
    \centering
    \begin{tabular}{|c | c |}
    \hline
  {VAR$(F)$} & {$t$} \\
      \hline
 1.00557 & 10. \\
 1.06288 & 7. \\
 1.32601 & 5. \\
 1.84472 & 4. \\
 4.11632 & 3. \\
 10.144 & 2.5 \\
 105.982 & 2. \\
      \hline
    \end{tabular}
    \caption{Variance of the truncated XEB estimate of $F$ against the truncation parameter $t$.}
    \label{tab:truncated_ce_variance}
\end{table}

\subsection{Randomness extractor}\label{sec:extractor}

Randomness extractors are functions that convert bits from a weak
source of randomness into near-uniform random bits~\cite{vadhan_pseudo_2012}.
In our protocol we apply a
randomness extractor to the output of a quantum computer,
which contains intrinsic randomness but is not uniformly distributed.
In this section, we describe the randomness extractor we
implemented and present some benchmark results of its running time.

For general sources of randomness, randomness extraction
is only possible if the
extractor is also given a small uniformly random input seed as
a catalyst. 

A weak random source has a distribution over $\{0,1\}^n$ which has some entropy. The most conservative estimate of the unpredictability of the outcomes is given by the min-entropy or equivalently the $\infty-$R{\'e}nyi entropy:
\begin{df}\label{def:min-entropy}
Let $X$ be a probability distribution on the hyper-cube $\{0,1\}^n$, and let $p_x$ be the probability of the string $x\in\{0,1\}^n$. The minimum entropy of $X$ is 
$$\min_x (-\log_2 p_x)=-\max_x \log_2 p_x = - \log_2 \max_x p_x $$
For an $n-$bit distribution $X$ with min-entropy $k$, we say that $X$ is an $(n,k)$ distribution.
\end{df}

We now formally define an extractor function. Let 
$\operatorname{Ext}: \{ 0, 1 \}^n \times \{0, 1 \}^d \to \{ 0, 1 \}^m$ be the function that takes as input samples from an $(n,k)$ distribution $X$ and a uniformly random $d-$bit string seed, and outputs an $m-$bit string that is $\varepsilon$-close to uniform. We say that the extractor is $(k,\varepsilon)$ if the output is $\varepsilon$ close to uniform random, where $\varepsilon$ is the statistical distance. In extracting randomness from random variables from the knowledge of just a lower-bound on the min-entropy, a key concept is  $k-$source.

\begin{df}
A random variable $X$ is called a $k-source$ if its min-entropy is at least $k$. That is, $\text{Pr}[X=x]\ge 2^{-k}$.
\end{df}

\subsubsection{Trevisan's extractor and HMAC}

We implemented a randomness extractor based on Trevisan's construction~\cite{trevisan2001extractors}.
Since this extractor is somewhat slow, we describe in
 App.~\ref{app:hmac}  an alternative construction using
the cryptographic primitive hash-based message authentication code (HMAC)
that is more efficient, though it is a heuristic, not a
theoretically proven extractor like Trevisan's.

We implemented Trevisan's extractor following primarily the construction of Raz, Reingold, and Vadhan~\cite{raz2002extracting} with some optimizations from~\cite{mauerer2012modular}. The initial extractor was implemented entirely in Python, and profiling was used to identify bottlenecks, which were then rewritten in C++. In our case, over 99\% of the running time was spent evaluating polynomials in a subroutine that computed Reed-Solomon codes. This code was migrated to a C++ library using NTL~\cite{NTL}.

For a fixed $\varepsilon$, the extractor uses $\bigO(\log^2 n)$ additional random bits. The theoretical optimal seed size for any seeded extractor, is $\log(n - k)  + 2 \log(2/\varepsilon)  + \bigO(1)$. For a fixed seed size, min-entropy, and $\varepsilon$, the extractor 
takes time linear in the size $n$ of the input. However, our inputs are several orders of magnitude larger than those considered by previous papers and previous benchmarked implementations of Trevisan's extractor, such as~~\cite{mauerer2012modular} and~\cite{ma2013postprocessing}.

\subsubsection{Benchmark results}

\begin{figure}[ht] 
    \centering
    \includegraphics[width=0.5\textwidth]{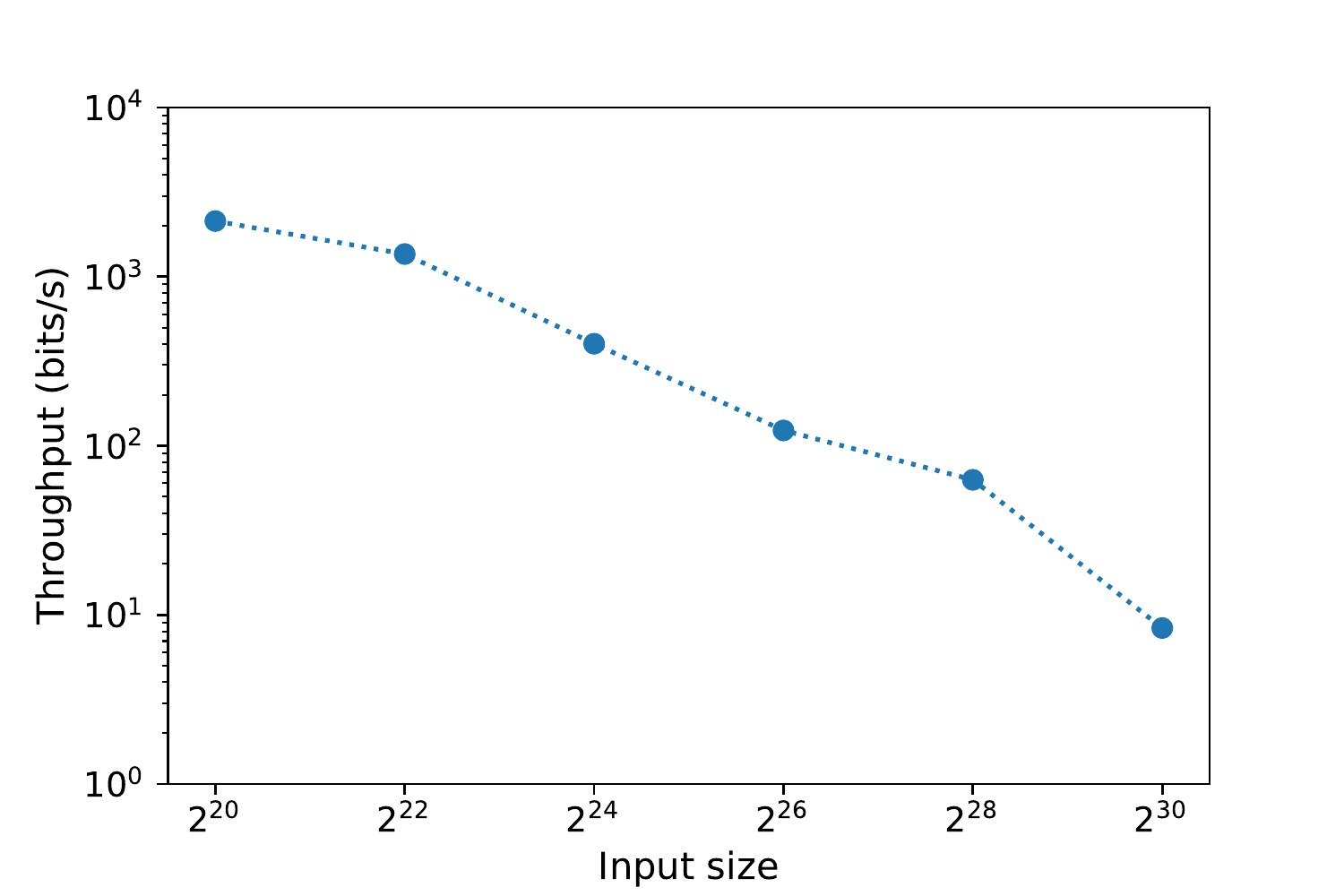}
    \caption{Throughput of the Trevisan randomness extractor for various
    input sizes, when the output length is 4096 bits. We used 64 threads. Each data point is the average of the
    values obtained from
    10 runs, and there are error bars of one standard deviation (too small
    to see).}
    \label{fig:extractor_throughput}
\end{figure}

We tested the performance of our implementation of Trevisan's randomness extractor
on various input sizes ranging from $2^{20}$ bits to $2^{30}$ bits.
We used a workstation with an Intel Xeon Gold 6154 3.00GHz CPU which has
18 cores, each with 4 threads. We used 64 threads for our benchmarking.

Part of the running time of our extractor is spent in
passing the input data from Python to C++. This conversion occurred at
a rate of about 44 Mbit/s. Once this conversion has taken place,
the extractor produces output bits at a constant rate,
which we term the {\it throughput}, when the
total length of the output is fixed.
In Figure \ref{fig:extractor_throughput}, we plot the throughput for the
various input sizes when the output length is fixed to 4096 bits.
At the input size of $2^{30}$, the throughput was
8.4 bits/s. As an example, at the input size of $2^{30}$, the Python to
C++ conversion took about 24 seconds, while the rest of the
extraction took about 490 seconds to produce the 4096 bits of output.
While this throughput is slow compared to, say, the computation of
a hash function like SHA-512, it is sufficient for many purposes.
For example, a high-security cryptographic key requiring 256 bits
of entropy may be used for days or weeks before needing to be refreshed.

\subsubsection{A faster randomness extractor using HMAC}
\label{app:hmac}
\newcommand{\rext}[1]{\ensuremath{\text{Ext}(#1)}}
\newcommand{\hmacexample}[1]{\ensuremath{\text{HMAC}_\text{SHA512}(#1)}}

Our implementation of Trevisan's randomness extractor suffers from
the disadvantage of being quite slow. In practice, theoretically
proven randomness extractors are rarely used, with common efficient heuristic
cryptographic primitives such as HMAC often used instead \cite{FIPS:198:1,HMAC:extractor}.
In this appendix, we explain how one can use an HMAC to construct
a heuristic randomness extractor that works in our setting.
Besides being a heuristic, our construction suffers from the
disadvantage of requiring a rather large seed size (linear in
the size of the output). Nevertheless, it may be of more practical
use in some situations than the Trevisan extractor.

The main obstacle to overcome in using an HMAC for randomness extraction
is that the output length is limited. For example,
a SHA512-based HMAC will only output 512 bits, even when
the input has many more bits of min-entropy.
Here, we show, using Lemma 6.38 in \cite{vadhan_pseudo_2012},
that one can extend the output length of a randomness extractor.
The lemma says the following:

\begin{lemma}\label{lem:ExtConcat}
    (Lemma 6.38 in~\cite{vadhan_pseudo_2012}): Suppose
    \(\text{Ext}_1\colon \{0,1\}^n \times \{0,1\}^{d_1} \rightarrow \{0,1\}^{m_1}\) is a $(k_1, \varepsilon_1)$ extractor and
    \(\text{Ext}_2\colon \{0,1\}^n \times \{0,1\}^{d_2} \rightarrow \{0,1\}^{m_2}\) is a $(k_2, \varepsilon_2)$ extractor
    for $k_2 = k_1 - m_1 - \log(1/\varepsilon_3)$.
    Then \(\text{Ext}'\colon \{0,1\}^n \times \{0,1\}^{d_1 + d_2} \rightarrow \{0,1\}^{m_1 + m_2}\)
    defined by \(\text{Ext}'(x, (y_1, y_2)) = (\text{Ext}_1(x, y_1), \text{Ext}_2(x, y_2))\)
    is a $(k_1, \varepsilon_1 + \varepsilon_2 + \varepsilon_3)$ extractor.
\end{lemma}

Changing notation, letting $\rext{X, S} := \text{Ext}_1 = \text{Ext}_2$,
$m := m_1 = m_2$, and 
$\varepsilon := \varepsilon_1 = \varepsilon_2$; and renaming the independent variable $\varepsilon_3$ as $\varepsilon_1$
we can restate this lemma as:
if $\rext{X, S} = Y$ is a $(k, \varepsilon)$-extractor with an $m$-bit output,
then $(\rext{X, S_1}, \rext{X, S_2})$ is a $(k + m + \log(1/\varepsilon_1), 2 \varepsilon + \varepsilon_1)$-extractor with a $2m$-bit output. 

Let us define $U_d$ to be a uniform random seed of size $d$ bits. As explained in \cite{vadhan_pseudo_2012}, $k_2 = k_1 - m_1 - \log(1/\varepsilon_3)$ in Lemma~\ref{lem:ExtConcat} arises because if one conditions a $k_1$-source on the output of $\text{Ext}_1(X, U_{d_1})$, then the source still has a conditional min-entropy of at least $k_1-m_1-\log(1/\varepsilon_3) = k_2$ except with probability $\varepsilon_3$. Therefore, $\text{Ext}_2(X, U_{d_2})$ can extract an additional $m_2$ almost-uniform bits.
We can also ensure that $\text{Ext}_2(X, U_{d_2})$ can extract an additional $m_2$ almost-uniform bits by instead requiring $X$ to be a $(k_2 + m_1 + \log(1/\varepsilon_3))$-source.

This analysis can be recursively applied.
\begin{multline}
    \text{Ext}''(X, (S_1, S_2, S_3, S_4)) \\
    \equiv (\rext{X, S_1}, \rext{X, S_2}, \rext{X, S_3}, \rext{X, S_4})
\end{multline}
can be considered to be the combination of two $(k + m + \log(1/\varepsilon_1), 2 \varepsilon + \varepsilon_1)$-extractors:
\begin{itemize}
    \item the first $(k + m + \log(1/\varepsilon_1), 2 \varepsilon + \varepsilon_1)$-extractor is $(\rext{X, S_1}, \rext{X, S_2})$ and
    \item the second $(k + m + \log(1/\varepsilon_1), 2 \varepsilon + \varepsilon_1)$-extractor is $(\rext{X, S_3}, \rext{X, S_4})$
\end{itemize}

Thus, selecting a new $\varepsilon_2$, $\text{Ext}''(X, (S_1, S_2, S_3, S_4))$ is a $(k' + m' + \log(1/\varepsilon_2), 2 \varepsilon' + \varepsilon_2)$-extractor, where
\begin{itemize}
    \item $k' = k + m + \log(1/\varepsilon_1)$,
    \item $m' = 2m$, and
    \item $\varepsilon' = 2 \varepsilon + \varepsilon_1$.
\end{itemize}
So we get a $\left(k + 3m + \log\left(\frac{1}{\varepsilon_1}\right)+\log\left(\frac{1}{ \varepsilon_2}\right), 4 \varepsilon + 2 \varepsilon_1 + \varepsilon_2\right)$-extractor.

In general, to output $2^t m$ bits, we can construct an extractor
\begin{multline}
    \text{Ext}^t(X, (S_1, S_2, ..., S_{2^t})) \\
    \equiv (\rext{X, S_1}, \rext{X, S_2}, ..., \rext{X, S_{2^t}}).
\end{multline}
This would be lead to the following extractor
$$\left(k + (2^t - 1) m + \sum_{i=1}^{t}\log(1/\varepsilon_i), 2^t \varepsilon + \sum_{i=1}^{t} 2^{t-i} \varepsilon_i\right)$$

This analysis demonstrates that given sufficient min-entropy in the input $X$, we can repeatedly apply the same randomness extractor with a fresh seed to extract the desired number of output bits that are statistically close to uniform.

\newpage
\onecolumngrid

\vspace{1em}
\begin{flushleft}
{\small Google Quantum AI and Collaborators}

\bigskip
{\small
\renewcommand{\author}[2]{#1$^\textrm{\scriptsize #2}$}
\renewcommand{\affiliation}[2]{$^\textrm{\scriptsize #1}$ #2 \\}

\newcommand{\corrauthora}[2]{#1$^{\textrm{\scriptsize #2}, \ddagger}$}
\newcommand{\corrauthorb}[2]{#1$^{\textrm{\scriptsize #2}, \mathsection}$}

\newcommand{\xGoogle}{\affiliation{1}{Google Research}}

\newcommand{\xNASA}{\affiliation{2}{Quantum Artificial Intelligence Laboratory, NASA Ames Research Center, Moffett Field, California 94035, USA}}

\newcommand{\xKBR}{\affiliation{3}{KBR, 601 Jefferson St., Houston, TX 77002, USA}}

\newcommand{\xUCONN}{\affiliation{4}{Department of Physics, University of Connecticut, Storrs, CT}}

\newcommand{\xNIST}{\affiliation{5}{National Institute of Standards and Technology (NIST), USA}}

\newcommand{\xUMass}{\affiliation{6}{Department of Electrical and Computer Engineering, University of Massachusetts, Amherst, MA}}

\newcommand{\xAU}{\affiliation{7}{Department of Electrical and Computer Engineering, Auburn University, Auburn, AL}}

\newcommand{\xCQCT}{\affiliation{8}{QSI, Faculty of Engineering and Information Technology, University of Technology Sydney, NSW, Australia}}

\newcommand{\xUCR}{\affiliation{9}{Department of Electrical and Computer Engineering, University of California, Riverside, CA}}

\newcommand{\xHU}{\affiliation{10}{Department of Chemistry and Chemical Biology, Harvard University, Cambridge, MA, USA}}

\newcommand{\xUOCA}{\affiliation{11}{Department of Physics and Astronomy, University of California, Riverside, CA}}

\newcommand{\Google}{1}
\newcommand{\NASA}{2}
\newcommand{\KBR}{3}
\newcommand{\UCONN}{4}
\newcommand{\NIST}{5}
\newcommand{\UMass}{6}
\newcommand{\AU}{7}
\newcommand{\CQCT}{8}
\newcommand{\UCR}{9}
\newcommand{\HU}{10}
\newcommand{\UOCA}{11}

\corrauthora{A. Morvan}{\Google},
\corrauthora{B. Villalonga}{\Google},
\corrauthora{X. Mi}{\Google},
\corrauthora{S. Mandr\`a}{\Google,\! \NASA,\! \KBR},
\author{A. Bengtsson}{\Google},
\author{P. V.~Klimov}{\Google},
\author{Z. Chen}{\Google},
\author{S. Hong}{\Google},
\author{C. Erickson}{\Google},
\author{I.~K.~Drozdov}{\Google,\! \UCONN},
\author{J. Chau}{\Google},
\author{G. Laun}{\Google},
\author{R. Movassagh}{\Google},
\author{A. Asfaw}{\Google},
% ...
\corrauthorb{K. Kechedzhi}{\Google},
\corrauthorb{Y. Chen}{\Google},
\corrauthorb{S. Boixo}{\Google},

\bigskip

\xGoogle
\xNASA
\xKBR
\xUCONN
\xNIST
\xUMass
\xAU
\xCQCT
\xUCR
\xHU
\xUOCA

{${}^\ddagger$ These authors contributed equally to this work.}\\
{${}^\mathsection$ Corresponding author: boixo@google.com}\\
{${}^\mathsection$ Corresponding author: bryanchen@google.com}\\
{${}^\mathsection$ Corresponding author: kostyantyn@google.com}

}
\end{flushleft}

\twocolumngrid

\break
\newpage
\bibliography{references}